\documentclass[5p,twocolumn,authoryear,nopreprintline]{elsarticle}
\usepackage[T1]{fontenc}
\usepackage{amsmath,amssymb}
\usepackage{newtxtext,newtxmath}
\usepackage{graphicx,booktabs,array}
\usepackage[table]{xcolor}
\usepackage{url}
\usepackage[hidelinks]{hyperref}
\hypersetup{pdftitle={Counterfactual Analysis of Extent Leakage from Annotation-Derived Prompts in Vision-Language Medical Image Segmentation}, pdfauthor={Niran Nataraj, Ankita Das, C. Krishna Mohan, Kenji Kawashima}, pdfkeywords={Evaluation methodology, Label leakage, Medical image segmentation, Shortcut learning, Vision-language models}}

\newcommand{\nAxisN}{746}
\newcommand{\nBIextCFShuf}{0.00}
\newcommand{\nBIextCFSizeBlind}{0.21}
\newcommand{\nBIextCFSizeOnly}{0.45}
\newcommand{\nBNullTF}{80.65}
\newcommand{\nBOracleOrig}{82.31}
\newcommand{\nBRcapShuf}{\ensuremath{-}0.01}
\newcommand{\nBRcapSizeBlind}{0.29}
\newcommand{\nBRcapSizeOnly}{0.66}
\newcommand{\nBRelyNoSize}{+8.43}
\newcommand{\nBRelyNoSizeHi}{+13.02}
\newcommand{\nBRelyNoSizeLo}{+4.38}
\newcommand{\nBRelySizeOnly}{+6.73}
\newcommand{\nBRelySizeOnlyHi}{+9.38}
\newcommand{\nBRelySizeOnlyLo}{+4.38}
\newcommand{\nBRelyTF}{+13.43}
\newcommand{\nBRelyTFHi}{+18.74}
\newcommand{\nBRelyTFLo}{+8.41}
\newcommand{\nBRfourEmptyEvalN}{39}
\newcommand{\nBRfourEmptyPosN}{2}
\newcommand{\nBRfourGain}{\ensuremath{-}1.13}
\newcommand{\nBRfourGainHi}{\ensuremath{-}0.10}
\newcommand{\nBRfourGainLes}{\ensuremath{-}0.33}
\newcommand{\nBRfourGainLesHi}{+0.10}
\newcommand{\nBRfourGainLesLo}{\ensuremath{-}0.86}
\newcommand{\nBRfourGainLo}{\ensuremath{-}2.52}
\newcommand{\nBRfourNormals}{\ensuremath{-}0.85}
\newcommand{\nBRfourNullDiff}{+0.54}
\newcommand{\nBRfourNullDiffHi}{+1.74}
\newcommand{\nBRfourNullDiffLo}{\ensuremath{-}0.46}
\newcommand{\nBRoneRatio}{1.3}
\newcommand{\nBRoneSpanNoSize}{0.11}
\newcommand{\nBRoneSpanOrig}{8.0}
\newcommand{\nBRthree}{49.9}
\newcommand{\nBRthreeHi}{60.7}
\newcommand{\nBRthreeLo}{37.6}
\newcommand{\nBRtwo}{39.1}
\newcommand{\nBRtwoHi}{58.3}
\newcommand{\nBRtwoLo}{21.9}
\newcommand{\nBRtwoTV}{28.5}
\newcommand{\nBRtwoTVHi}{40.7}
\newcommand{\nBRtwoTVLo}{16.1}
\newcommand{\nBRzeroGap}{\ensuremath{-}0.47}
\newcommand{\nBRzeroMean}{85.26}
\newcommand{\nBandCutHi}{17.9}
\newcommand{\nBandCutLo}{7.0}
\newcommand{\nBandLarge}{+2.68}
\newcommand{\nBandLargeHi}{+6.31}
\newcommand{\nBandLargeLo}{\ensuremath{-}0.79}
\newcommand{\nBandMid}{+4.91}
\newcommand{\nBandSmall}{+19.13}
\newcommand{\nBandSmallHi}{+21.26}
\newcommand{\nBandSmallLo}{+17.02}
\newcommand{\nBusiDiagN}{30}
\newcommand{\nBusiLesionTVN}{130}
\newcommand{\nBusiNec}{57.2}
\newcommand{\nBusiNoLesionTestN}{13}
\newcommand{\nBusiOracle}{85.73}
\newcommand{\nBusiSuff}{30.6}
\newcommand{\nBusiSuffClass}{69.7}
\newcommand{\nBusiTrainEmptyN}{107}
\newcommand{\nCalibNull}{12.3}
\newcommand{\nCalibShuf}{5.7}
\newcommand{\nCalibTrue}{6.8}
\newcommand{\nCvcCeiling}{0.24}
\newcommand{\nCvcN}{60}
\newcommand{\nCvcSmallN}{58}
\newcommand{\nDescEqOneD}{\ensuremath{-}0.19}
\newcommand{\nDescEqOneHi}{+0.12}
\newcommand{\nDescEqOneLo}{\ensuremath{-}0.58}
\newcommand{\nDescriberAll}{72.38}
\newcommand{\nDescriberAllD}{\ensuremath{-}4.84}
\newcommand{\nDescriberAllHi}{\ensuremath{-}3.80}
\newcommand{\nDescriberAllLo}{\ensuremath{-}5.91}
\newcommand{\nDescriberDG}{69.71}
\newcommand{\nEdgeCutAll}{73.84}
\newcommand{\nEdgeCutAllD}{\ensuremath{-}3.39}
\newcommand{\nEdgeCutAllHi}{\ensuremath{-}2.39}
\newcommand{\nEdgeCutAllLo}{\ensuremath{-}4.45}
\newcommand{\nEdgeCutDG}{71.41}
\newcommand{\nEdgeLotoD}{\ensuremath{-}0.26}
\newcommand{\nEdgeLotoHi}{+0.09}
\newcommand{\nEdgeLotoLo}{\ensuremath{-}0.66}
\newcommand{\nEdgeQata}{75.66}
\newcommand{\nEdgeQataBest}{76.22}
\newcommand{\nEdgeQataHigherN}{3}
\newcommand{\nGbest}{5.99}
\newcommand{\nGbestHi}{6.90}
\newcommand{\nGbestLo}{5.10}
\newcommand{\nGibOneAbs}{1.55}
\newcommand{\nIextBkaiTe}{0.28}
\newcommand{\nIextBkaiTr}{0.18}
\newcommand{\nIextBusiTe}{0.47}
\newcommand{\nIextBusiTr}{0.48}
\newcommand{\nIextCFSizeBlind}{0.16}
\newcommand{\nIextCFSizeOnly}{0.56}
\newcommand{\nIextCFTF}{0.00}
\newcommand{\nIextClinicTe}{0.48}
\newcommand{\nIextClinicTr}{0.44}
\newcommand{\nIextColonTe}{0.35}
\newcommand{\nIextCvcThreeP}{0.31}
\newcommand{\nIextCvcThreeTe}{0.02}
\newcommand{\nIextKvasirTe}{0.57}
\newcommand{\nIextKvasirTr}{0.56}
\newcommand{\nIextQataTe}{0.52}
\newcommand{\nIextQataTr}{0.50}
\newcommand{\nLocNec}{11.3}
\newcommand{\nLocNecHi}{15.1}
\newcommand{\nLocNecLo}{7.9}
\newcommand{\nNextHiInt}{77}
\newcommand{\nNextLoInt}{52}
\newcommand{\nParaFrameLossHi}{24.0}
\newcommand{\nParaFrameLossLo}{16.1}
\newcommand{\nParaFrameVsEmptyMax}{4.7}
\newcommand{\nParaMaxShift}{2.1}
\newcommand{\nParaMaxShiftTerse}{1.5}
\newcommand{\nPolypGain}{14.02}
\newcommand{\nPolypOracle}{77.23}
\newcommand{\nPolypPrior}{39.7}
\newcommand{\nPolypPriorHi}{50.8}
\newcommand{\nPolypPriorLo}{27.3}
\newcommand{\nPoudelRows}{3852}
\newcommand{\nQIextCFZoneBlind}{0.19}
\newcommand{\nQIextCFZoneOnly}{0.47}
\newcommand{\nQNullBlind}{79.17}
\newcommand{\nQNullTF}{79.20}
\newcommand{\nQPairsBlindMinAbs}{6.45}
\newcommand{\nQRcapZoneBlind}{0.47}
\newcommand{\nQRcapZoneOnly}{0.71}
\newcommand{\nQRelyBlind}{+11.28}
\newcommand{\nQRelyBlindHi}{+12.00}
\newcommand{\nQRelyBlindHierHi}{+14.59}
\newcommand{\nQRelyBlindHierLo}{+6.77}
\newcommand{\nQRelyBlindLo}{+10.55}
\newcommand{\nQRelyZoneOnly}{\ensuremath{-}0.39}
\newcommand{\nQRelyZoneOnlyHi}{\ensuremath{-}0.07}
\newcommand{\nQRelyZoneOnlyHierHi}{+4.15}
\newcommand{\nQRelyZoneOnlyHierLo}{\ensuremath{-}5.24}
\newcommand{\nQRelyZoneOnlyLo}{\ensuremath{-}0.70}
\newcommand{\nQRoneRatio}{0.0}
\newcommand{\nQRoneSpanBlind}{0.00}
\newcommand{\nQRoneSpanOrig}{7.9}
\newcommand{\nQRthree}{91.0}
\newcommand{\nQRthreeHi}{93.2}
\newcommand{\nQRthreeHierHi}{133.1}
\newcommand{\nQRthreeHierLo}{64.8}
\newcommand{\nQRthreeLo}{88.8}
\newcommand{\nQRthreePairsMet}{6}
\newcommand{\nQRtwo}{9.0}
\newcommand{\nQRtwoHi}{11.0}
\newcommand{\nQRtwoHierHi}{13.1}
\newcommand{\nQRtwoHierLo}{6.3}
\newcommand{\nQRtwoLo}{7.1}
\newcommand{\nQRzeroGap}{+0.25}
\newcommand{\nQRzeroMean}{83.66}
\newcommand{\nQTBOrig}{15.77}
\newcommand{\nQTBOrigSeedMax}{19.10}
\newcommand{\nQTBOrigSeedMin}{10.85}
\newcommand{\nQataAblLeakN}{443}
\newcommand{\nQataClfAcc}{65.1}
\newcommand{\nQataClfFeatures}{106}
\newcommand{\nQataClfHi}{69.1}
\newcommand{\nQataClfLo}{61.0}
\newcommand{\nQataCountCC}{98.4}
\newcommand{\nQataJaccard}{0.77}
\newcommand{\nQataKappaMax}{0.91}
\newcommand{\nQataKappaMin}{0.56}
\newcommand{\nQataLaterality}{92.7}
\newcommand{\nQataNec}{63.4}
\newcommand{\nQataNull}{60.24}
\newcommand{\nQataOracle}{83.41}
\newcommand{\nQataOtherNecMax}{1.8}
\newcommand{\nQataSideLeft}{4.7}
\newcommand{\nQataSideRight}{87.5}
\newcommand{\nQataTau}{10}
\newcommand{\nQataZoneAreaSpearman}{0.89}
\newcommand{\nQataZoneMajority}{32.0}
\newcommand{\nQataZoneRule}{51.5}
\newcommand{\nQataZoneRuleHi}{55.6}
\newcommand{\nQataZoneRuleLo}{47.3}
\newcommand{\nQuantN}{2633}
\newcommand{\nQuantOff}{16}
\newcommand{\nQuantOk}{2617}
\newcommand{\nQuantTol}{0.006}
\newcommand{\nRcapCFShuf}{\ensuremath{-}0.01}
\newcommand{\nRcapCFSizeBlind}{0.37}
\newcommand{\nRcapCFSizeOnly}{0.76}
\newcommand{\nRcapCFTF}{0.00}
\newcommand{\nRcapPermTeMax}{0.00}
\newcommand{\nRcapPermTrMax}{0.00}
\newcommand{\nRcapTeBkai}{0.59}
\newcommand{\nRcapTeBusi}{0.60}
\newcommand{\nRcapTeClinic}{0.54}
\newcommand{\nRcapTeColon}{0.71}
\newcommand{\nRcapTeCvcThree}{0.24}
\newcommand{\nRcapTeKvasir}{0.70}
\newcommand{\nRcapTeMaxNoCvc}{0.77}
\newcommand{\nRcapTeMinNoCvc}{0.54}
\newcommand{\nRcapTeQata}{0.77}
\newcommand{\nRcapTrBkai}{0.48}
\newcommand{\nRcapTrBusi}{0.72}
\newcommand{\nRcapTrClinic}{0.68}
\newcommand{\nRcapTrKvasir}{0.77}
\newcommand{\nRcapTrQata}{0.74}
\newcommand{\nRecDiceDel}{78.82}
\newcommand{\nRecDiceDelLvit}{76.70}
\newcommand{\nRecDiceOneZone}{75.99}
\newcommand{\nRecDiceOracle}{83.87}
\newcommand{\nRecDiceSixZones}{71.54}
\newcommand{\nRecDiceZoneOnlyLvit}{83.43}
\newcommand{\nRecNecHi}{60.2}
\newcommand{\nRecNecLo}{55.8}
\newcommand{\nRecNecReg}{43.6}
\newcommand{\nRecNecRegHi}{45.8}
\newcommand{\nRecNecRegLo}{41.3}
\newcommand{\nRecPublished}{85.22}
\newcommand{\nRecRepro}{85.21}
\newcommand{\nRecSuffHi}{89.4}
\newcommand{\nRecSuffLo}{86.1}
\newcommand{\nRelEmptyAll}{63.21}
\newcommand{\nRelEmptyAllD}{\ensuremath{-}14.02}
\newcommand{\nRelEmptyAllHi}{\ensuremath{-}12.41}
\newcommand{\nRelEmptyAllLo}{\ensuremath{-}15.63}
\newcommand{\nRelEmptyDG}{60.14}
\newcommand{\nRelLeakAll}{77.23}
\newcommand{\nRelLeakDG}{75.00}
\newcommand{\nRelyNoSize}{+2.39}
\newcommand{\nRelyNoSizeHi}{+3.16}
\newcommand{\nRelyNoSizeHierHi}{+3.67}
\newcommand{\nRelyNoSizeHierLo}{+1.12}
\newcommand{\nRelyNoSizeLo}{+1.62}
\newcommand{\nRelyShuf}{+3.14}
\newcommand{\nRelyShufHi}{+3.86}
\newcommand{\nRelyShufLo}{+2.41}
\newcommand{\nRelySizeOnly}{\ensuremath{-}0.09}
\newcommand{\nRelySizeOnlyHi}{+0.48}
\newcommand{\nRelySizeOnlyLo}{\ensuremath{-}0.66}
\newcommand{\nRelyTF}{+2.82}
\newcommand{\nRelyTFBest}{+2.49}
\newcommand{\nRelyTFBestHi}{+3.15}
\newcommand{\nRelyTFBestLo}{+1.84}
\newcommand{\nRelyTFHi}{+3.53}
\newcommand{\nRelyTFHierHi}{+4.11}
\newcommand{\nRelyTFHierLo}{+1.54}
\newcommand{\nRelyTFLo}{+2.11}
\newcommand{\nRetLeakAll}{76.00}
\newcommand{\nRetLeakAllD}{\ensuremath{-}1.22}
\newcommand{\nRetLeakAllHi}{\ensuremath{-}0.44}
\newcommand{\nRetLeakAllLo}{\ensuremath{-}2.01}
\newcommand{\nRetLeakDG}{73.65}
\newcommand{\nRfourGain}{\ensuremath{-}0.53}
\newcommand{\nRfourGainHi}{\ensuremath{-}0.28}
\newcommand{\nRfourGainLo}{\ensuremath{-}0.78}
\newcommand{\nRfourNullDiff}{+0.31}
\newcommand{\nRfourNullDiffHi}{+0.80}
\newcommand{\nRfourNullDiffLo}{\ensuremath{-}0.18}
\newcommand{\nRock}{+0.22}
\newcommand{\nRoneRatio}{1.2}
\newcommand{\nRoneSpanNoSize}{0.13}
\newcommand{\nRoneSpanOrig}{11.1}
\newcommand{\nRthree}{89.7}
\newcommand{\nRthreeBest}{100.1}
\newcommand{\nRthreeBestHi}{111.3}
\newcommand{\nRthreeBestLo}{90.3}
\newcommand{\nRthreeHi}{99.8}
\newcommand{\nRthreeHierHi}{132.9}
\newcommand{\nRthreeHierLo}{50.7}
\newcommand{\nRthreeLo}{80.4}
\newcommand{\nRthreePairsMet}{6}
\newcommand{\nRtwo}{37.4}
\newcommand{\nRtwoBest}{36.2}
\newcommand{\nRtwoBestHi}{45.6}
\newcommand{\nRtwoBestLo}{26.8}
\newcommand{\nRtwoHi}{46.8}
\newcommand{\nRtwoHierHi}{51.3}
\newcommand{\nRtwoHierLo}{26.0}
\newcommand{\nRtwoLo}{28.5}
\newcommand{\nRzeroGap}{\ensuremath{-}1.22}
\newcommand{\nRzeroGapAbs}{1.22}
\newcommand{\nRzeroMean}{76.00}
\newcommand{\nSextHiInt}{98}
\newcommand{\nSextLoInt}{84}
\newcommand{\nSizeOnlyDistinct}{19}
\newcommand{\nTBNoSize}{2.31}
\newcommand{\nTBOrig}{6.18}
\newcommand{\nTBOrigSeedMax}{7.53}
\newcommand{\nTBOrigSeedMin}{5.15}
\newcommand{\nTBSizeOnlyMinSeed}{3.1}
\newcommand{\nTabTwoBkaiEmpty}{54.55}
\newcommand{\nTabTwoBkaiG}{24.61}
\newcommand{\nTabTwoBkaiMatch}{79.16}
\newcommand{\nTabTwoBkaiMism}{73.75}
\newcommand{\nTabTwoBkaiNec}{77.2}
\newcommand{\nTabTwoBkaiPi}{78.0}
\newcommand{\nTabTwoBkaiSuff}{92.0}
\newcommand{\nTabTwoBusiEmpty}{71.73}
\newcommand{\nTabTwoBusiG}{14.00}
\newcommand{\nTabTwoBusiMatch}{85.73}
\newcommand{\nTabTwoBusiMism}{59.50}
\newcommand{\nTabTwoBusiPi}{\ensuremath{-}87.4}
\newcommand{\nTabTwoClinicEmpty}{69.92}
\newcommand{\nTabTwoClinicG}{10.87}
\newcommand{\nTabTwoClinicMatch}{80.80}
\newcommand{\nTabTwoClinicMism}{76.06}
\newcommand{\nTabTwoClinicNec}{67.4}
\newcommand{\nTabTwoClinicPi}{56.5}
\newcommand{\nTabTwoClinicSuff}{84.1}
\newcommand{\nTabTwoColonEmpty}{59.17}
\newcommand{\nTabTwoColonG}{12.72}
\newcommand{\nTabTwoColonMatch}{71.89}
\newcommand{\nTabTwoColonMism}{60.96}
\newcommand{\nTabTwoColonNec}{68.2}
\newcommand{\nTabTwoColonPi}{14.1}
\newcommand{\nTabTwoColonSuff}{98.3}
\newcommand{\nTabTwoCvcThreeEmpty}{65.32}
\newcommand{\nTabTwoCvcThreeG}{15.53}
\newcommand{\nTabTwoCvcThreeMatch}{80.84}
\newcommand{\nTabTwoCvcThreeMism}{79.49}
\newcommand{\nTabTwoCvcThreeNec}{67.2}
\newcommand{\nTabTwoCvcThreePi}{91.3}
\newcommand{\nTabTwoCvcThreeSuff}{85.9}
\newcommand{\nTabTwoKvasirEmpty}{81.04}
\newcommand{\nTabTwoKvasirG}{9.11}
\newcommand{\nTabTwoKvasirMatch}{90.15}
\newcommand{\nTabTwoKvasirMism}{81.04}
\newcommand{\nTabTwoKvasirNec}{51.8}
\newcommand{\nTabTwoKvasirPi}{\ensuremath{-}0.1}
\newcommand{\nTabTwoKvasirSuff}{87.7}
\newcommand{\nTabTwoQataEmpty}{60.24}
\newcommand{\nTabTwoQataG}{23.17}
\newcommand{\nTabTwoQataMatch}{83.41}
\newcommand{\nTabTwoQataMism}{70.67}
\newcommand{\nTabTwoQataPi}{45.0}
\newcommand{\nTabTwoQataSuff}{90.9}
\newcommand{\nTabTwoRecEmpty}{70.53}
\newcommand{\nTabTwoRecG}{14.68}
\newcommand{\nTabTwoRecMatch}{85.21}
\newcommand{\nTabTwoRecMism}{72.66}
\newcommand{\nTabTwoRecNec}{58.0}
\newcommand{\nTabTwoRecPi}{14.5}
\newcommand{\nTabTwoRecSuff}{87.8}
\newcommand{\nTextFreeAll}{72.65}
\newcommand{\nTextFreeAllD}{\ensuremath{-}4.57}
\newcommand{\nTextFreeAllHi}{\ensuremath{-}3.52}
\newcommand{\nTextFreeAllLo}{\ensuremath{-}5.67}
\newcommand{\nTextFreeDG}{70.14}
\newcommand{\nThrN}{578}
\newcommand{\nThrPerImage}{70.98}
\newcommand{\nThrShare}{44.7}
\newcommand{\nThrShareHi}{51.7}
\newcommand{\nThrShareLo}{37.7}
\newcommand{\nThrSizeOnly}{74.49}
\newcommand{\nUnseenN}{581}
\newcommand{\nXSAll}{74.59}
\newcommand{\nXSAllD}{\ensuremath{-}2.63}
\newcommand{\nXSAllHi}{\ensuremath{-}1.68}
\newcommand{\nXSAllLo}{\ensuremath{-}3.62}
\newcommand{\nXSDG}{72.26}
\newcommand{\nXSDGD}{\ensuremath{-}2.74}
\newcommand{\nXSDGHi}{\ensuremath{-}1.61}
\newcommand{\nXSDGLo}{\ensuremath{-}3.84}
\newcommand{\nXSgap}{2.63}
\newcommand{\nXSgapEst}{1.26}
\newcommand{\nXSnoAAll}{73.85}
\newcommand{\nXSnoAAllD}{\ensuremath{-}3.38}
\newcommand{\nXSnoAAllHi}{\ensuremath{-}2.35}
\newcommand{\nXSnoAAllLo}{\ensuremath{-}4.39}
\newcommand{\nXSnoADG}{71.42}
\newcommand{\nXSnoADGD}{\ensuremath{-}3.58}
\newcommand{\nXSnoADGHi}{\ensuremath{-}2.36}
\newcommand{\nXSnoADGLo}{\ensuremath{-}4.76}
\newcommand{\nXSnoEAll}{74.09}
\newcommand{\nXSnoEAllD}{\ensuremath{-}3.14}
\newcommand{\nXSnoEAllHi}{\ensuremath{-}2.10}
\newcommand{\nXSnoEAllLo}{\ensuremath{-}4.19}
\newcommand{\nXSnoEDG}{71.67}
\newcommand{\nXStrueAll}{75.85}
\newcommand{\nXStrueAllD}{\ensuremath{-}1.38}
\newcommand{\nXStrueAllHi}{\ensuremath{-}0.50}
\newcommand{\nXStrueAllLo}{\ensuremath{-}2.27}
\newcommand{\nXStrueDG}{73.64}
\newcommand{\nXStrueVsRetAll}{\ensuremath{-}0.16}
\newcommand{\nXStrueVsRetAllHi}{+0.52}
\newcommand{\nXStrueVsRetAllLo}{\ensuremath{-}0.84}
\newcommand{\nXSvsNoADG}{+0.84}
\newcommand{\nXSvsNoADGAbs}{0.84}
\newcommand{\nXSvsNoADGHi}{+1.43}
\newcommand{\nXSvsNoADGLo}{+0.23}
\newcommand{\nXSvsNoEDG}{+0.59}
\newcommand{\nXSvsNoEDGAbs}{0.59}
\newcommand{\nXSvsNoEDGHi}{+1.18}
\newcommand{\nXSvsNoEDGLo}{\ensuremath{-}0.02}
\newcommand{\nXSvsNoEDGPairs}{7}
\newcommand{\nXSvsTFAllAbs}{1.94}
\newcommand{\nXSvsTFAllHi}{+2.55}
\newcommand{\nXSvsTFAllLo}{+1.36}
\newcommand{\sBRelyNoSizeHierHi}{+17.65}
\newcommand{\sBRelyNoSizeHierLo}{+1.12}
\newcommand{\sBRelyShufHierHi}{+23.10}
\newcommand{\sBRelyShufHierLo}{+5.81}
\newcommand{\sBRelySizeOnlyHierHi}{+17.01}
\newcommand{\sBRelySizeOnlyHierLo}{\ensuremath{-}3.01}
\newcommand{\sBRelyTFHierHi}{+23.00}
\newcommand{\sBRelyTFHierLo}{+5.73}
\newcommand{\sBRthreeHierHi}{124.4}
\newcommand{\sBRthreeHierLo}{12.7}
\newcommand{\sBRthreePairsMet}{2}
\newcommand{\sBRtwoHierHi}{79.7}
\newcommand{\sBRtwoHierLo}{18.1}
\newcommand{\sBRtwoPairsMet}{6}
\newcommand{\sBkaiNoPolypN}{3}
\newcommand{\sBusiDedupG}{10.13}
\newcommand{\sBusiDedupGDelta}{\ensuremath{-}0.51}
\newcommand{\sBusiDedupGHi}{15.37}
\newcommand{\sBusiDedupGLo}{5.32}
\newcommand{\sBusiDedupN}{50}
\newcommand{\sBusiDedupNec}{67.7}
\newcommand{\sBusiDedupNecDelta}{+10.6}
\newcommand{\sBusiDedupNecHi}{138.6}
\newcommand{\sBusiDedupNecLo}{35.4}
\newcommand{\sBusiDedupP}{\ensuremath{-}0.63}
\newcommand{\sBusiDedupPDelta}{+3.27}
\newcommand{\sBusiDedupPHi}{+5.94}
\newcommand{\sBusiDedupPLo}{\ensuremath{-}7.51}
\newcommand{\sBusiDedupPi}{\ensuremath{-}6.3}
\newcommand{\sBusiDedupPiDelta}{+30.5}
\newcommand{\sBusiDedupPiHi}{54.2}
\newcommand{\sBusiDedupPiLo}{\ensuremath{-}101.1}
\newcommand{\sBusiDedupRelyNoSize}{+7.08}
\newcommand{\sBusiDedupRelyNoSizeDelta}{\ensuremath{-}1.34}
\newcommand{\sBusiDedupRelyNoSizeHi}{+12.21}
\newcommand{\sBusiDedupRelyNoSizeLo}{+2.51}
\newcommand{\sBusiDedupRelyShuf}{+12.38}
\newcommand{\sBusiDedupRelyShufDelta}{\ensuremath{-}1.19}
\newcommand{\sBusiDedupRelyShufHi}{+18.28}
\newcommand{\sBusiDedupRelyShufLo}{+6.92}
\newcommand{\sBusiDedupRelySizeOnly}{+5.61}
\newcommand{\sBusiDedupRelySizeOnlyDelta}{\ensuremath{-}1.12}
\newcommand{\sBusiDedupRelySizeOnlyHi}{+8.47}
\newcommand{\sBusiDedupRelySizeOnlyLo}{+3.03}
\newcommand{\sBusiDedupRelyTF}{+12.23}
\newcommand{\sBusiDedupRelyTFDelta}{\ensuremath{-}1.20}
\newcommand{\sBusiDedupRelyTFHi}{+18.37}
\newcommand{\sBusiDedupRelyTFLo}{+6.51}
\newcommand{\sBusiDedupRthree}{52.7}
\newcommand{\sBusiDedupRthreeDelta}{+2.8}
\newcommand{\sBusiDedupRthreeHi}{65.6}
\newcommand{\sBusiDedupRthreeLo}{37.5}
\newcommand{\sBusiDedupRtwo}{43.2}
\newcommand{\sBusiDedupRtwoDelta}{+4.1}
\newcommand{\sBusiDedupRtwoHi}{68.1}
\newcommand{\sBusiDedupRtwoLo}{21.4}
\newcommand{\sBusiDedupSuff}{20.4}
\newcommand{\sBusiDedupSuffDelta}{\ensuremath{-}10.2}
\newcommand{\sBusiDedupSuffHi}{63.2}
\newcommand{\sBusiDedupSuffLo}{\ensuremath{-}53.2}
\newcommand{\sBusiDupTest}{20}
\newcommand{\sBusiDupTestLesion}{15}
\newcommand{\sClusterDescriberAllDHi}{\ensuremath{-}2.91}
\newcommand{\sClusterDescriberAllDLo}{\ensuremath{-}7.71}
\newcommand{\sClusterEdgeCutAllDHi}{\ensuremath{-}1.73}
\newcommand{\sClusterEdgeCutAllDLo}{\ensuremath{-}4.73}
\newcommand{\sClusterFiveDescriberAllDHi}{\ensuremath{-}2.85}
\newcommand{\sClusterFiveDescriberAllDLo}{\ensuremath{-}7.97}
\newcommand{\sClusterFiveEdgeCutAllDHi}{\ensuremath{-}1.89}
\newcommand{\sClusterFiveEdgeCutAllDLo}{\ensuremath{-}4.66}
\newcommand{\sClusterFiveGHi}{16.80}
\newcommand{\sClusterFiveGLo}{11.11}
\newcommand{\sClusterFiveNClusters}{266}
\newcommand{\sClusterFiveNecHi}{77.2}
\newcommand{\sClusterFiveNecLo}{61.7}
\newcommand{\sClusterFivePHi}{+9.56}
\newcommand{\sClusterFivePLo}{\ensuremath{-}0.08}
\newcommand{\sClusterFivePiHi}{59.6}
\newcommand{\sClusterFivePiLo}{\ensuremath{-}0.7}
\newcommand{\sClusterFiveRelEmptyAllDHi}{\ensuremath{-}11.11}
\newcommand{\sClusterFiveRelEmptyAllDLo}{\ensuremath{-}16.80}
\newcommand{\sClusterFiveRelyNoSizeHi}{+3.39}
\newcommand{\sClusterFiveRelyNoSizeLo}{+0.75}
\newcommand{\sClusterFiveRelyShufHi}{+4.00}
\newcommand{\sClusterFiveRelyShufLo}{+1.98}
\newcommand{\sClusterFiveRelySizeOnlyHi}{+0.94}
\newcommand{\sClusterFiveRelySizeOnlyLo}{\ensuremath{-}1.17}
\newcommand{\sClusterFiveRelyTFHi}{+3.61}
\newcommand{\sClusterFiveRelyTFLo}{+1.67}
\newcommand{\sClusterFiveRetLeakAllDHi}{+0.26}
\newcommand{\sClusterFiveRetLeakAllDLo}{\ensuremath{-}2.51}
\newcommand{\sClusterFiveRthreeHi}{101.9}
\newcommand{\sClusterFiveRthreeLo}{79.2}
\newcommand{\sClusterFiveRtwoHi}{51.4}
\newcommand{\sClusterFiveRtwoLo}{26.7}
\newcommand{\sClusterFiveSuffHi}{100.1}
\newcommand{\sClusterFiveSuffLo}{88.6}
\newcommand{\sClusterFiveTextFreeAllDHi}{\ensuremath{-}3.08}
\newcommand{\sClusterFiveTextFreeAllDLo}{\ensuremath{-}6.64}
\newcommand{\sClusterFiveXSAllDHi}{\ensuremath{-}1.38}
\newcommand{\sClusterFiveXSAllDLo}{\ensuremath{-}4.04}
\newcommand{\sClusterFiveXSnoAAllDHi}{\ensuremath{-}2.25}
\newcommand{\sClusterFiveXSnoAAllDLo}{\ensuremath{-}4.87}
\newcommand{\sClusterFiveXSnoEAllDHi}{\ensuremath{-}1.82}
\newcommand{\sClusterFiveXSnoEAllDLo}{\ensuremath{-}5.04}
\newcommand{\sClusterFiveXStrueAllDHi}{\ensuremath{-}0.17}
\newcommand{\sClusterFiveXStrueAllDLo}{\ensuremath{-}2.46}
\newcommand{\sClusterG}{14.02}
\newcommand{\sClusterGHi}{16.57}
\newcommand{\sClusterGImgHi}{15.63}
\newcommand{\sClusterGImgLo}{12.41}
\newcommand{\sClusterGLo}{11.32}
\newcommand{\sClusterNClusters}{273}
\newcommand{\sClusterNec}{68.8}
\newcommand{\sClusterNecHi}{77.1}
\newcommand{\sClusterNecImgHi}{74.2}
\newcommand{\sClusterNecImgLo}{63.3}
\newcommand{\sClusterNecLo}{62.2}
\newcommand{\sClusterP}{+5.56}
\newcommand{\sClusterPHi}{+9.17}
\newcommand{\sClusterPImgHi}{+7.61}
\newcommand{\sClusterPImgLo}{+3.53}
\newcommand{\sClusterPLo}{+0.59}
\newcommand{\sClusterPi}{39.7}
\newcommand{\sClusterPiHi}{58.6}
\newcommand{\sClusterPiImgHi}{50.8}
\newcommand{\sClusterPiImgLo}{27.3}
\newcommand{\sClusterPiLo}{4.9}
\newcommand{\sClusterRelEmptyAllDHi}{\ensuremath{-}11.32}
\newcommand{\sClusterRelEmptyAllDLo}{\ensuremath{-}16.57}
\newcommand{\sClusterRelyNoSize}{+2.39}
\newcommand{\sClusterRelyNoSizeHi}{+3.46}
\newcommand{\sClusterRelyNoSizeImgHi}{+3.16}
\newcommand{\sClusterRelyNoSizeImgLo}{+1.62}
\newcommand{\sClusterRelyNoSizeLo}{+0.79}
\newcommand{\sClusterRelyShuf}{+3.14}
\newcommand{\sClusterRelyShufHi}{+3.96}
\newcommand{\sClusterRelyShufImgHi}{+3.86}
\newcommand{\sClusterRelyShufImgLo}{+2.41}
\newcommand{\sClusterRelyShufLo}{+2.06}
\newcommand{\sClusterRelySizeOnly}{\ensuremath{-}0.09}
\newcommand{\sClusterRelySizeOnlyHi}{+0.93}
\newcommand{\sClusterRelySizeOnlyImgHi}{+0.48}
\newcommand{\sClusterRelySizeOnlyImgLo}{\ensuremath{-}0.66}
\newcommand{\sClusterRelySizeOnlyLo}{\ensuremath{-}1.00}
\newcommand{\sClusterRelyTF}{+2.82}
\newcommand{\sClusterRelyTFHi}{+3.62}
\newcommand{\sClusterRelyTFImgHi}{+3.53}
\newcommand{\sClusterRelyTFImgLo}{+2.11}
\newcommand{\sClusterRelyTFLo}{+1.66}
\newcommand{\sClusterRetLeakAllDHi}{+0.25}
\newcommand{\sClusterRetLeakAllDLo}{\ensuremath{-}2.61}
\newcommand{\sClusterRthree}{89.7}
\newcommand{\sClusterRthreeHi}{101.8}
\newcommand{\sClusterRthreeLo}{77.9}
\newcommand{\sClusterRtwo}{37.4}
\newcommand{\sClusterRtwoHi}{51.5}
\newcommand{\sClusterRtwoImgHi}{46.8}
\newcommand{\sClusterRtwoImgLo}{28.5}
\newcommand{\sClusterRtwoLo}{26.3}
\newcommand{\sClusterSevenDescriberAllDHi}{\ensuremath{-}3.10}
\newcommand{\sClusterSevenDescriberAllDLo}{\ensuremath{-}7.33}
\newcommand{\sClusterSevenEdgeCutAllDHi}{\ensuremath{-}1.75}
\newcommand{\sClusterSevenEdgeCutAllDLo}{\ensuremath{-}4.94}
\newcommand{\sClusterSevenGHi}{16.39}
\newcommand{\sClusterSevenGLo}{11.52}
\newcommand{\sClusterSevenNClusters}{298}
\newcommand{\sClusterSevenNecHi}{75.3}
\newcommand{\sClusterSevenNecLo}{62.1}
\newcommand{\sClusterSevenPHi}{+8.92}
\newcommand{\sClusterSevenPLo}{+1.20}
\newcommand{\sClusterSevenPiHi}{58.0}
\newcommand{\sClusterSevenPiLo}{10.0}
\newcommand{\sClusterSevenRelEmptyAllDHi}{\ensuremath{-}11.52}
\newcommand{\sClusterSevenRelEmptyAllDLo}{\ensuremath{-}16.39}
\newcommand{\sClusterSevenRelyNoSizeHi}{+3.67}
\newcommand{\sClusterSevenRelyNoSizeLo}{+0.82}
\newcommand{\sClusterSevenRelyShufHi}{+4.11}
\newcommand{\sClusterSevenRelyShufLo}{+2.00}
\newcommand{\sClusterSevenRelySizeOnlyHi}{+0.97}
\newcommand{\sClusterSevenRelySizeOnlyLo}{\ensuremath{-}0.96}
\newcommand{\sClusterSevenRelyTFHi}{+3.77}
\newcommand{\sClusterSevenRelyTFLo}{+1.71}
\newcommand{\sClusterSevenRetLeakAllDHi}{+0.22}
\newcommand{\sClusterSevenRetLeakAllDLo}{\ensuremath{-}2.54}
\newcommand{\sClusterSevenRthreeHi}{101.4}
\newcommand{\sClusterSevenRthreeLo}{78.0}
\newcommand{\sClusterSevenRtwoHi}{53.1}
\newcommand{\sClusterSevenRtwoLo}{25.0}
\newcommand{\sClusterSevenSuffHi}{99.5}
\newcommand{\sClusterSevenSuffLo}{88.7}
\newcommand{\sClusterSevenTextFreeAllDHi}{\ensuremath{-}3.20}
\newcommand{\sClusterSevenTextFreeAllDLo}{\ensuremath{-}6.24}
\newcommand{\sClusterSevenXSAllDHi}{\ensuremath{-}1.45}
\newcommand{\sClusterSevenXSAllDLo}{\ensuremath{-}3.86}
\newcommand{\sClusterSevenXSnoAAllDHi}{\ensuremath{-}2.20}
\newcommand{\sClusterSevenXSnoAAllDLo}{\ensuremath{-}4.73}
\newcommand{\sClusterSevenXSnoEAllDHi}{\ensuremath{-}1.77}
\newcommand{\sClusterSevenXSnoEAllDLo}{\ensuremath{-}4.77}
\newcommand{\sClusterSevenXStrueAllDHi}{\ensuremath{-}0.26}
\newcommand{\sClusterSevenXStrueAllDLo}{\ensuremath{-}2.37}
\newcommand{\sClusterSuff}{93.5}
\newcommand{\sClusterSuffHi}{99.4}
\newcommand{\sClusterSuffImgHi}{97.0}
\newcommand{\sClusterSuffImgLo}{90.0}
\newcommand{\sClusterSuffLo}{88.9}
\newcommand{\sClusterTextFreeAllDHi}{\ensuremath{-}3.17}
\newcommand{\sClusterTextFreeAllDLo}{\ensuremath{-}6.54}
\newcommand{\sClusterXSAllDHi}{\ensuremath{-}1.46}
\newcommand{\sClusterXSAllDLo}{\ensuremath{-}3.95}
\newcommand{\sClusterXSnoAAllDHi}{\ensuremath{-}2.29}
\newcommand{\sClusterXSnoAAllDLo}{\ensuremath{-}4.89}
\newcommand{\sClusterXSnoEAllDHi}{\ensuremath{-}1.85}
\newcommand{\sClusterXSnoEAllDLo}{\ensuremath{-}4.91}
\newcommand{\sClusterXStrueAllDHi}{\ensuremath{-}0.31}
\newcommand{\sClusterXStrueAllDLo}{\ensuremath{-}2.38}
\newcommand{\sDedupDropN}{55}
\newcommand{\sDedupG}{13.92}
\newcommand{\sDedupGDG}{14.84}
\newcommand{\sDedupGDGDelta}{\ensuremath{-}0.03}
\newcommand{\sDedupGDGHi}{16.79}
\newcommand{\sDedupGDGLo}{12.90}
\newcommand{\sDedupGDelta}{\ensuremath{-}0.10}
\newcommand{\sDedupGHi}{15.69}
\newcommand{\sDedupGLo}{12.20}
\newcommand{\sDedupN}{626}
\newcommand{\sDedupNDG}{526}
\newcommand{\sDedupNec}{69.0}
\newcommand{\sDedupNecDG}{71.0}
\newcommand{\sDedupNecDGDelta}{+0.4}
\newcommand{\sDedupNecDGHi}{77.1}
\newcommand{\sDedupNecDGLo}{64.7}
\newcommand{\sDedupNecDelta}{+0.2}
\newcommand{\sDedupNecHi}{74.6}
\newcommand{\sDedupNecLo}{63.1}
\newcommand{\sDedupP}{+4.86}
\newcommand{\sDedupPDG}{+5.78}
\newcommand{\sDedupPDGDelta}{\ensuremath{-}0.74}
\newcommand{\sDedupPDGHi}{+8.25}
\newcommand{\sDedupPDGLo}{+3.30}
\newcommand{\sDedupPDelta}{\ensuremath{-}0.71}
\newcommand{\sDedupPHi}{+7.03}
\newcommand{\sDedupPLo}{+2.65}
\newcommand{\sDedupPi}{34.9}
\newcommand{\sDedupPiDG}{39.0}
\newcommand{\sDedupPiDGDelta}{\ensuremath{-}4.9}
\newcommand{\sDedupPiDGHi}{51.6}
\newcommand{\sDedupPiDGLo}{24.2}
\newcommand{\sDedupPiDelta}{\ensuremath{-}4.8}
\newcommand{\sDedupPiHi}{47.0}
\newcommand{\sDedupPiLo}{20.7}
\newcommand{\sDedupRelyNoSize}{+2.10}
\newcommand{\sDedupRelyNoSizeDG}{+2.39}
\newcommand{\sDedupRelyNoSizeDGDelta}{\ensuremath{-}0.31}
\newcommand{\sDedupRelyNoSizeDGHi}{+3.32}
\newcommand{\sDedupRelyNoSizeDGLo}{+1.45}
\newcommand{\sDedupRelyNoSizeDelta}{\ensuremath{-}0.29}
\newcommand{\sDedupRelyNoSizeHi}{+2.93}
\newcommand{\sDedupRelyNoSizeLo}{+1.27}
\newcommand{\sDedupRelyShuf}{+2.98}
\newcommand{\sDedupRelyShufDG}{+3.20}
\newcommand{\sDedupRelyShufDGDelta}{\ensuremath{-}0.17}
\newcommand{\sDedupRelyShufDGHi}{+4.12}
\newcommand{\sDedupRelyShufDGLo}{+2.32}
\newcommand{\sDedupRelyShufDelta}{\ensuremath{-}0.16}
\newcommand{\sDedupRelyShufHi}{+3.78}
\newcommand{\sDedupRelyShufLo}{+2.19}
\newcommand{\sDedupRelySizeOnly}{\ensuremath{-}0.02}
\newcommand{\sDedupRelySizeOnlyDG}{+0.07}
\newcommand{\sDedupRelySizeOnlyDGDelta}{+0.09}
\newcommand{\sDedupRelySizeOnlyDGHi}{+0.77}
\newcommand{\sDedupRelySizeOnlyDGLo}{\ensuremath{-}0.61}
\newcommand{\sDedupRelySizeOnlyDelta}{+0.07}
\newcommand{\sDedupRelySizeOnlyHi}{+0.59}
\newcommand{\sDedupRelySizeOnlyLo}{\ensuremath{-}0.61}
\newcommand{\sDedupRelyTF}{+2.68}
\newcommand{\sDedupRelyTFDG}{+2.95}
\newcommand{\sDedupRelyTFDGDelta}{\ensuremath{-}0.15}
\newcommand{\sDedupRelyTFDGHi}{+3.82}
\newcommand{\sDedupRelyTFDGLo}{+2.10}
\newcommand{\sDedupRelyTFDelta}{\ensuremath{-}0.15}
\newcommand{\sDedupRelyTFHi}{+3.47}
\newcommand{\sDedupRelyTFLo}{+1.92}
\newcommand{\sDedupRthree}{90.1}
\newcommand{\sDedupRthreeDG}{89.1}
\newcommand{\sDedupRthreeDGDelta}{+0.4}
\newcommand{\sDedupRthreeDGHi}{100.4}
\newcommand{\sDedupRthreeDGLo}{79.1}
\newcommand{\sDedupRthreeDelta}{+0.4}
\newcommand{\sDedupRthreeHi}{100.6}
\newcommand{\sDedupRthreeLo}{80.6}
\newcommand{\sDedupRtwo}{39.2}
\newcommand{\sDedupRtwoDG}{36.7}
\newcommand{\sDedupRtwoDGDelta}{+1.8}
\newcommand{\sDedupRtwoDGHi}{46.8}
\newcommand{\sDedupRtwoDGLo}{27.1}
\newcommand{\sDedupRtwoDelta}{+1.8}
\newcommand{\sDedupRtwoHi}{49.2}
\newcommand{\sDedupRtwoLo}{30.1}
\newcommand{\sDedupSuff}{94.3}
\newcommand{\sDedupSuffDG}{95.0}
\newcommand{\sDedupSuffDGDelta}{+0.9}
\newcommand{\sDedupSuffDGHi}{99.0}
\newcommand{\sDedupSuffDGLo}{91.1}
\newcommand{\sDedupSuffDelta}{+0.8}
\newcommand{\sDedupSuffHi}{97.9}
\newcommand{\sDedupSuffLo}{90.7}
\newcommand{\sDedupTabFiveMaxShift}{0.33}
\newcommand{\sDedupXSAll}{74.12}
\newcommand{\sDedupXSAllD}{\ensuremath{-}2.70}
\newcommand{\sDedupXSAllDHi}{\ensuremath{-}1.68}
\newcommand{\sDedupXSAllDLo}{\ensuremath{-}3.73}
\newcommand{\sDupThreshold}{0.98}
\newcommand{\sEmptyPredLargeSmall}{1 of 30}
\newcommand{\sEmptyPredMatched}{22}
\newcommand{\sEmptyPredMatchedColon}{21}
\newcommand{\sEmptyPredN}{681}
\newcommand{\sIextCFShuf}{0.00}
\newcommand{\sNoPolypG}{14.03}
\newcommand{\sNoPolypGDelta}{+0.01}
\newcommand{\sNoPolypGHi}{15.71}
\newcommand{\sNoPolypGLo}{12.40}
\newcommand{\sNoPolypN}{678}
\newcommand{\sNoPolypNec}{69.1}
\newcommand{\sNoPolypNecDelta}{+0.3}
\newcommand{\sNoPolypNecHi}{74.4}
\newcommand{\sNoPolypNecLo}{63.5}
\newcommand{\sNoPolypP}{+5.62}
\newcommand{\sNoPolypPDelta}{+0.06}
\newcommand{\sNoPolypPHi}{+7.65}
\newcommand{\sNoPolypPLo}{+3.54}
\newcommand{\sNoPolypPi}{40.1}
\newcommand{\sNoPolypPiDelta}{+0.4}
\newcommand{\sNoPolypPiHi}{51.0}
\newcommand{\sNoPolypPiLo}{27.4}
\newcommand{\sNoPolypSuff}{94.1}
\newcommand{\sNoPolypSuffDelta}{+0.6}
\newcommand{\sNoPolypSuffHi}{97.4}
\newcommand{\sNoPolypSuffLo}{90.7}
\newcommand{\sNsdBkaiEmpty}{55.55}
\newcommand{\sNsdBkaiG}{25.37}
\newcommand{\sNsdBkaiGHi}{30.92}
\newcommand{\sNsdBkaiGLo}{20.10}
\newcommand{\sNsdBkaiMatch}{80.92}
\newcommand{\sNsdBkaiMism}{75.41}
\newcommand{\sNsdBkaiNec}{77.5}
\newcommand{\sNsdBkaiNecHi}{87.0}
\newcommand{\sNsdBkaiNecLo}{67.5}
\newcommand{\sNsdBkaiPi}{78.3}
\newcommand{\sNsdBkaiPiHi}{90.4}
\newcommand{\sNsdBkaiPiLo}{65.7}
\newcommand{\sNsdBkaiSuff}{92.2}
\newcommand{\sNsdBkaiSuffHi}{96.9}
\newcommand{\sNsdBkaiSuffLo}{86.7}
\newcommand{\sNsdBusiEmpty}{74.33}
\newcommand{\sNsdBusiG}{14.03}
\newcommand{\sNsdBusiGHi}{20.50}
\newcommand{\sNsdBusiGLo}{8.36}
\newcommand{\sNsdBusiMatch}{88.36}
\newcommand{\sNsdBusiMism}{61.92}
\newcommand{\sNsdBusiNec}{58.8}
\newcommand{\sNsdBusiNecHi}{111.0}
\newcommand{\sNsdBusiNecLo}{33.0}
\newcommand{\sNsdBusiPi}{\ensuremath{-}88.4}
\newcommand{\sNsdBusiPiHi}{\ensuremath{-}22.1}
\newcommand{\sNsdBusiPiLo}{\ensuremath{-}208.4}
\newcommand{\sNsdBusiSuff}{30.1}
\newcommand{\sNsdBusiSuffHi}{64.6}
\newcommand{\sNsdBusiSuffLo}{\ensuremath{-}29.5}
\newcommand{\sNsdClinicEmpty}{73.20}
\newcommand{\sNsdClinicG}{11.96}
\newcommand{\sNsdClinicGHi}{18.30}
\newcommand{\sNsdClinicGLo}{6.16}
\newcommand{\sNsdClinicMatch}{85.16}
\newcommand{\sNsdClinicMism}{80.20}
\newcommand{\sNsdClinicNec}{69.7}
\newcommand{\sNsdClinicNecHi}{86.1}
\newcommand{\sNsdClinicNecLo}{43.1}
\newcommand{\sNsdClinicPi}{58.6}
\newcommand{\sNsdClinicPiHi}{91.4}
\newcommand{\sNsdClinicPiLo}{8.9}
\newcommand{\sNsdClinicSuff}{84.1}
\newcommand{\sNsdClinicSuffHi}{95.2}
\newcommand{\sNsdClinicSuffLo}{64.5}
\newcommand{\sNsdColonEmpty}{61.36}
\newcommand{\sNsdColonG}{13.59}
\newcommand{\sNsdColonGHi}{15.92}
\newcommand{\sNsdColonGLo}{11.25}
\newcommand{\sNsdColonMatch}{74.95}
\newcommand{\sNsdColonMism}{63.74}
\newcommand{\sNsdColonNec}{68.5}
\newcommand{\sNsdColonNecHi}{76.6}
\newcommand{\sNsdColonNecLo}{60.0}
\newcommand{\sNsdColonPi}{17.5}
\newcommand{\sNsdColonPiHi}{37.2}
\newcommand{\sNsdColonPiLo}{\ensuremath{-}7.3}
\newcommand{\sNsdColonSuff}{97.6}
\newcommand{\sNsdColonSuffHi}{103.3}
\newcommand{\sNsdColonSuffLo}{92.1}
\newcommand{\sNsdCvcThreeEmpty}{68.07}
\newcommand{\sNsdCvcThreeG}{16.20}
\newcommand{\sNsdCvcThreeGHi}{20.18}
\newcommand{\sNsdCvcThreeGLo}{12.43}
\newcommand{\sNsdCvcThreeMatch}{84.28}
\newcommand{\sNsdCvcThreeMism}{82.94}
\newcommand{\sNsdCvcThreeNec}{67.3}
\newcommand{\sNsdCvcThreeNecHi}{79.2}
\newcommand{\sNsdCvcThreeNecLo}{54.6}
\newcommand{\sNsdCvcThreePi}{91.8}
\newcommand{\sNsdCvcThreePiHi}{100.9}
\newcommand{\sNsdCvcThreePiLo}{81.4}
\newcommand{\sNsdCvcThreeSuff}{85.8}
\newcommand{\sNsdCvcThreeSuffHi}{92.9}
\newcommand{\sNsdCvcThreeSuffLo}{78.0}
\newcommand{\sNsdKvasirEmpty}{82.84}
\newcommand{\sNsdKvasirG}{9.49}
\newcommand{\sNsdKvasirGHi}{13.16}
\newcommand{\sNsdKvasirGLo}{5.88}
\newcommand{\sNsdKvasirMatch}{92.33}
\newcommand{\sNsdKvasirMism}{83.10}
\newcommand{\sNsdKvasirNec}{52.3}
\newcommand{\sNsdKvasirNecHi}{69.8}
\newcommand{\sNsdKvasirNecLo}{32.1}
\newcommand{\sNsdKvasirPi}{2.7}
\newcommand{\sNsdKvasirPiHi}{36.7}
\newcommand{\sNsdKvasirPiLo}{\ensuremath{-}55.0}
\newcommand{\sNsdKvasirSuff}{88.1}
\newcommand{\sNsdKvasirSuffHi}{95.3}
\newcommand{\sNsdKvasirSuffLo}{78.5}
\newcommand{\sNsdPolypDGEmpty}{62.30}
\newcommand{\sNsdPolypDGG}{15.72}
\newcommand{\sNsdPolypDGGHi}{17.67}
\newcommand{\sNsdPolypDGGLo}{13.81}
\newcommand{\sNsdPolypDGMatch}{78.01}
\newcommand{\sNsdPolypDGMism}{69.46}
\newcommand{\sNsdPolypDGNec}{71.0}
\newcommand{\sNsdPolypDGNecHi}{76.4}
\newcommand{\sNsdPolypDGNecLo}{65.4}
\newcommand{\sNsdPolypDGPi}{45.6}
\newcommand{\sNsdPolypDGPiHi}{56.6}
\newcommand{\sNsdPolypDGPiLo}{33.1}
\newcommand{\sNsdPolypDGSuff}{93.8}
\newcommand{\sNsdPolypDGSuffHi}{97.3}
\newcommand{\sNsdPolypDGSuffLo}{90.1}
\newcommand{\sNsdPolypEmpty}{65.31}
\newcommand{\sNsdPolypG}{14.80}
\newcommand{\sNsdPolypGHi}{16.50}
\newcommand{\sNsdPolypGLo}{13.10}
\newcommand{\sNsdPolypMatch}{80.12}
\newcommand{\sNsdPolypMism}{71.46}
\newcommand{\sNsdPolypNec}{69.2}
\newcommand{\sNsdPolypNecHi}{74.5}
\newcommand{\sNsdPolypNecLo}{63.7}
\newcommand{\sNsdPolypPi}{41.6}
\newcommand{\sNsdPolypPiHi}{52.3}
\newcommand{\sNsdPolypPiLo}{29.6}
\newcommand{\sNsdPolypSuff}{93.2}
\newcommand{\sNsdPolypSuffHi}{96.6}
\newcommand{\sNsdPolypSuffLo}{89.8}
\newcommand{\sNsdQataEmpty}{66.23}
\newcommand{\sNsdQataG}{22.94}
\newcommand{\sNsdQataGHi}{23.89}
\newcommand{\sNsdQataGLo}{22.00}
\newcommand{\sNsdQataMatch}{89.17}
\newcommand{\sNsdQataMism}{75.80}
\newcommand{\sNsdQataNec}{62.4}
\newcommand{\sNsdQataNecHi}{65.0}
\newcommand{\sNsdQataNecLo}{59.9}
\newcommand{\sNsdQataPi}{41.7}
\newcommand{\sNsdQataPiHi}{45.5}
\newcommand{\sNsdQataPiLo}{37.7}
\newcommand{\sNsdQataSuff}{89.5}
\newcommand{\sNsdQataSuffHi}{91.2}
\newcommand{\sNsdQataSuffLo}{87.7}
\newcommand{\sNsdRelLeakAll}{80.12}
\newcommand{\sNsdXSAll}{77.36}
\newcommand{\sParaBkaiClinical}{75.64}
\newcommand{\sParaBkaiClinicalFrame}{54.18}
\newcommand{\sParaBkaiEmpty}{54.55}
\newcommand{\sParaBkaiList}{77.12}
\newcommand{\sParaBkaiListFrame}{60.02}
\newcommand{\sParaBkaiMatch}{79.16}
\newcommand{\sParaBkaiN}{100}
\newcommand{\sParaBkaiTerse}{79.37}
\newcommand{\sParaBkaiVerbose}{79.26}
\newcommand{\sParaBusiClinical}{81.52}
\newcommand{\sParaBusiClinicalFrame}{71.91}
\newcommand{\sParaBusiEmpty}{71.73}
\newcommand{\sParaBusiList}{78.93}
\newcommand{\sParaBusiListFrame}{73.73}
\newcommand{\sParaBusiMatch}{85.73}
\newcommand{\sParaBusiN}{78}
\newcommand{\sParaBusiTerse}{85.39}
\newcommand{\sParaBusiVerbose}{85.41}
\newcommand{\sParaClinicClinical}{81.18}
\newcommand{\sParaClinicClinicalFrame}{68.73}
\newcommand{\sParaClinicEmpty}{69.92}
\newcommand{\sParaClinicList}{80.85}
\newcommand{\sParaClinicListFrame}{72.22}
\newcommand{\sParaClinicMatch}{80.80}
\newcommand{\sParaClinicN}{61}
\newcommand{\sParaClinicTerse}{80.17}
\newcommand{\sParaClinicVerbose}{80.34}
\newcommand{\sParaColonClinical}{72.81}
\newcommand{\sParaColonClinicalFrame}{57.25}
\newcommand{\sParaColonEmpty}{59.17}
\newcommand{\sParaColonList}{73.33}
\newcommand{\sParaColonListFrame}{61.98}
\newcommand{\sParaColonMatch}{71.89}
\newcommand{\sParaColonN}{360}
\newcommand{\sParaColonTerse}{71.86}
\newcommand{\sParaColonVerbose}{72.68}
\newcommand{\sParaCvcThreeClinical}{77.14}
\newcommand{\sParaCvcThreeClinicalFrame}{63.08}
\newcommand{\sParaCvcThreeEmpty}{65.32}
\newcommand{\sParaCvcThreeList}{78.27}
\newcommand{\sParaCvcThreeListFrame}{67.76}
\newcommand{\sParaCvcThreeMatch}{80.84}
\newcommand{\sParaCvcThreeN}{60}
\newcommand{\sParaCvcThreeTerse}{80.84}
\newcommand{\sParaCvcThreeVerbose}{80.95}
\newcommand{\sParaKvasirClinical}{88.97}
\newcommand{\sParaKvasirClinicalFrame}{80.52}
\newcommand{\sParaKvasirEmpty}{81.04}
\newcommand{\sParaKvasirList}{89.45}
\newcommand{\sParaKvasirListFrame}{83.73}
\newcommand{\sParaKvasirMatch}{90.15}
\newcommand{\sParaKvasirN}{100}
\newcommand{\sParaKvasirTerse}{90.04}
\newcommand{\sParaKvasirVerbose}{90.04}
\newcommand{\sParaPooledClinical}{80.81}
\newcommand{\sParaPooledClinicalD}{\ensuremath{-}1.20}
\newcommand{\sParaPooledClinicalDHi}{\ensuremath{-}0.91}
\newcommand{\sParaPooledClinicalDLo}{\ensuremath{-}1.48}
\newcommand{\sParaPooledClinicalDT}{\ensuremath{-}0.62}
\newcommand{\sParaPooledClinicalDTHi}{\ensuremath{-}0.34}
\newcommand{\sParaPooledClinicalDTLo}{\ensuremath{-}0.90}
\newcommand{\sParaPooledClinicalFrame}{65.95}
\newcommand{\sParaPooledClinicalFrameD}{\ensuremath{-}16.06}
\newcommand{\sParaPooledEmpty}{61.25}
\newcommand{\sParaPooledEmptyD}{\ensuremath{-}20.75}
\newcommand{\sParaPooledList}{79.94}
\newcommand{\sParaPooledListD}{\ensuremath{-}2.06}
\newcommand{\sParaPooledListDHi}{\ensuremath{-}1.73}
\newcommand{\sParaPooledListDLo}{\ensuremath{-}2.39}
\newcommand{\sParaPooledListDT}{\ensuremath{-}1.49}
\newcommand{\sParaPooledListDTHi}{\ensuremath{-}1.18}
\newcommand{\sParaPooledListDTLo}{\ensuremath{-}1.80}
\newcommand{\sParaPooledListFrame}{57.96}
\newcommand{\sParaPooledListFrameD}{\ensuremath{-}24.04}
\newcommand{\sParaPooledMatch}{82.01}
\newcommand{\sParaPooledN}{2872}
\newcommand{\sParaPooledTerse}{81.43}
\newcommand{\sParaPooledTerseD}{\ensuremath{-}0.58}
\newcommand{\sParaPooledTerseDHi}{\ensuremath{-}0.45}
\newcommand{\sParaPooledTerseDLo}{\ensuremath{-}0.70}
\newcommand{\sParaPooledVerbose}{81.42}
\newcommand{\sParaPooledVerboseD}{\ensuremath{-}0.59}
\newcommand{\sParaPooledVerboseDHi}{\ensuremath{-}0.35}
\newcommand{\sParaPooledVerboseDLo}{\ensuremath{-}0.82}
\newcommand{\sParaPooledVerboseDT}{\ensuremath{-}0.01}
\newcommand{\sParaPooledVerboseDTHi}{+0.25}
\newcommand{\sParaPooledVerboseDTLo}{\ensuremath{-}0.26}
\newcommand{\sParaQataClinical}{82.10}
\newcommand{\sParaQataClinicalFrame}{67.08}
\newcommand{\sParaQataEmpty}{60.24}
\newcommand{\sParaQataList}{80.81}
\newcommand{\sParaQataListFrame}{54.69}
\newcommand{\sParaQataMatch}{83.41}
\newcommand{\sParaQataN}{2113}
\newcommand{\sParaQataTerse}{82.65}
\newcommand{\sParaQataVerbose}{82.50}
\newcommand{\sParaSetMaxLoss}{6.8}
\newcommand{\sPolypDupTestN}{681}
\newcommand{\sPolypDupTotal}{1}
\newcommand{\sQClusterPiHi}{49.4}
\newcommand{\sQClusterPiLo}{40.2}
\newcommand{\sQClusterRelyBlindHi}{+12.24}
\newcommand{\sQClusterRelyBlindLo}{+10.31}
\newcommand{\sQClusterRelyZoneOnlyHiThree}{+0.002}
\newcommand{\sQClusterRelyZoneOnlyLo}{\ensuremath{-}0.79}
\newcommand{\sQClusterRthreeHi}{93.7}
\newcommand{\sQClusterRthreeLo}{88.5}
\newcommand{\sQClusterRtwoHi}{11.4}
\newcommand{\sQClusterRtwoLo}{6.8}
\newcommand{\sQRecClusterPiHi}{22.8}
\newcommand{\sQRecClusterPiLo}{5.1}
\newcommand{\sQRecClusterSuffHi}{90.1}
\newcommand{\sQRecClusterSuffLo}{85.3}
\newcommand{\sQSubjImagesN}{2113}
\newcommand{\sQSubjN}{474}
\newcommand{\sQTrainNoIdN}{2354}
\newcommand{\sQValNoIdN}{597}
\newcommand{\sRecDistinctCaptions}{191}
\newcommand{\sRecGSameRegister}{13.34}
\newcommand{\sRecIouEmpty}{57.49}
\newcommand{\sRecIouG}{19.50}
\newcommand{\sRecIouGHi}{20.33}
\newcommand{\sRecIouGLo}{18.64}
\newcommand{\sRecIouMatch}{77.00}
\newcommand{\sRecIouMism}{62.03}
\newcommand{\sRecIouNec}{59.2}
\newcommand{\sRecIouNecHi}{61.3}
\newcommand{\sRecIouNecLo}{57.2}
\newcommand{\sRecIouPi}{23.3}
\newcommand{\sRecIouPiHi}{27.6}
\newcommand{\sRecIouPiLo}{18.6}
\newcommand{\sRecIouSuff}{87.1}
\newcommand{\sRecIouSuffHi}{88.6}
\newcommand{\sRecIouSuffLo}{85.5}
\newcommand{\sRecPiMixedExclSameText}{0.0}
\newcommand{\sRecPiMixedHi}{19.8}
\newcommand{\sRecPiMixedLo}{8.9}
\newcommand{\sRecPiSameRegister}{15.9}
\newcommand{\sRecPiSameRegisterHi}{21.7}
\newcommand{\sRecPiSameRegisterLo}{9.9}
\newcommand{\sRecSameTextFrac}{10.9}
\newcommand{\sRecSameTextN}{230}
\newcommand{\sRuleBkaiAgree}{62.96}
\newcommand{\sRuleBkaiBest}{81.59}
\newcommand{\sRuleBkaiCompact}{55.25}
\newcommand{\sRuleBkaiContrast}{75.86}
\newcommand{\sRuleBkaiDrop}{77.62}
\newcommand{\sRuleBkaiEmpty}{54.55}
\newcommand{\sRuleBkaiGDrop}{77.76}
\newcommand{\sRuleBkaiGrad}{77.19}
\newcommand{\sRuleBkaiGradComp}{75.78}
\newcommand{\sRuleBkaiGradImg}{52.66}
\newcommand{\sRuleBkaiGradWin}{65.77}
\newcommand{\sRuleBkaiGtAreaCut}{79.96}
\newcommand{\sRuleBkaiHalf}{62.98}
\newcommand{\sRuleBkaiMatch}{79.16}
\newcommand{\sRuleBkaiOtsu}{51.08}
\newcommand{\sRuleBkaiPlateau}{5.08}
\newcommand{\sRuleBusiAgree}{62.92}
\newcommand{\sRuleBusiBest}{86.40}
\newcommand{\sRuleBusiCompact}{36.47}
\newcommand{\sRuleBusiContrast}{60.21}
\newcommand{\sRuleBusiDrop}{62.84}
\newcommand{\sRuleBusiEmpty}{71.73}
\newcommand{\sRuleBusiGDrop}{63.66}
\newcommand{\sRuleBusiGrad}{63.95}
\newcommand{\sRuleBusiGradComp}{65.81}
\newcommand{\sRuleBusiGradImg}{57.01}
\newcommand{\sRuleBusiGradWin}{62.56}
\newcommand{\sRuleBusiGtAreaCut}{85.30}
\newcommand{\sRuleBusiHalf}{63.06}
\newcommand{\sRuleBusiMatch}{85.73}
\newcommand{\sRuleBusiOtsu}{62.68}
\newcommand{\sRuleBusiPlateau}{4.07}
\newcommand{\sRuleClinicAgree}{75.13}
\newcommand{\sRuleClinicBest}{87.12}
\newcommand{\sRuleClinicCompact}{62.55}
\newcommand{\sRuleClinicContrast}{69.73}
\newcommand{\sRuleClinicDrop}{74.44}
\newcommand{\sRuleClinicEmpty}{69.92}
\newcommand{\sRuleClinicGDrop}{76.28}
\newcommand{\sRuleClinicGrad}{77.64}
\newcommand{\sRuleClinicGradComp}{75.91}
\newcommand{\sRuleClinicGradImg}{56.98}
\newcommand{\sRuleClinicGradWin}{75.94}
\newcommand{\sRuleClinicGtAreaCut}{85.83}
\newcommand{\sRuleClinicHalf}{75.79}
\newcommand{\sRuleClinicMatch}{80.80}
\newcommand{\sRuleClinicOtsu}{70.55}
\newcommand{\sRuleClinicPlateau}{4.44}
\newcommand{\sRuleColonAgree}{65.02}
\newcommand{\sRuleColonBest}{75.86}
\newcommand{\sRuleColonCompact}{60.94}
\newcommand{\sRuleColonContrast}{60.08}
\newcommand{\sRuleColonDrop}{63.25}
\newcommand{\sRuleColonEmpty}{59.17}
\newcommand{\sRuleColonGDrop}{65.05}
\newcommand{\sRuleColonGrad}{68.27}
\newcommand{\sRuleColonGradComp}{68.60}
\newcommand{\sRuleColonGradImg}{43.15}
\newcommand{\sRuleColonGradWin}{66.94}
\newcommand{\sRuleColonGtAreaCut}{73.75}
\newcommand{\sRuleColonHalf}{64.84}
\newcommand{\sRuleColonMatch}{71.89}
\newcommand{\sRuleColonOtsu}{58.25}
\newcommand{\sRuleColonPlateau}{7.07}
\newcommand{\sRuleCvcThreeAgree}{70.88}
\newcommand{\sRuleCvcThreeBest}{83.90}
\newcommand{\sRuleCvcThreeCompact}{68.95}
\newcommand{\sRuleCvcThreeContrast}{78.92}
\newcommand{\sRuleCvcThreeDrop}{77.42}
\newcommand{\sRuleCvcThreeEmpty}{65.32}
\newcommand{\sRuleCvcThreeGDrop}{76.06}
\newcommand{\sRuleCvcThreeGrad}{74.30}
\newcommand{\sRuleCvcThreeGradComp}{74.97}
\newcommand{\sRuleCvcThreeGradImg}{37.66}
\newcommand{\sRuleCvcThreeGradWin}{73.17}
\newcommand{\sRuleCvcThreeGtAreaCut}{82.47}
\newcommand{\sRuleCvcThreeHalf}{71.01}
\newcommand{\sRuleCvcThreeMatch}{80.84}
\newcommand{\sRuleCvcThreeOtsu}{65.94}
\newcommand{\sRuleCvcThreePlateau}{6.35}
\newcommand{\sRuleKvasirAgree}{85.11}
\newcommand{\sRuleKvasirBest}{91.74}
\newcommand{\sRuleKvasirCompact}{73.26}
\newcommand{\sRuleKvasirContrast}{79.03}
\newcommand{\sRuleKvasirDrop}{82.42}
\newcommand{\sRuleKvasirEmpty}{81.04}
\newcommand{\sRuleKvasirGDrop}{85.11}
\newcommand{\sRuleKvasirGrad}{87.93}
\newcommand{\sRuleKvasirGradComp}{88.21}
\newcommand{\sRuleKvasirGradImg}{75.67}
\newcommand{\sRuleKvasirGradWin}{86.38}
\newcommand{\sRuleKvasirGtAreaCut}{91.20}
\newcommand{\sRuleKvasirHalf}{85.03}
\newcommand{\sRuleKvasirMatch}{90.15}
\newcommand{\sRuleKvasirOtsu}{82.28}
\newcommand{\sRuleKvasirPlateau}{2.67}
\newcommand{\sRulePolypAgreeAll}{69.09}
\newcommand{\sRulePolypAgreeDG}{66.33}
\newcommand{\sRulePolypBestAll}{80.75}
\newcommand{\sRulePolypBestDG}{78.86}
\newcommand{\sRulePolypCompactAll}{62.77}
\newcommand{\sRulePolypCompactDG}{60.96}
\newcommand{\sRulePolypContrastAll}{67.71}
\newcommand{\sRulePolypContrastDG}{65.76}
\newcommand{\sRulePolypDropAll}{70.42}
\newcommand{\sRulePolypDropDG}{68.36}
\newcommand{\sRulePolypEmptyAll}{63.21}
\newcommand{\sRulePolypEmptyDG}{60.14}
\newcommand{\sRulePolypGDropAll}{71.84}
\newcommand{\sRulePolypGDropDG}{69.56}
\newcommand{\sRulePolypGradAll}{73.84}
\newcommand{\sRulePolypGradCompAll}{73.75}
\newcommand{\sRulePolypGradCompDG}{71.26}
\newcommand{\sRulePolypGradDG}{71.41}
\newcommand{\sRulePolypGradImgAll}{50.07}
\newcommand{\sRulePolypGradImgDG}{45.67}
\newcommand{\sRulePolypGradWinAll}{70.98}
\newcommand{\sRulePolypGradWinDG}{68.33}
\newcommand{\sRulePolypGtAreaCutAll}{79.08}
\newcommand{\sRulePolypGtAreaCutDG}{76.99}
\newcommand{\sRulePolypHalfAll}{69.06}
\newcommand{\sRulePolypHalfDG}{66.31}
\newcommand{\sRulePolypMatchDG}{75.00}
\newcommand{\sRulePolypOtsuAll}{62.50}
\newcommand{\sRulePolypOtsuDG}{59.10}
\newcommand{\sRulePolypPlateauAll}{5.83}
\newcommand{\sRulePolypPlateauDG}{6.38}
\newcommand{\sRuleQataAgree}{76.13}
\newcommand{\sRuleQataBest}{83.72}
\newcommand{\sRuleQataDrop}{75.05}
\newcommand{\sRuleQataGDrop}{75.40}
\newcommand{\sRuleQataGrad}{75.66}
\newcommand{\sRuleQataGtAreaCut}{81.93}
\newcommand{\sRuleQataHalf}{76.04}
\newcommand{\sRuleQataMatch}{83.41}
\newcommand{\sRuleQataOtsu}{76.22}
\newcommand{\sRuleQataPlateau}{5.29}
\newcommand{\sSameTextBusiPiExcl}{\ensuremath{-}106.6}
\newcommand{\sSameTextColonMaxShift}{3.1}
\newcommand{\sSameTextColonN}{16}
\newcommand{\sSameTextPolypPiExcl}{36.6}
\newcommand{\sSameTextPolypPiExclHi}{48.2}
\newcommand{\sSameTextPolypPiExclLo}{23.1}
\newcommand{\sSameTextQataPiExcl}{36.2}
\newcommand{\sSeedBusiNoSizeLesEmptyOne}{76.05}
\newcommand{\sSeedBusiNoSizeLesEmptyThree}{75.97}
\newcommand{\sSeedBusiNoSizeLesEmptyTwo}{74.92}
\newcommand{\sSeedBusiNoSizeLesGOne}{4.73}
\newcommand{\sSeedBusiNoSizeLesGThree}{5.80}
\newcommand{\sSeedBusiNoSizeLesGTwo}{7.16}
\newcommand{\sSeedBusiNoSizeLesMatchOne}{80.78}
\newcommand{\sSeedBusiNoSizeLesMatchThree}{81.76}
\newcommand{\sSeedBusiNoSizeLesMatchTwo}{82.09}
\newcommand{\sSeedBusiOrigLesEmptyOne}{74.20}
\newcommand{\sSeedBusiOrigLesEmptyThree}{59.71}
\newcommand{\sSeedBusiOrigLesEmptyTwo}{67.75}
\newcommand{\sSeedBusiOrigLesGOne}{7.97}
\newcommand{\sSeedBusiOrigLesGThree}{22.23}
\newcommand{\sSeedBusiOrigLesGTwo}{15.08}
\newcommand{\sSeedBusiOrigLesMatchOne}{82.17}
\newcommand{\sSeedBusiOrigLesMatchThree}{81.94}
\newcommand{\sSeedBusiOrigLesMatchTwo}{82.83}
\newcommand{\sSeedBusiShufLesEmptyOne}{81.71}
\newcommand{\sSeedBusiShufLesEmptyThree}{80.09}
\newcommand{\sSeedBusiShufLesEmptyTwo}{80.57}
\newcommand{\sSeedBusiShufLesGOne}{\ensuremath{-}0.31}
\newcommand{\sSeedBusiShufLesGThree}{\ensuremath{-}0.83}
\newcommand{\sSeedBusiShufLesGTwo}{0.16}
\newcommand{\sSeedBusiShufLesMatchOne}{81.39}
\newcommand{\sSeedBusiShufLesMatchThree}{79.26}
\newcommand{\sSeedBusiShufLesMatchTwo}{80.73}
\newcommand{\sSeedBusiSizeOnlyLesEmptyOne}{77.67}
\newcommand{\sSeedBusiSizeOnlyLesEmptyThree}{66.13}
\newcommand{\sSeedBusiSizeOnlyLesEmptyTwo}{78.05}
\newcommand{\sSeedBusiSizeOnlyLesGOne}{2.70}
\newcommand{\sSeedBusiSizeOnlyLesGThree}{15.56}
\newcommand{\sSeedBusiSizeOnlyLesGTwo}{4.33}
\newcommand{\sSeedBusiSizeOnlyLesMatchOne}{80.37}
\newcommand{\sSeedBusiSizeOnlyLesMatchThree}{81.69}
\newcommand{\sSeedBusiSizeOnlyLesMatchTwo}{82.38}
\newcommand{\sSeedBusiTFLesEmptyOne}{81.04}
\newcommand{\sSeedBusiTFLesEmptyThree}{80.48}
\newcommand{\sSeedBusiTFLesEmptyTwo}{80.43}
\newcommand{\sSeedBusiTFLesGOne}{0.00}
\newcommand{\sSeedBusiTFLesGThree}{0.08}
\newcommand{\sSeedBusiTFLesGTwo}{\ensuremath{-}0.06}
\newcommand{\sSeedBusiTFLesMatchOne}{81.04}
\newcommand{\sSeedBusiTFLesMatchThree}{80.56}
\newcommand{\sSeedBusiTFLesMatchTwo}{80.36}
\newcommand{\sSeedKvasirNoSizeAllEmptyOne}{72.18}
\newcommand{\sSeedKvasirNoSizeAllEmptyThree}{72.67}
\newcommand{\sSeedKvasirNoSizeAllEmptyTwo}{71.79}
\newcommand{\sSeedKvasirNoSizeAllGOne}{2.36}
\newcommand{\sSeedKvasirNoSizeAllGThree}{2.21}
\newcommand{\sSeedKvasirNoSizeAllGTwo}{2.37}
\newcommand{\sSeedKvasirNoSizeAllMatchOne}{74.54}
\newcommand{\sSeedKvasirNoSizeAllMatchThree}{74.88}
\newcommand{\sSeedKvasirNoSizeAllMatchTwo}{74.16}
\newcommand{\sSeedKvasirOrigAllEmptyOne}{69.54}
\newcommand{\sSeedKvasirOrigAllEmptyThree}{70.74}
\newcommand{\sSeedKvasirOrigAllEmptyTwo}{69.20}
\newcommand{\sSeedKvasirOrigAllGOne}{7.53}
\newcommand{\sSeedKvasirOrigAllGThree}{5.15}
\newcommand{\sSeedKvasirOrigAllGTwo}{5.84}
\newcommand{\sSeedKvasirOrigAllMatchOne}{77.07}
\newcommand{\sSeedKvasirOrigAllMatchThree}{75.90}
\newcommand{\sSeedKvasirOrigAllMatchTwo}{75.04}
\newcommand{\sSeedKvasirShufAllEmptyOne}{72.86}
\newcommand{\sSeedKvasirShufAllEmptyThree}{72.90}
\newcommand{\sSeedKvasirShufAllEmptyTwo}{73.13}
\newcommand{\sSeedKvasirShufAllGOne}{\ensuremath{-}0.44}
\newcommand{\sSeedKvasirShufAllGThree}{\ensuremath{-}0.84}
\newcommand{\sSeedKvasirShufAllGTwo}{\ensuremath{-}0.30}
\newcommand{\sSeedKvasirShufAllMatchOne}{72.42}
\newcommand{\sSeedKvasirShufAllMatchThree}{72.06}
\newcommand{\sSeedKvasirShufAllMatchTwo}{72.83}
\newcommand{\sSeedKvasirSizeOnlyAllEmptyOne}{69.79}
\newcommand{\sSeedKvasirSizeOnlyAllEmptyThree}{70.49}
\newcommand{\sSeedKvasirSizeOnlyAllEmptyTwo}{68.94}
\newcommand{\sSeedKvasirSizeOnlyAllGOne}{6.09}
\newcommand{\sSeedKvasirSizeOnlyAllGThree}{3.06}
\newcommand{\sSeedKvasirSizeOnlyAllGTwo}{7.47}
\newcommand{\sSeedKvasirSizeOnlyAllMatchOne}{75.88}
\newcommand{\sSeedKvasirSizeOnlyAllMatchThree}{73.54}
\newcommand{\sSeedKvasirSizeOnlyAllMatchTwo}{76.41}
\newcommand{\sSeedKvasirTFAllEmptyOne}{73.36}
\newcommand{\sSeedKvasirTFAllEmptyThree}{72.33}
\newcommand{\sSeedKvasirTFAllEmptyTwo}{72.27}
\newcommand{\sSeedKvasirTFAllGOne}{\ensuremath{-}0.03}
\newcommand{\sSeedKvasirTFAllGThree}{\ensuremath{-}0.12}
\newcommand{\sSeedKvasirTFAllGTwo}{0.01}
\newcommand{\sSeedKvasirTFAllMatchOne}{73.32}
\newcommand{\sSeedKvasirTFAllMatchThree}{72.21}
\newcommand{\sSeedKvasirTFAllMatchTwo}{72.28}
\newcommand{\sSeedQataBlindEmptyOne}{79.06}
\newcommand{\sSeedQataBlindEmptyThree}{79.38}
\newcommand{\sSeedQataBlindEmptyTwo}{79.06}
\newcommand{\sSeedQataBlindGOne}{1.43}
\newcommand{\sSeedQataBlindGThree}{1.27}
\newcommand{\sSeedQataBlindGTwo}{1.55}
\newcommand{\sSeedQataBlindMatchOne}{80.50}
\newcommand{\sSeedQataBlindMatchThree}{80.65}
\newcommand{\sSeedQataBlindMatchTwo}{80.62}
\newcommand{\sSeedQataOnlyEmptyOne}{69.82}
\newcommand{\sSeedQataOnlyEmptyThree}{68.89}
\newcommand{\sSeedQataOnlyEmptyTwo}{63.80}
\newcommand{\sSeedQataOnlyGOne}{11.16}
\newcommand{\sSeedQataOnlyGThree}{13.26}
\newcommand{\sSeedQataOnlyGTwo}{18.65}
\newcommand{\sSeedQataOnlyMatchOne}{80.98}
\newcommand{\sSeedQataOnlyMatchThree}{82.15}
\newcommand{\sSeedQataOnlyMatchTwo}{82.45}
\newcommand{\sSeedQataOrigEmptyOne}{66.42}
\newcommand{\sSeedQataOrigEmptyThree}{64.63}
\newcommand{\sSeedQataOrigEmptyTwo}{72.61}
\newcommand{\sSeedQataOrigGOne}{17.36}
\newcommand{\sSeedQataOrigGThree}{19.10}
\newcommand{\sSeedQataOrigGTwo}{10.85}
\newcommand{\sSeedQataOrigMatchOne}{83.78}
\newcommand{\sSeedQataOrigMatchThree}{83.73}
\newcommand{\sSeedQataOrigMatchTwo}{83.46}
\newcommand{\sSeedQataTFEmptyOne}{79.17}
\newcommand{\sSeedQataTFEmptyThree}{79.04}
\newcommand{\sSeedQataTFEmptyTwo}{79.39}
\newcommand{\sSeedQataTFGOne}{0.01}
\newcommand{\sSeedQataTFGThree}{0.01}
\newcommand{\sSeedQataTFGTwo}{0.01}
\newcommand{\sSeedQataTFMatchOne}{79.18}
\newcommand{\sSeedQataTFMatchThree}{79.05}
\newcommand{\sSeedQataTFMatchTwo}{79.40}
\newcommand{\sSeedXSAllPredOne}{75.07}
\newcommand{\sSeedXSAllPredThree}{74.59}
\newcommand{\sSeedXSAllPredTwo}{74.11}
\newcommand{\sSeedXSAllTrueOne}{76.07}
\newcommand{\sSeedXSAllTrueThree}{76.07}
\newcommand{\sSeedXSAllTrueTwo}{75.39}
\newcommand{\sSeedXSnoAAllPredOne}{73.92}
\newcommand{\sSeedXSnoAAllPredThree}{73.83}
\newcommand{\sSeedXSnoAAllPredTwo}{73.79}
\newcommand{\sSeedXSnoEAllPredOne}{73.85}
\newcommand{\sSeedXSnoEAllPredThree}{74.26}
\newcommand{\sSeedXSnoEAllPredTwo}{74.15}
\newcommand{\sSizeBusiHMin}{316}
\newcommand{\sSizeBusiWMin}{319}
\newcommand{\sSizeKvasirHMax}{1070}
\newcommand{\sSizeKvasirWMax}{1348}
\newcommand{\sSplitBkaiTest}{100}
\newcommand{\sSplitBkaiTotal}{999}
\newcommand{\sSplitBkaiTrain}{799}
\newcommand{\sSplitBkaiVal}{100}
\newcommand{\sSplitBusiTest}{78}
\newcommand{\sSplitBusiTotal}{780}
\newcommand{\sSplitBusiTrain}{624}
\newcommand{\sSplitBusiVal}{78}
\newcommand{\sSplitClinicTest}{61}
\newcommand{\sSplitClinicTotal}{612}
\newcommand{\sSplitClinicTrain}{490}
\newcommand{\sSplitClinicVal}{61}
\newcommand{\sSplitColonTest}{360}
\newcommand{\sSplitColonTotal}{380}
\newcommand{\sSplitColonTrain}{20}
\newcommand{\sSplitCvcThreeTest}{60}
\newcommand{\sSplitCvcThreeTotal}{66}
\newcommand{\sSplitCvcThreeTrain}{6}
\newcommand{\sSplitKvasirTest}{100}
\newcommand{\sSplitKvasirTotal}{1000}
\newcommand{\sSplitKvasirTrain}{800}
\newcommand{\sSplitKvasirVal}{100}
\newcommand{\sSplitQataTest}{2113}
\newcommand{\sSplitQataTotal}{9258}
\newcommand{\sSplitQataTrain}{5716}
\newcommand{\sSplitQataVal}{1429}
\newcommand{\sXSBinZeroN}{40}
\newcommand{\sXSBinZeroTrainN}{800}
\newcommand{\nw}[1]{\ifcase#1 zero\or one\or two\or three\or four\or five\or six\or seven\or eight\or nine\else #1\fi}

\definecolor{leak}{HTML}{B8332A}
\definecolor{img}{HTML}{2B6A9E}
\definecolor{okgreen}{HTML}{00795A}
\newcommand{\metyes}{\textcolor{okgreen}{\textbf{yes}}}
\newcommand{\metno}{no}
\newcommand{\metpart}{consistent}

\newcommand{\tstar}{t^{\star}}
\newcommand{\tnull}{t_{\varnothing}}
\newcommand{\D}{\mathrm{D}}
\newcommand{\E}{\mathbb{E}}
\newcommand{\capt}[1]{\textit{``#1''}}
\newcommand{\ci}[2]{[#1,\,#2]}



\begin{document}

\begin{frontmatter}

\title{Counterfactual Analysis of Extent Leakage from Annotation-Derived Prompts in Vision-Language Medical Image Segmentation}

\author[ut]{Niran Nataraj\corref{cor1}}
\ead{niran@g.ecc.u-tokyo.ac.jp}
\author[iith]{Ankita Das}
\ead{cs23resch01003@iith.ac.in}
\author[iith]{C. Krishna Mohan}
\ead{ckm@cse.iith.ac.in}
\author[ut]{Kenji Kawashima}
\ead{kkawa729@g.ecc.u-tokyo.ac.jp}
\cortext[cor1]{Corresponding author.}

\affiliation[ut]{organization={Department of Information Physics and Computing, Graduate School of Information Science and Technology, The University of Tokyo},
            addressline={7-3-1 Hongo, Bunkyo-ku},
            city={Tokyo},
            postcode={113-8656},
            country={Japan}}
\affiliation[iith]{organization={Department of Computer Science and Engineering, Indian Institute of Technology Hyderabad},
            addressline={Kandi},
            city={Sangareddy},
            state={Telangana},
            postcode={502284},
            country={India}}

\begin{abstract}
Text-guided segmentation models receive an image and a short caption about the lesion, and on public benchmarks the caption improves accuracy considerably. On colonoscopy and breast ultrasound, however, the size, count and location words of each caption were computed from the mask that later scores the prediction, and the chest X-ray captions were written by benchmark annotators. We tested what two published models take from them. On colonoscopy and chest X-ray, it is mostly the lesion's extent. Keeping only the size word or the lung-zone phrase recovers \nSextLoInt{} to \nSextHiInt{} percent of the improvement over an empty caption, and removing it loses \nNextLoInt{} to \nNextHiInt{} percent. The rest matters little, and a polyp called a rock is segmented just as well. We call this extent leakage. On breast ultrasound, nearly half the captions also name the diagnosis, which carries part of the benefit. Retraining on three datasets shows that models pick up this habit from their training captions, though only partly on breast ultrasound, and that it weakens their segmentation from the image alone. The leak can be measured without training a model, since caption words alone predict lesion area (coefficient of determination \nRcapTeMinNoCvc{} to \nRcapTeMaxNoCvc{} on six of seven test sets). Without the test mask, the best method we tested, which estimates extent from the image, reaches \nXSAll{} Dice on colonoscopy against the published \nRelLeakAll{}. We propose an evaluation protocol that keeps what a caption says about the image apart from what it copies from the label.
\end{abstract}


\end{frontmatter}

\begin{figure}[t]
\centering
\includegraphics[width=\columnwidth]{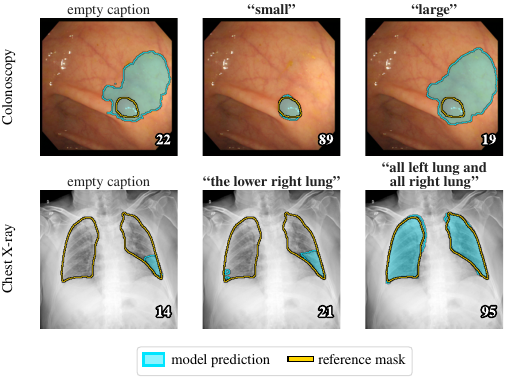}
\caption{A published text-guided segmenter reads the lesion's extent from its caption. Each image is shown with the empty caption and with its own caption in which only the extent has been changed. Top: the size word of a colonoscopy caption, which is computed from the mask. Bottom: the lung-zone phrase of a chest X-ray caption, which was written by annotators. Numbers are Dice scores.}
\label{fig:teaser}
\end{figure}

\begin{figure*}[t]
\centering
\includegraphics[width=\textwidth]{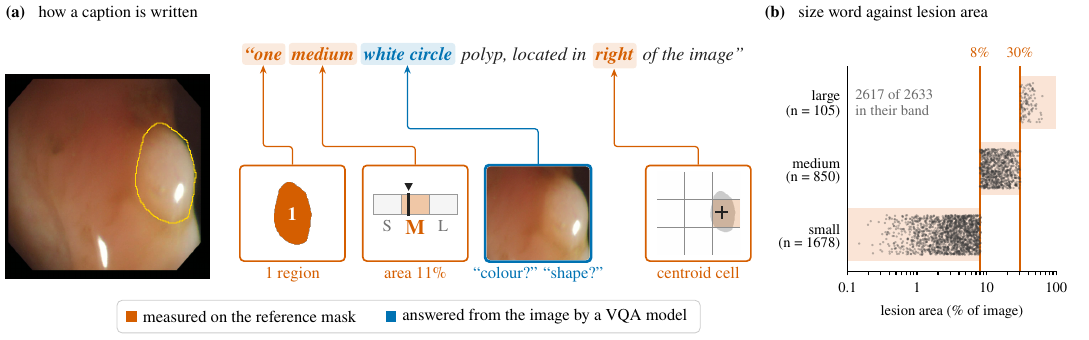}
\caption{How a benchmark caption is written. (a) A CVC-ColonDB test image and its caption. A template fills in the count, size and location words from the reference mask, and the colour and shape words from visual-question-answering (VQA) queries on the image \citep{poudel2024}. (b) The size word of each single-lesion caption of Section~\ref{sec:quantiser} against the true lesion area. Shading marks the band that the 8\% and 30\% thresholds of Eq.~\eqref{eq:quantiser} give each word; almost every caption lies in its band.}
\label{fig:generation}
\end{figure*}

\begin{figure*}[t]
\centering
\includegraphics[width=\textwidth]{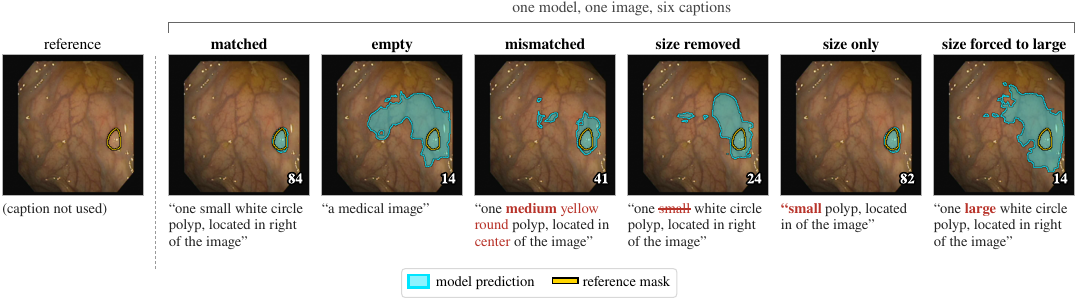}
\caption{The caption edits of Section~\ref{sec:m-interv} on one CVC-ClinicDB test image, with the published MedCLIPSeg checkpoint. Words that differ from the matched caption are in red; a struck-through word has been deleted. \emph{Size removed}, \emph{size only} and \emph{size forced to large} are the ablated, isolated and forced captions for the size word. Numbers are Dice scores.}
\label{fig:protocol}
\end{figure*}

\section{Introduction}

Tell a published text-guided segmenter that the polyp in a colonoscopy frame is \emph{small}, and it outlines a small polyp almost exactly. Tell it that the same polyp is \emph{large}, and it buries the polyp in false positives (Fig.~\ref{fig:teaser}). A model that follows its prompt is exactly what text-guided segmentation sets out to build \citep{reclmis,medclipsamv2,medclipseg,lvit,guidedecoder}. Alongside the image, the model receives a short text prompt describing the target, such as the number of lesions, their size and their position; we call such prompts \emph{captions}. Captions provide information about the target and guide the model toward accurate segmentation. Reported results suggest that text can substantially influence segmentation. MedCLIPSeg, for example, reports that an under-specified caption lowers its overall Dice score by 26.9 points \citep{medclipseg}. On its published checkpoints, we find that an empty caption costs \nTabTwoKvasirG{} to \nTabTwoBkaiG{} points, depending on the test set.

\emph{Where do these captions come from?} For chest X-ray, the authors of LViT \citep{lvit} annotated the QaTa-COV19 dataset \citep{degerli2021qata,qata} with the affected side of the chest, the number of affected regions and the lung zones involved. For colonoscopy and breast ultrasound, the captions released with MedCLIPSeg are, image for image, the prompts that \citet{poudel2024} generated with a fixed template. The template computes the count, size and location words from the reference mask (Fig.~\ref{fig:generation}). A typical caption reads \capt{one small pink circle polyp, located in right of the image}, where \emph{small} indicates that the mask covers less than 8\% of the image. The same captions are given to the model at test time. Its input therefore contains information computed from the very mask that its output is scored against, information that does not exist in clinical use, where the mask is what the model is asked to produce.

So what information do the models actually use from these captions? To find out, we edit the captions provided to the published checkpoints of MedCLIPSeg and RecLMIS, one part at a time, across seven test sets from three imaging modalities (Fig.~\ref{fig:protocol}; RecLMIS on chest X-ray only). These counterfactual edits show what each part of a caption contributes. To see where the behaviour comes from, we retrain MedCLIPSeg with edited captions on three training sets and evaluate the results against criteria that we fixed in advance (\ref{app:registration}). Finally, we measure how much the captions alone reveal about the masks, and we test methods that never see the test annotation.

On colonoscopy and chest X-ray, the models take mainly one thing: the lesion's extent. We call the part of the caption that states it the \emph{extent attribute}: the size word for colonoscopy and breast ultrasound, and the lung-zone phrase for chest X-ray. On these two modalities, the extent attribute alone recovers most of the segmentation benefit provided by the caption. If the captions improved segmentation by describing the lesion, their other words should also contribute. They hardly do: on colonoscopy, colour, shape and count each recover little or none of the benefit, and calling the polyp a \emph{rock} does not reduce performance (Section~\ref{sec:rq1}). Retraining shows that the training captions teach this: a model trained without the extent attribute ignores it, and on colonoscopy and chest X-ray a model trained on the extent attribute alone behaves like one trained on the full captions (Section~\ref{sec:rq2}). The lesson is not free. Models trained on the full captions segment less accurately when the caption is withheld, whereas models trained on captions that do not state the image's own extent show no such loss (Section~\ref{sec:rq3}). The models resemble Clever Hans, the horse that seemed to do arithmetic but read its answers from cues its questioners gave without knowing it \citep{pfungst1911}: they answer correctly by reading a cue that the benchmark supplies. Most medical-imaging studies do not test for such Clever Hans predictors \citep{lapuschkin2019,vasquez2024}. We call this form of label leakage \emph{extent leakage}.

Extent leakage is visible in the captions themselves, before any model is trained: a regression that reads only the caption explains most of the variation in lesion area on every test set except CVC-300 (Section~\ref{sec:res-score}). How far can a model get without the test mask? Our best answer is ExtentSeg, an extent-aware segmenter built for this study, which estimates the extent from the image instead of reading it from a caption. It comes within \nXSgap{} Dice points of the published colonoscopy result and gains \nXSvsTFAllAbs{} points over models trained the same way without text. Given the true extent, it matches our retrained models that read the mask-derived caption (Section~\ref{sec:rq4}). The benefit of text on these benchmarks has therefore been overestimated, and the overestimate can be measured. Our contributions are:
\begin{enumerate}
\item caption edits on two published models and three modalities, showing that on six of the seven test sets (colonoscopy and chest X-ray) the benefit of text is carried by an extent attribute that was computed from the mask or written by the benchmark's annotators, and that on breast ultrasound the diagnosis word shares this benefit;
\item a retraining study on three training sets, with criteria fixed in advance, showing that the training captions create this dependence and that it weakens segmentation from the image;
\item an evaluation protocol, including a caption-only leakage score that needs no segmentation model, and reference results without the mask, among them those of ExtentSeg.
\end{enumerate}
Code and the edited caption sets will be made public on acceptance (see Data availability). A related manuscript by the first author, in preparation, studies how well the training set covers lesion sizes in image-only polyp segmenters; it uses the same public splits and shares no results with this paper.

\section{Related work}

\subsection{Text-guided medical image segmentation}
\label{sec:tgs}
Existing text-guided medical image segmentation methods incorporate language into segmentation networks in different ways. LViT \citep{lvit} fuses text and image features in a hybrid convolutional and Transformer network, whereas RecLMIS \citep{reclmis} adds a cross-modal reconstruction objective. MedCLIPSeg \citep{medclipseg} adds probabilistic vision-language adapters to frozen encoders from contrastive language-image pretraining \citep[CLIP;][]{clip}, in this case UniMed-CLIP \citep{unimedclip}. The chest X-ray captions studied here are the descriptions that the LViT authors wrote for QaTa-COV19; they also described the MosMedData+ computed tomography dataset \citep{morozov2020}, which we do not use. \citet{guidedecoder} split these annotations into three parts of increasing positional detail and found that the most detailed part, the lung zones, gives their model the same Dice as the full annotation (89.8 points, against 84.1 without text). They read this as the value of detailed position information. Our Zone-only retraining agrees with their result; we add that the zone phrase is an annotation of the lesion's extent, and that relying on it costs accuracy when segmenting from the image. Later segmenters adopted these annotations as their text \citep{vitexnet,causalclipseg,tmc,tgcnet,ddcmd,fmiseg}; other methods use text only during training \citep{stpnet,sgseg,prolearn,cap2seg,zhao2025kd} or train with incomplete prompts \citep{arseg}.

Promptable models such as the Segment Anything Model (SAM; \citealp{sam}) and MedSAM \citep{medsam} are trained and evaluated with box or point prompts simulated from the reference annotation \citep{mazurowski2023}. There, the simulated prompt stands for a real user action: at deployment, a user draws the box. A caption computed from the mask, by contrast, stands for a clinical description that is uncertain and often not recorded (Section~\ref{sec:d-measure}). Large text-prompted medical segmenters take the name of the target, such as an object type or anatomical term, as the prompt \citep{biomedparse,sat}; such a prompt says what to segment but nothing about the particular mask.

Prompts computed from the mask have been used before. \citet{poudel2024} computed the number, size and location words of their prompts from the mask. They found that replacing these words at inference barely changes one fine-tuned model (CLIPSeg) but degrades another (CRIS), most for size and location, and read this as evidence of which models use language. Their prompts are the colonoscopy and breast-ultrasound captions of the MedCLIPSeg benchmarks, and adapters for CLIP-based segmenters have been trained and evaluated on them \citep{vlsmadapter}. \citet{textsens2609} trained segmenters with five vision-language encoders on these captions and removed or altered the text at test time, much as we do. They concluded that the effect of text depends on the dataset and acts on the whole image rather than on particular regions, judging this through a decoder that applies text by modulating whole feature channels; they did not examine how the captions were built. We show that (i) the dependence rests on the extent attribute, shared on breast ultrasound with the diagnosis word; (ii) the extent does not merely shift a global threshold but removes false positives away from the lesion; (iii) the training captions teach this dependence at a cost to segmentation from the image; and (iv) a caption-only score measures it without training a segmenter.

\subsection{Shortcuts, leakage and partial-input baselines}
Shortcuts are decision rules that work on standard benchmarks but fail in real-world conditions \citep{geirhos2020}. They are well documented in medical imaging, where they can inflate reported performance \citep{degrave2021,ongly2024,zech2018}, and leakage of information that is available only at evaluation time is one of their most direct causes \citep{kapoor2023,kaufman2012}. Evaluation pitfalls in medical imaging have also been documented for challenge rankings \citep{maierhein2018} and metrics \citep{maierhein2024,reinke2024}. Partial-input baselines, such as hypothesis-only \citep{gururangan2018,poliak2018} and question-only \citep{goyal2017} models, exposed annotation artefacts in natural language inference and language priors in VQA \citep{agrawal2018}. They now measure how much of a multimodal benchmark can be solved without the image \citep{tst2025,mmstar}, including medical multiple-choice benchmarks, where text alone reaches most of the full accuracy \citep{medqamm}. Such baselines give only a lower bound on the information available \citep{feng2019}, and a high score does not show that a model uses it \citep{srikanth2022}; we therefore pair our caption-only score with retraining. Referring-expression models have also been shown to ignore the structure of their text \citep{cirik2018}, and multimodal classifiers often gain little from combining modalities \citep{hessel2020}. Counterfactual benchmarks for segmentation edit the image so that the prompted object is absent \citep{hallusegbench}, or pair an unchanged mask with a misleading prompt \citep{cafe2026}, and find that models often produce a mask anyway. Our edits instead change single attributes of the caption, at test time and in training.

\subsection{Modality imbalance and size priors}
Networks trained on several modalities together can under-learn one of them, sometimes ending up worse than a single-modality network \citep{du2023,huang2022,wang2020,wu2022}. Gradient modulation was proposed to counter this \citep{peng2022}, and modality dropout \citep{moddrop} to counter the fragility of multimodal Transformers when a modality is missing \citep{ma2022}. In these studies every modality is a legitimate input. Here, one input carries the extent of the label. Learning using privileged information \citep{vapnik2015} describes the role such a caption ought to play: information available in training but not at test time. Lesion size is also a familiar prior in segmentation \citep{kervadec2019}. A network can regress the size of the target from the image to constrain its segmentation \citep{kervadec2019curr}, a known or predicted size can constrain segmentation in a new domain \citep{bateson2021}, and a class-ratio prior can guide source-free adaptation \citep{bateson2022}. TGANet \citep{tganet} predicts the polyp count and a three-level size label from the image to weight its attribute embeddings, and FANetv2 \citep{fanetv2} predicts the same two attributes as an auxiliary task. ExtentSeg (Section~\ref{sec:m-extentseg}, Fig.~\ref{fig:extentseg}) belongs to this family. Feeding a conditioned segmenter a wrong value of its condition is an established test: \citet{lemay2021} gave a U-Net conditioned by feature-wise linear modulation (FiLM) each tumour type in turn and found that a wrong type changed the predicted tumour core, most strongly when the type implied a different size. \citet{cao2026} tested phase-conditioned echocardiography segmenters with the true, an estimated and a random phase, and called the gap between true and deployable conditioning \emph{conditioning-availability bias}. We test ExtentSeg with the true extent and with its own estimate in the same spirit.

\section{Benchmarks, captions and models}
\label{sec:setup}

\subsection{Benchmarks}
\label{sec:data}
We use the seven captioned test sets released with MedCLIPSeg (Table~\ref{tab:data}; \citealp{medclipseg_data}). The colonoscopy model is trained on Kvasir-SEG \citep{kvasirseg} and tested on its test split and on four datasets from other centres, which we call the \emph{unseen centres}: CVC-ClinicDB \citep{clinicdb}, CVC-300 \citep{cvc300}, CVC-ColonDB \citep{bernal2012,colondb} and BKAI-IGH \citep{bkai}. The ultrasound model is trained and tested on the Breast Ultrasound Images dataset (BUSI; \citealp{busi}) and the chest X-ray model on QaTa-COV19 \citep{degerli2021qata,qata,lvit}, both with their released splits. All seven datasets are public; no new data were collected from human subjects.

\begin{table}[t]
\centering
\caption{Benchmarks and the extent attribute of their captions.}
\label{tab:data}
\footnotesize
\setlength{\tabcolsep}{3pt}
\begin{tabular}{@{}llrrrrl@{}}
\toprule
Dataset & Modality & Total & Train & Val & Test & Extent attribute$^\ast$\\
\midrule
Kvasir-SEG & colonoscopy & \sSplitKvasirTotal & \sSplitKvasirTrain & \sSplitKvasirVal & \sSplitKvasirTest & size word\\
CVC-ClinicDB & colonoscopy & \sSplitClinicTotal & (\sSplitClinicTrain) & (\sSplitClinicVal) & \sSplitClinicTest & size word\\
CVC-300 & colonoscopy & \sSplitCvcThreeTotal & (\sSplitCvcThreeTrain) & - & \sSplitCvcThreeTest & size word\\
CVC-ColonDB & colonoscopy & \sSplitColonTotal & (\sSplitColonTrain) & - & \sSplitColonTest & size word\\
BKAI-IGH & colonoscopy & \sSplitBkaiTotal & (\sSplitBkaiTrain) & (\sSplitBkaiVal) & \sSplitBkaiTest & size word\\
BUSI & ultrasound & \sSplitBusiTotal & \sSplitBusiTrain & \sSplitBusiVal & \sSplitBusiTest & size word\\
QaTa-COV19 & chest X-ray & \sSplitQataTotal & \sSplitQataTrain & \sSplitQataVal & \sSplitQataTest & lung zones\\
\bottomrule
\end{tabular}\\[3pt]
\begin{minipage}{\columnwidth}\footnotesize Image counts of the MedCLIPSeg release. Of the colonoscopy datasets, only Kvasir-SEG is used for training; splits in parentheses are not used. $^\ast$Size word: the mask area thresholded by Eq.~\eqref{eq:quantiser}; lung zones: written by annotators (Section~\ref{sec:quantiser}).\end{minipage}
\end{table}

Test images rarely duplicate training images, except on BUSI. Searching for identical or near-identical images (a correlation above \sDupThreshold{} between $64\times64$ grey-level thumbnails), we found \nw{\sPolypDupTotal} near-duplicate (in Kvasir-SEG) among the \sPolypDupTestN{} colonoscopy test images and none in the unseen centres. On BUSI, \sBusiDupTest{} of the \sSplitBusiTest{} test images (\sBusiDupTestLesion{} with a lesion) nearly duplicate a training or validation image, probably from the same patient. No QaTa-COV19 test subject appears in training or validation, although \sQTrainNoIdN{} training and \sQValNoIdN{} validation images lack a subject identifier. Two further issues concern the test data alone: \sDedupDropN{} of the \sSplitCvcThreeTest{} CVC-300 test images also appear in CVC-ColonDB, so pooled colonoscopy results count them twice, and the captions of \nw{\sBkaiNoPolypN} BKAI-IGH test images say \capt{no polyp} although their masks contain one. \ref{app:sens} repeats the analyses without these images; for the BKAI-IGH images it does so only for the published models.

\subsection{Construction of the captions}
\label{sec:quantiser}
Fig.~\ref{fig:generation} shows how a colonoscopy caption is written. The count, size and location words are computed from the reference mask $y$, and the colour and shape words come from VQA queries on the image \citep{poudel2024}. Every colonoscopy and BUSI caption released with MedCLIPSeg is one of the prompts that \citet{poudel2024} released for the same image (their template P6; \citealp{medvlsm_code}; see \ref{app:impl}). The size word sorts the lesion's area fraction $a(y)=|y|/|\Omega|$, the share of the image domain $\Omega$ covered by the mask, into three levels:
\begin{equation}
w(y)=\begin{cases}\textit{small}, & a(y)<0.08,\\ \textit{medium}, & 0.08\le a(y)<0.30,\\ \textit{large}, & a(y)\ge 0.30.\end{cases}
\label{eq:quantiser}
\end{equation}
In the five caption files we checked (training captions of Kvasir-SEG, CVC-ClinicDB, BKAI-IGH and BUSI, test captions of CVC-ColonDB), this rule reproduces \nQuantOk{} of the \nQuantN{} single-lesion captions; the other \nQuantOff{} lie within \nQuantTol{} of a threshold, as expected when areas are measured on resized masks. The size word is therefore a function of the reference mask. The MedCLIPSeg supplement lists the attributes of its templates but no size attribute or threshold, although its example captions contain size words \citep{medclipseg}; we recovered Eq.~\eqref{eq:quantiser} from the released captions.

In QaTa-COV19, the list of involved lung zones (e.g., \capt{all left lung and all right lung}) plays the role of the size word. LViT's annotators wrote it for the benchmark with the help of professionals and checked it against the masks \citep{lvit}. It is not a fixed function of the mask: the best rule based on how much of each lung zone the mask covers reproduces the named zones for only \nQataZoneRule\% of test images, and a classifier on mask-shape features for \nQataClfAcc\%, whereas the stated side and number of regions agree with the mask far more often (\ref{app:qata}). Sides are named by image position, the opposite of radiological convention, so the phrase is not taken from a radiology report, as \citet{guidedecoder} also note. MedCLIPSeg's captions reword the LViT annotations without changing their content, so LViT, RecLMIS and MedCLIPSeg receive the same extent information.

\begin{figure*}[!t]
\centering
\includegraphics[width=\textwidth]{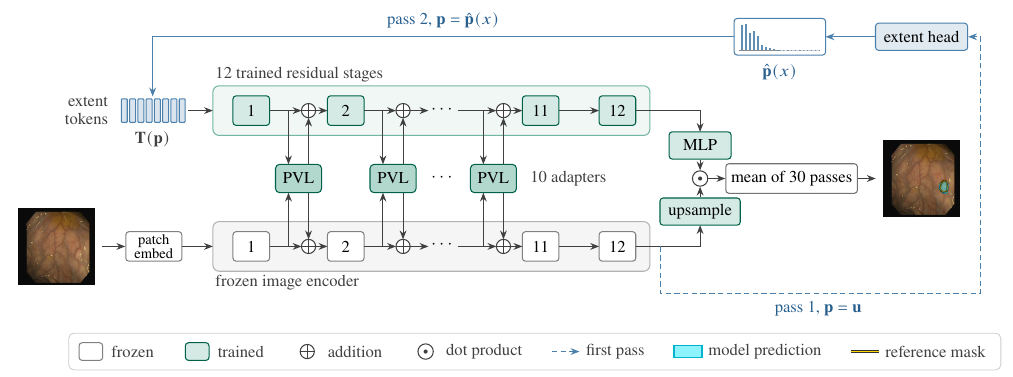}
\caption{ExtentSeg. Eight extent tokens, Eq.~\eqref{eq:tokens}, pass through twelve trained residual stages and reach the frozen image encoder through ten probabilistic vision-language (PVL) adapters. A first pass with the uniform extent $\mathbf u$ feeds the extent head. Its estimate $\hat{\mathbf p}(x)$ builds the tokens of the second pass, which produces the segmentation. Nothing derived from the mask enters at test time.}
\label{fig:extentseg}
\end{figure*}

\subsection{Models and evaluation}
MedCLIPSeg keeps a UniMed-CLIP \citep{unimedclip} image encoder (a Vision Transformer, ViT-B/16; \citealp{vit}) and a PubMedBERT \citep{pubmedbert} text encoder frozen, and trains probabilistic adapters inserted into ten encoder layers. With the released test captions, its published checkpoints for Kvasir-SEG, BUSI and QaTa-COV19 \citep{medclipseg_ckpt} reproduce the reported Dice on all seven test sets to within 0.02 points (Table~\ref{tab:released}). Our second model, RecLMIS \citep{reclmis}, was developed independently; its published checkpoint also reproduces its reported result (\nRecRepro{} against \nRecPublished). We use \emph{published} for the authors' trained checkpoints and \emph{released} for their captions and code.

We follow the released evaluation code. The probability map is averaged over 30 stochastic forward passes, thresholded at 0.5 and compared with the mask by the Dice coefficient \citep{dice1945}, in points from 0 to 100; an empty prediction on an empty mask scores 100. Every image is resized, without cropping, to $224\times224$ pixels, the input size of the frozen encoders. The prediction is scaled back to the native resolution of the mask (nearest neighbour) before scoring, so every Dice value in this paper, including those printed on example images, is computed at native resolution. Native test images range from $\sSizeBusiWMin\times\sSizeBusiHMin$ (BUSI) to $\sSizeKvasirWMax\times\sSizeKvasirHMax$ pixels (Kvasir-SEG); QaTa-COV19 images are natively $224\times224$. Example images in the figures are shown as released, resized but not cropped.

RecLMIS is run with its released test script (one forward pass in evaluation mode, threshold 0.5, native $224\times224$ resolution); the small changes needed to run it are listed in \ref{app:impl}. Its matched caption is the LViT wording it was trained on \citep{lvit_code}. With MedCLIPSeg's rewording of the same content, it gains \sRecGSameRegister{} points instead of \nTabTwoRecG. Its mismatched captions are those given to MedCLIPSeg, in MedCLIPSeg's wording.

\section{Methodology}
\label{sec:method}
Sections~\ref{sec:m-interv} to~\ref{sec:m-span} define the caption edits and the measures computed from them, Sections~\ref{sec:m-cf} and~\ref{sec:m-rely} the retraining, and Sections~\ref{sec:m-score} to~\ref{sec:m-extentseg} the methods that measure or avoid the leakage without the test mask.

\subsection{Caption edits}
\label{sec:m-interv}
Let $f(x,t)\in[0,1]^{H\times W}$ be the probability map predicted for image $x$ with caption $t$, and let $\D_i(t)$ be the Dice score of test image $x_i$, $i=1,\dots,n$, with caption $t$. A caption is made of attributes: the count, size, colour, shape and location words, the diagnosis word that some BUSI captions contain, and, on QaTa-COV19, the side, the number of regions and the zone phrase. Starting from each image's released caption $\tstar_i$, we build the following captions by editing the text (Fig.~\ref{fig:protocol}):
\begin{itemize}\setlength{\itemsep}{0pt}
\item \emph{matched}: $\tstar_i$ itself;
\item \emph{empty}: $\tnull$, the fixed text \capt{a medical image};
\item \emph{mismatched}: $\tstar_{\sigma(i)}$, the caption of another test image, where $\sigma$ is a random reordering of the test images in which no image keeps its own caption, drawn once per test set and shared by all models;
\item \emph{ablated}: $\tstar_i\setminus a$, with attribute $a$ deleted;
\item \emph{isolated}: $\tstar_i|_a$, with every attribute except $a$ deleted;
\item \emph{forced}: $\tstar_i[a\leftarrow v]$, with $a$ replaced by the value $v$.
\end{itemize}
The same edits produce the training captions of Section~\ref{sec:m-cf}. Because QaTa-COV19 has only \sRecDistinctCaptions{} distinct test captions, \sRecSameTextN{} of its \sSplitQataTest{} test images (\sRecSameTextFrac\%) receive a mismatched caption identical to their own.

\subsection{Decomposing the benefit of the caption}
\label{sec:m-decomp}
\begin{figure}[!t]
\centering
\includegraphics[width=\columnwidth]{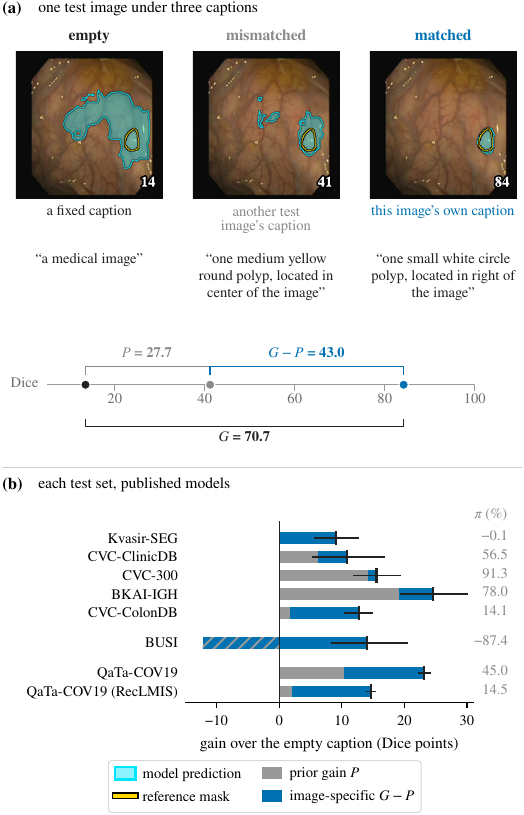}
\caption{Decomposition of the text gain. (a) The test image of Fig.~\ref{fig:protocol} with the empty, mismatched and matched captions. Another image's caption already gives the prior gain $P$, Eq.~\eqref{eq:prior}; only the remainder $G-P$ needs the image's own caption. (b) The same decomposition for each test set, with the text gain $G$ (tick), its 95\% confidence interval (CI) and the prior share $\pi$. On BUSI, $P$ is negative (hatched).}
\label{fig:decomp}
\end{figure}

The quantity usually reported is the improvement of the matched caption over the empty caption, which we call the \emph{text gain},
\begin{equation}
G=\frac{1}{n}\sum_{i=1}^{n}\bigl[\D_i(\tstar_i)-\D_i(\tnull)\bigr].
\label{eq:gain}
\end{equation}
Part of this gain need not depend on the image at all, because any caption from the benchmark may push the prediction towards typical lesion sizes. We estimate this dataset-level part, the \emph{prior gain} $P$, with the mismatched caption, and express it as a fraction of the text gain, the \emph{prior share} $\pi$:
\begin{equation}
P=\frac{1}{n}\sum_{i=1}^{n}\bigl[\D_i(\tstar_{\sigma(i)})-\D_i(\tnull)\bigr],\qquad \pi=\frac{P}{G}.
\label{eq:prior}
\end{equation}
The prior share is negative when another image's caption is worse than no caption. The rest of the gain, $G-P$, is specific to the image (Fig.~\ref{fig:decomp}(a)). For an attribute $a$, the \emph{sufficiency} $S_a$ is the fraction of $G$ recovered when the caption keeps only $a$, and the \emph{necessity} $N_a$ is the fraction lost when only $a$ is deleted:
\begin{align}
S_a&=\frac{1}{nG}\sum_{i=1}^{n}\bigl[\D_i(\tstar_i|_a)-\D_i(\tnull)\bigr],\label{eq:suff}\\
N_a&=\frac{1}{nG}\sum_{i=1}^{n}\bigl[\D_i(\tstar_i)-\D_i(\tstar_i\setminus a)\bigr].\label{eq:nec}
\end{align}
If other parts of the caption partly repeat an attribute's information, its necessity is lower than its sufficiency. We write $S_{\mathrm{ext}}$ and $N_{\mathrm{ext}}$ for the sufficiency and necessity of the extent attribute (the size word or the lung-zone phrase).

\subsection{Response to the size word}
\label{sec:m-span}
Let $A_i(t)=|\{p\in\Omega: f(x_i,t)_p>0.5\}|/|\Omega|$ be the fraction of image $x_i$ predicted as lesion with caption $t$. We write $w$ both for the size-word attribute and, as in Eq.~\eqref{eq:quantiser}, for its value. The \emph{size response} is the mean change in this fraction when the size word is forced from \emph{small} to \emph{large}:
\begin{equation}
\Delta_{\mathrm{size}}=\frac{1}{n}\sum_{i=1}^{n}\Bigl[A_i\bigl(\tstar_i[w\leftarrow\textit{large}]\bigr)-A_i\bigl(\tstar_i[w\leftarrow\textit{small}]\bigr)\Bigr].
\label{eq:span}
\end{equation}
On QaTa-COV19 the zone phrase takes the place of the size word, and the size response compares the one-zone phrase \capt{the lower right lung} with the six-zone phrase \capt{all left lung and all right lung}.

\subsection{Retraining with edited captions}
\label{sec:m-cf}
Editing captions at test time describes a trained model; retraining shows where its behaviour comes from. We retrain MedCLIPSeg on Kvasir-SEG, BUSI and QaTa-COV19 with the published recipe: 100 epochs of Adam \citep{kingma2015}, a learning rate decayed from $3\times10^{-4}$ to $10^{-4}$ on a cosine schedule \citep{loshchilov2017}, and batch size 24. We use five sets of training captions:
\begin{itemize}\setlength{\itemsep}{0pt}
\item \emph{Original}: the released captions;
\item \emph{Size-blind}: $\tstar_i\setminus w$, the size word deleted;
\item \emph{Size-only}: $\tstar_i|_w$, only the size word kept;
\item \emph{Mismatched}: each image paired with another image's caption, by one fixed reordering as in Section~\ref{sec:m-interv};
\item \emph{Text-free}: the empty caption $\tnull$ for every image.
\end{itemize}
On BUSI the Size-only caption also drops the diagnosis word, \emph{benign} or \emph{malignant}, and the \nBusiTrainEmptyN{} training images without a lesion keep their caption \capt{No tumor in the breast ultrasound image.} in the Original, Size-blind and Size-only sets. On QaTa-COV19 the zone phrase plays the role of the size word (\emph{Zone-blind}, \emph{Zone-only}); the Zone-blind captions still state the side and the number of affected regions, and no Mismatched models were trained.

All retrained models are evaluated at their last epoch, with the released test captions: the colonoscopy models on the five colonoscopy test sets ($n=681$), the QaTa-COV19 models on its \sSplitQataTest{} test images, and the BUSI models on all \sSplitBusiTest{} BUSI test images. For BUSI, criteria R1-R3 and $\rho$ (below) use only the 65 test images with a lesion, because only their captions have a size word. $G_{\mathcal T}$ is the text gain of the models trained on caption set $\mathcal T$, with each image's scores averaged over seeds.

We fixed five criteria in advance. For BUSI and QaTa-COV19 they were committed to version control before training began; for Kvasir-SEG they were written on the day the first model was trained and committed before the three-seed results that decide them (\ref{app:registration}):
\begin{itemize}
\item[\textbf{R0}] \emph{Reproduction}: with the matched caption, the mean Dice of the Original models lies within 1.5 points of the published checkpoint.
\item[\textbf{R1}] \emph{Size response}: the size response of the Size-blind models is below 25\% of that of the Original models.
\item[\textbf{R2}] \emph{Necessity}: the Size-blind models keep at most half of the Original text gain, $G_{\text{Size-blind}}/G_{\text{Original}}\le 50\%$, with the upper confidence bound below 50\%.
\item[\textbf{R3}] \emph{Sufficiency}: the Size-only models keep at least three quarters of it, $G_{\text{Size-only}}/G_{\text{Original}}\ge 75\%$, with the lower confidence bound above 75\%.
\item[\textbf{R4}] \emph{Uninformative captions}: the text gain of the Mismatched models, and the difference between their empty-caption Dice and that of the Text-free models, both lie within $\pm1$ point.
\end{itemize}
Before any BUSI or QaTa-COV19 model was trained, we amended the criteria for these two datasets in three ways: R4 also requires both 95\% CIs to lie within $\pm2$ points; $\rho_{\text{Size-blind}}>0$ (Section~\ref{sec:m-rely}), with an interval excluding zero, becomes a sixth criterion; and a result whose point estimate is on the required side but whose interval reaches the bar is called \emph{consistent}.

\subsection{Segmentation without a caption}
\label{sec:m-rely}
A model that reads the extent from the caption may learn less about it from the image. We therefore compare, with the empty caption, the models trained on each caption set $\mathcal T$ with the Original models:
\begin{equation}
\rho_{\mathcal T}=\frac{1}{n}\sum_{i=1}^{n}\bigl[\bar{\D}^{\,\mathcal T}_i(\tnull)-\bar{\D}^{\,\text{Original}}_i(\tnull)\bigr],
\label{eq:reliance}
\end{equation}
where $\bar{\D}^{\,\mathcal T}_i$ is averaged over seeds. A positive \emph{empty-caption difference} $\rho_{\mathcal T}$ means that the models trained on $\mathcal T$ segment better from the image alone.

\subsection{Caption-only leakage score}
\label{sec:m-score}
How much a set of captions reveals about the masks can be measured without any segmentation model. For each image with a lesion, a ridge regression \citep{hoerl1970} predicts the log lesion area $z_i=\log a(y_i)$ from the words and word pairs of its caption, with five-fold cross-validation \citep[scikit-learn;][]{sklearn}: each image's prediction $\hat z_i$ comes from a regression fitted on the other four folds. We report the coefficient of determination of these predictions,
\begin{equation}
R^2_{\mathrm{cap}}=1-\frac{\sum_i(z_i-\hat z_i)^2}{\sum_i(z_i-\bar z)^2},
\label{eq:r2}
\end{equation}
where $\bar z$ is the mean of the $z_i$. Shuffling the captions across images gives the chance level. The folds are drawn over images, not over videos or subjects.

Because $R^2_{\mathrm{cap}}$ gives a fit but no significance test, we add a model-free measure, $I_{\mathrm{ext}}=(\hat I-\E_{\mathrm{perm}}\hat I)/\hat H(b)$. Here $b$ is the tercile of the lesion area (whether the lesion is among the smallest, middle or largest third of the lesions scored), $\hat I$ is the mutual information between $b$ and the caption's extent token $e$ \citep[with the Miller-Madow correction;][]{miller1955}, $\E_{\mathrm{perm}}\hat I$ is its mean over 200 random permutations of $e$, and $\hat H(b)$ is the entropy of the terciles. In words, $I_{\mathrm{ext}}$ is the fraction of the uncertainty about the tercile that the extent token removes beyond chance: about 0 when the token says nothing about the area, and close to 1 when it fixes the tercile. The same permutations give a one-sided $p$-value, whose smallest possible value is $1/201\approx0.005$. The extent token is the size word or, for caption sets without one (QaTa-COV19 and Size-blind captions) and for the edited BUSI and QaTa-COV19 training-caption sets of Section~\ref{sec:m-cf}, the whole caption.

\subsection{Annotation-free methods}
\label{sec:m-free}
We call a method \emph{annotation-free} if nothing derived from the test annotation enters its input. We test four such methods.

\emph{Edge-aligned threshold.} The published model is run with three fixed captions, \capt{a small region in the image.} and its \emph{medium} and \emph{large} variants, and the three probability maps are averaged into a score map $s(x)$. Let $\Omega_k$ be the $k$ highest-scoring pixels. Among 96 candidate areas between 0.1\% and 60\% of the image, the rule picks the one whose boundary lies where $s$ changes most steeply:
\begin{equation}
k^\ast=\arg\max_k\frac{1}{|\partial\Omega_k|}\sum_{p\in\partial\Omega_k}\bigl\|\nabla s(p)\bigr\|,
\label{eq:edgecut}
\end{equation}
where $\nabla$ is a $3\times3$ Sobel gradient and $\partial\Omega_k$ is the boundary of $\Omega_k$; if $s\le0.5$ everywhere, the mask is empty. On QaTa-COV19, $s$ averages seven fixed captions that cover one to six lung zones and three size words (\ref{app:impl}). This rule was the best of twelve label-free rules on the pooled colonoscopy test sets, though not on every single test set or on BUSI (Table~\ref{tab:rules}), so choosing it used the test labels. If the rule for each test set is instead chosen on the remaining test sets, pooled Dice changes by only \nEdgeLotoD{} \ci{\nEdgeLotoLo}{\nEdgeLotoHi}. On QaTa-COV19, \nw{\nEdgeQataHigherN} of the seven rules evaluated there score higher than this one (at best \nEdgeQataBest{} against \nEdgeQata).

\emph{Image-only describer.} This method predicts the caption from the image and gives the predicted caption to the published model. The prediction uses the image features of the published model run with the empty caption. A ridge regression estimates the log area, which is turned into \emph{small}, \emph{medium} or \emph{large} at the terciles of the Kvasir-SEG training lesion areas, and logistic classifiers predict the count and location words; colour and shape are not predicted. All three are fitted on the Kvasir-SEG training split. Cutting at the thresholds of Eq.~\eqref{eq:quantiser} instead changes its Dice by only \nDescEqOneD{} \ci{\nDescEqOneLo}{\nDescEqOneHi}.

The other two methods are the Text-free models of Section~\ref{sec:m-cf} and ExtentSeg.

\subsection{An extent-aware segmenter}
\label{sec:m-extentseg}
If the useful content of the caption is the extent, a segmenter can estimate the extent from the image and condition on that estimate instead. ExtentSeg replaces MedCLIPSeg's text branch with learned extent tokens and keeps the image encoder, adapters and mask head (Fig.~\ref{fig:extentseg}). Extent is divided into $K=20$ bins at the quantiles of the Kvasir-SEG training lesion areas, and a probability distribution $\mathbf p\in\Delta^{K-1}$ over the bins is mapped to $M=8$ tokens
\begin{equation}
\mathbf T(\mathbf p)=\mathbf B+\mathbf 1_M\,(\mathbf p^{\top}\mathbf E),\label{eq:tokens}
\end{equation}
where $\mathbf 1_M$ is a vector of $M$ ones, $\mathbf E\in\mathbb R^{K\times768}$ holds one learned embedding per bin, $\mathbf B\in\mathbb R^{M\times768}$ is a learned base, and $\Delta^{K-1}$ is the probability simplex. Each token is thus a learned base vector plus the average of the bin embeddings, weighted by $\mathbf p$. In place of the language layers, twelve residual stages $\mathbf T\leftarrow\mathbf T+\mathrm{Lin}_\ell(\mathrm{LN}(\mathbf T))$ refine the tokens; each is a layer normalisation \citep[LN;][]{ba2016} followed by a zero-initialised linear layer, so every stage starts as the identity.

An extent head estimates $\hat{\mathbf p}(x)$ from the image alone, using a first forward pass conditioned on the uniform distribution $\mathbf u$; a second pass, conditioned on $\hat{\mathbf p}(x)$, produces the segmentation. The head's loss does not flow back into the image features, so it updates only the head. During training, each sample is conditioned on the true extent (slightly blurred), on the model's own estimate, or on no extent information:

\begin{equation}
\mathbf p=\begin{cases}\tilde{\mathbf q}(k_i) & \text{with probability }0.45,\\ \operatorname{sg}[\hat{\mathbf p}(x_i)] & \text{with probability }0.45,\\ \mathbf u & \text{with probability }0.10,\end{cases}
\label{eq:sched}
\end{equation}
where $k_i$ is the bin of training image $i$'s mask, $\tilde{\mathbf q}(k)$ is a Gaussian over bins centred at $k$ (standard deviation one bin), and $\operatorname{sg}$ is a stop-gradient. The loss combines Dice and binary cross-entropy (BCE) segmentation losses, a cross-entropy (CE) term for the extent head and an area term:
\begin{equation}
\begin{split}
\mathcal L={}&\tfrac12\mathcal L_{\mathrm{Dice}}+\tfrac12\mathcal L_{\mathrm{BCE}}+\lambda_e\,\mathrm{CE}\bigl(\tilde{\mathbf q}(k_i),\hat{\mathbf p}\bigr)\\
&+\lambda_a\bigl(\log\bar m_i-\log\bar a_i\bigr)^2,
\end{split}
\label{eq:loss}
\end{equation}
where $\bar m_i$ is the mean predicted probability over image $i$ and $\bar a_i=\tilde{\mathbf q}(k_i)^{\top}\mathbf c$ is the expected area under the soft target, with $\mathbf c$ the bin centres (\ref{app:impl}). The area term is used only when a sample is conditioned on the true extent and its lesion is larger than the first bin edge. We set $\lambda_e=\lambda_a=0.1$; the variant without the area loss sets $\lambda_a=0$. The \emph{extent-free control} is ExtentSeg with its extent input removed; its area term uses the true extent of every training image whose lesion is larger than the first bin edge. All variants use the training recipe of Section~\ref{sec:m-cf}.

At test time ExtentSeg is conditioned on its own estimate $\hat{\mathbf p}(x)$. Conditioning it instead on $\tilde{\mathbf q}(k)$ for the bin $k$ of the test mask gives an upper reference that is not annotation-free.

\subsection{Statistical analysis}
\label{sec:stats}
CIs are 95\% percentile intervals from a paired bootstrap over test images \citep{efron1994} with $10^4$ resamples (random seed 0); for ratios, numerator and denominator are resampled together. Every retrained model, as well as ExtentSeg and its two variants, was trained with three seeds (1, 2 and 3), and each image's scores are averaged over the seeds before resampling. Where stated, \emph{seed-resampled} intervals also resample the three seeds of each model in every bootstrap draw. Pooled results are means over all images of the pooled test sets. Differences and ratios are computed from unrounded values, so they can differ in the last digit from calculations on the printed values.

Not all test images are independent: frames from one colonoscopy video are correlated, and QaTa-COV19 contains several images per subject. As sensitivity analyses, two cluster bootstraps resample whole video sequences (\sClusterNClusters{} clusters of the 681 colonoscopy test images, built from the images alone) or the \sQSubjN{} QaTa-COV19 test subjects instead of images; \ref{app:sens} describes the clustering and reports the results.

\section{Results}
\label{sec:results}

\subsection{What the models take from the caption}
\label{sec:rq1}

\begin{figure}[t]
\centering
\includegraphics[width=\columnwidth]{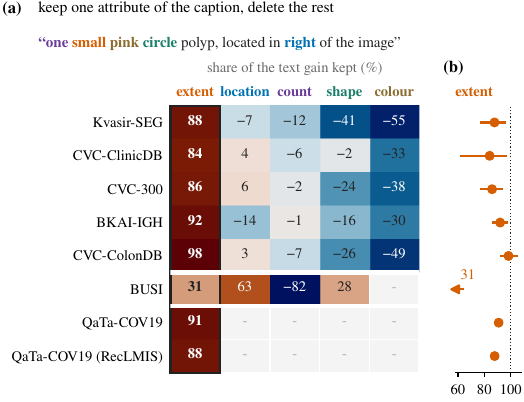}
\caption{Which caption attribute carries the text gain. (a) Sufficiency of each attribute, Eq.~\eqref{eq:suff}: the share of the text gain recovered when every test caption keeps only that attribute (100: as good as the whole caption; 0: no better than the empty caption). For QaTa-COV19 the extent attribute is the lung-zone phrase; -: not in the template. BUSI: the 65 images with a lesion. (b) The extent column with 95\% CIs.}
\label{fig:attributes}
\end{figure}

\begin{figure*}[t]
\centering
\includegraphics[width=\textwidth]{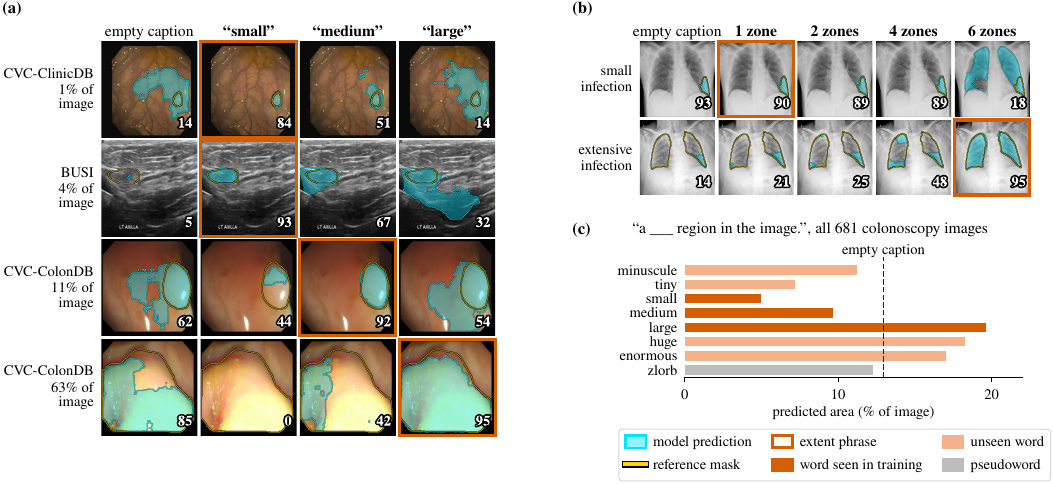}
\caption{Predictions of the published models when only the extent in the caption is changed. (a) Colonoscopy and breast ultrasound, with the size word replaced (percentage: true lesion area). (b) Chest X-ray, with the lung-zone phrase replaced by one covering 1, 2, 4 or 6 zones. Orange frames mark the image's true extent; numbers are Dice scores. (c) Mean predicted area over the 681 colonoscopy test images for one caption with different size words, including words not seen in training and the made-up word \emph{zlorb}.}
\label{fig:steering}
\end{figure*}

\subsubsection{Dataset-level and image-specific benefit}
Part of the reported benefit is not specific to the image. The text gains of the published checkpoints range from \nTabTwoKvasirG{} to \nTabTwoBkaiG{} points (Table~\ref{tab:released}, Fig.~\ref{fig:decomp}). On CVC-ClinicDB, CVC-300 and BKAI-IGH, a caption from a different image already recovers most of this gain; on Kvasir-SEG, the training domain, it recovers none. Pooled over the five colonoscopy test sets, the prior share is \nPolypPrior\% \ci{\nPolypPriorLo}{\nPolypPriorHi}. This interval treats video frames as independent; when whole video sequences are resampled it widens to \ci{\sClusterPiLo}{\sClusterPiHi}, and under the coarsest clustering it includes zero (Table~\ref{tab:cluster}).

The prior gain mainly corrects how much of the image the model labels as lesion. Without a caption, the colonoscopy model labels on average \nCalibNull\% of an unseen-centre image as polyp, nearly twice the true \nCalibTrue\%; with another image's caption, it labels \nCalibShuf\%. On BUSI the prior share is negative, because \nBusiNoLesionTestN{} test images contain no lesion, and another image's caption may describe a lesion that is absent or deny one that is present. Since $\pi$ ranges from strongly negative to over 90\%, a gain over the empty caption alone says little about how much of the benefit is specific to the image. The picture is the same on the normalised surface Dice \citep{nsd}, a boundary-based score (Table~\ref{tab:nsd}), and without the duplicated test images (Table~\ref{tab:dedup}). Without the QaTa-COV19 images whose mismatched caption is identical to their own (Section~\ref{sec:m-interv}), $\pi$ falls from \nTabTwoQataPi\% to \sSameTextQataPiExcl\% for MedCLIPSeg and from \nTabTwoRecPi\% to \sRecPiMixedExclSameText\% for RecLMIS.

\begin{table}[!t]
\centering
\caption{Benefit of the caption for the published checkpoints.}
\label{tab:released}
\footnotesize
\setlength{\tabcolsep}{2.6pt}
\begin{tabular}{@{}lrrrrrrrr@{}}
\toprule
Test set & Empty & Mism. & Match. & $G$ & $\pi$ & $S_{\mathrm{ext}}$ & $N_{\mathrm{ext}}$ & $R^2_{\mathrm{cap}}$\\
\midrule
\multicolumn{9}{@{}l}{\emph{MedCLIPSeg}}\\
Kvasir-SEG & \nTabTwoKvasirEmpty & \nTabTwoKvasirMism & \nTabTwoKvasirMatch & \nTabTwoKvasirG & \nTabTwoKvasirPi & \nTabTwoKvasirSuff & \nTabTwoKvasirNec & \nRcapTeKvasir\\
CVC-ClinicDB & \nTabTwoClinicEmpty & \nTabTwoClinicMism & \nTabTwoClinicMatch & \nTabTwoClinicG & \nTabTwoClinicPi & \nTabTwoClinicSuff & \nTabTwoClinicNec & \nRcapTeClinic\\
CVC-300 & \nTabTwoCvcThreeEmpty & \nTabTwoCvcThreeMism & \nTabTwoCvcThreeMatch & \nTabTwoCvcThreeG & \nTabTwoCvcThreePi & \nTabTwoCvcThreeSuff & \nTabTwoCvcThreeNec & \nRcapTeCvcThree\\
BKAI-IGH & \nTabTwoBkaiEmpty & \nTabTwoBkaiMism & \nTabTwoBkaiMatch & \nTabTwoBkaiG & \nTabTwoBkaiPi & \nTabTwoBkaiSuff & \nTabTwoBkaiNec & \nRcapTeBkai\\
CVC-ColonDB & \nTabTwoColonEmpty & \nTabTwoColonMism & \nTabTwoColonMatch & \nTabTwoColonG & \nTabTwoColonPi & \nTabTwoColonSuff & \nTabTwoColonNec & \nRcapTeColon\\
BUSI$^\dagger$ & \nTabTwoBusiEmpty & \nTabTwoBusiMism & \nTabTwoBusiMatch & \nTabTwoBusiG & \nTabTwoBusiPi & \nBusiSuff & \nBusiNec & \nRcapTeBusi\\
QaTa-COV19 & \nTabTwoQataEmpty & \nTabTwoQataMism & \nTabTwoQataMatch & \nTabTwoQataG & \nTabTwoQataPi$^\ddagger$ & \nTabTwoQataSuff & \nQataNec$^\S$ & \nRcapTeQata\\
\addlinespace
\multicolumn{9}{@{}l}{\emph{RecLMIS}}\\
QaTa-COV19$^\P$ & \nTabTwoRecEmpty & \nTabTwoRecMism & \nTabTwoRecMatch & \nTabTwoRecG & \nTabTwoRecPi & \nTabTwoRecSuff & \nTabTwoRecNec & -\\
\bottomrule
\end{tabular}\\[2pt]
\begin{minipage}{\columnwidth}\footnotesize Dice with the empty, mismatched and matched captions; $G$, Eq.~\eqref{eq:gain}; $\pi$, $S_{\mathrm{ext}}$ and $N_{\mathrm{ext}}$, Eqs.~\eqref{eq:prior}-\eqref{eq:nec}, in percent; $R^2_{\mathrm{cap}}$, Eq.~\eqref{eq:r2}, of the test captions (at most \nRcapPermTeMax{} with shuffled captions). $^\dagger$$S_{\mathrm{ext}}$ and $N_{\mathrm{ext}}$ on the 65 images with a lesion. $^\ddagger$\sSameTextQataPiExcl\% without the images whose mismatched caption equals their own. $^\S$Zone phrase removed from every caption directly, because deleting it through the template misses it in \nQataAblLeakN{} captions. $^\P$LViT wording; other conditions in \ref{app:sens}.\end{minipage}
\end{table}

\subsubsection{Which attribute carries the benefit}
The image-specific part of the gain is carried by the extent attribute. On every colonoscopy test set, the extent attribute alone recovers \nTabTwoClinicSuff-\nTabTwoColonSuff\% of the text gain, and deleting it loses \nTabTwoKvasirNec-\nTabTwoBkaiNec\% (Table~\ref{tab:released}). No other attribute recovers much on its own on these test sets (Fig.~\ref{fig:attributes}), and colour, the word that most plainly describes the lesion's appearance, is worse than no caption at all. The count and location words are also computed from the mask, yet the models barely use them: deleting the location word costs \nLocNec\% \ci{\nLocNecLo}{\nLocNecHi} of the gain, and deleting the count word costs nothing.

On QaTa-COV19, the zone phrase alone recovers \nTabTwoQataSuff\% of the gain for MedCLIPSeg and \nTabTwoRecSuff\% for RecLMIS, while deleting the stated count or side costs at most \nQataOtherNecMax\% (no other attribute was tested alone). Deleting the zone phrase costs RecLMIS \nTabTwoRecNec\% of its gain in its own wording. Our registered test deleted it from MedCLIPSeg's wording instead; there the cost is only \nRecNecReg\% \ci{\nRecNecRegLo}{\nRecNecRegHi}, below the 50\% bar set in advance (Table~\ref{tab:criteria}).

On BUSI, the caption leaks a second label: \nBusiDiagN{} of the 65 lesion captions name the diagnosis, \emph{benign} or \emph{malignant}, always correctly. Extent and diagnosis share the gain. On these 65 images, deleting the size word loses \nBusiNec\% of the gain; the size word alone recovers \nBusiSuff\%, and together with the diagnosis \nBusiSuffClass\%. (The other BUSI cells of Fig.~\ref{fig:attributes}(a) also keep the diagnosis word where the caption has one, so they are not comparable with the colonoscopy cells.)

Three controls show that the models respond to this content, not to medical vocabulary. Replacing \emph{polyp} with \emph{rock} (\capt{one small pink circle rock, located in right of the image}) costs nothing on the colonoscopy test sets (\nRock{} points). A made-up word (\capt{a zlorb region in the image.}) recovers only \nGibOneAbs{} of the \nPolypGain{} points of text gain. Rewriting the captions in other styles, from bare lists to clinical sentences, changes the pooled result by at most \nParaMaxShift{} points (up to \sParaSetMaxLoss{} on single test sets; Table~\ref{tab:para}); only the list style, which is not a sentence, misses our registered bar of 1.0 point from the terse rewrite (Table~\ref{tab:criteria}). Keeping the new wording but emptying its content loses \nParaFrameLossLo{} to \nParaFrameLossHi{} points and leaves the result within \nParaFrameVsEmptyMax{} points of the empty caption.

\begin{figure}[t]
\centering
\includegraphics[width=\columnwidth]{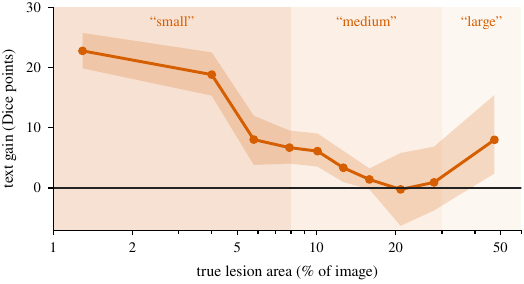}
\caption{Text gain of the published models against lesion area, over the \nAxisN{} colonoscopy and BUSI test images with a lesion (deciles of the Kvasir-SEG training areas; band: 95\% CI). Shading marks the size word that the template assigns to each area. The gain is largest where the caption says \emph{small}.}
\label{fig:axis}
\end{figure}

\begin{figure*}[t]
\centering
\includegraphics[width=\textwidth]{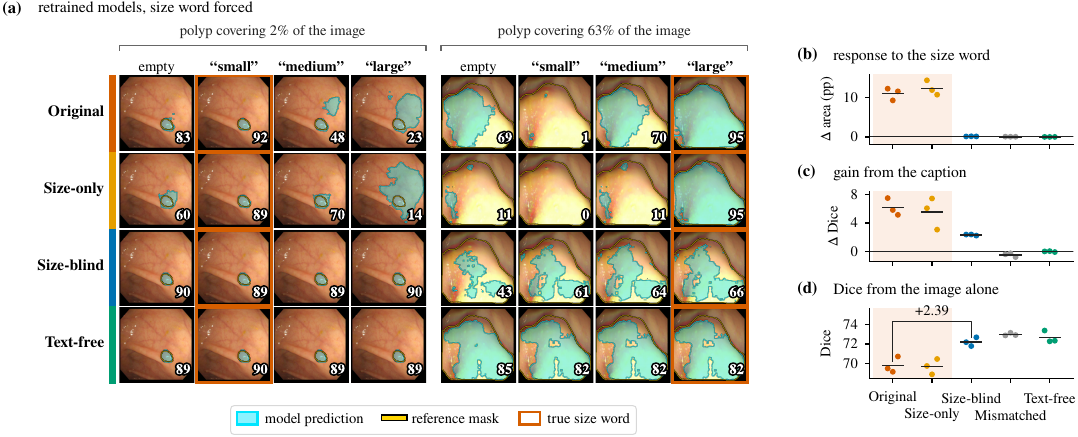}
\caption{Retraining MedCLIPSeg on edited Kvasir-SEG captions. (a) Four retrained models (seed 1); each image is given its own caption with the size word forced. (b)-(d) All five models on the five colonoscopy test sets (dots: seeds; lines: means; shaded: models trained with the size word): (b) size response, Eq.~\eqref{eq:span}; (c) text gain; (d) empty-caption Dice and $\rho$, Eq.~\eqref{eq:reliance}.}
\label{fig:retraining}
\end{figure*}

\begin{table*}[t]
\centering
\caption{Retraining with edited captions, against the registered criteria.}
\label{tab:cf}
\footnotesize
\setlength{\tabcolsep}{3pt}
\begin{tabular}{@{}l>{\raggedright\arraybackslash}p{3.3cm}>{\raggedright\arraybackslash}p{4.3cm}>{\raggedright\arraybackslash}p{4.3cm}>{\raggedright\arraybackslash}p{3.7cm}@{}}
\toprule
& Criterion & Kvasir-SEG ($n=681$) & BUSI ($n=\sSplitBusiTest$; R1-R3, $\rho$: 65) & QaTa-COV19 ($n=\sSplitQataTest$)\\
\midrule
R0 & Original within $\pm1.5$ of the published checkpoint & \nRzeroMean{} vs.\ \nPolypOracle{} (\nRzeroGap) \metyes & \nBRzeroMean{} vs.\ \nBusiOracle{} (\nBRzeroGap) \metyes & \nQRzeroMean{} vs.\ \nQataOracle{} (\nQRzeroGap) \metyes\\
R1 & Size-blind size response $<25\%$ of Original & \nRoneRatio\% (\nRoneSpanNoSize{} vs.\ \nRoneSpanOrig{} pp) \metyes & \nBRoneRatio\% (\nBRoneSpanNoSize{} vs.\ \nBRoneSpanOrig{} pp) \metyes & \nQRoneRatio\% (\nQRoneSpanBlind{} vs.\ \nQRoneSpanOrig{} pp) \metyes\\
R2 & Size-blind retains $\le50\%$ of $G$ (CI) & \nRtwo\% \ci{\nRtwoLo}{\nRtwoHi} \metyes & \nBRtwo\% \ci{\nBRtwoLo}{\nBRtwoHi} \metpart & \nQRtwo\% \ci{\nQRtwoLo}{\nQRtwoHi} \metyes\\
R3 & Size-only retains $\ge75\%$ of $G$ (CI) & \nRthree\% \ci{\nRthreeLo}{\nRthreeHi} \metyes & \nBRthree\% \ci{\nBRthreeLo}{\nBRthreeHi} \metno & \nQRthree\% \ci{\nQRthreeLo}{\nQRthreeHi} \metyes\\
R4 & Mismatched: $|G|\le1$; empty-caption score within $\pm1$ of Text-free & $G=\nRfourGain$ \ci{\nRfourGainLo}{\nRfourGainHi}; $\nRfourNullDiff$ \ci{\nRfourNullDiffLo}{\nRfourNullDiffHi} \metyes & $G=\nBRfourGain$ \ci{\nBRfourGainLo}{\nBRfourGainHi}; $\nBRfourNullDiff$ \ci{\nBRfourNullDiffLo}{\nBRfourNullDiffHi} \metno & -\\
$\rho$ & Size-blind better from the image alone ($\rho>0$, CI) & \nRelyNoSize{} \ci{\nRelyNoSizeLo}{\nRelyNoSizeHi} (not registered) & \nBRelyNoSize{} \ci{\nBRelyNoSizeLo}{\nBRelyNoSizeHi} \metyes & \nQRelyBlind{} \ci{\nQRelyBlindLo}{\nQRelyBlindHi} \metyes\\
\bottomrule
\end{tabular}\\[2pt]
\begin{minipage}{\textwidth}\footnotesize MedCLIPSeg, three seeds per model; the Kvasir-SEG models are tested on the five colonoscopy test sets. On QaTa-COV19 the extent attribute is the zone phrase (Zone-blind and Zone-only models). CI: image bootstrap over seed-averaged scores, as registered; seed-resampled intervals are given in \ref{app:sens}. pp: percentage points of image area.\end{minipage}
\end{table*}

\subsubsection{How the size word acts on the prediction}
In effect, the size word tells the model how much of the image to segment (Fig.~\ref{fig:steering}). Told \emph{small}, the model segments a small polyp almost exactly; told \emph{large}, it buries the same polyp in false positives. A polyp covering most of the image can vanish when the caption says \emph{small}, although this is rare (\sEmptyPredLargeSmall{} lesions that the template calls \emph{large}). The leaked size word can also erase a lesion: with the matched caption the prediction is empty for \sEmptyPredMatched{} of the \sEmptyPredN{} colonoscopy lesions (\sEmptyPredMatchedColon{} in CVC-ColonDB), against none with the empty caption. On colonoscopy, the mean predicted area roughly doubles from \emph{small} to \emph{medium} and again to \emph{large}, and words never seen in training, such as \emph{tiny}, \emph{huge} and \emph{enormous}, fall on the same scale; only \emph{minuscule} is out of place (Fig.~\ref{fig:steering}(c)). The chest X-ray model responds to the lung zones in the same way (Fig.~\ref{fig:steering}(b)).

This is not just a shift of the decision threshold. Choosing a threshold for each image from an estimate of lesion volume is an established remedy for volume bias \citep{tapas}, but it does not explain the effect here. Adding a fixed offset to the output logits for each size word (in effect, one threshold per word), fitted on the training images, recovers only \nThrShare\% \ci{\nThrShareLo}{\nThrShareHi} of the word's benefit on the unseen centres. Even an offset tuned for each test image on its own mask reaches only \nThrPerImage{} Dice, against \nThrSizeOnly{} with the size word alone (${n=\nThrN}$, without the \nw{\sBkaiNoPolypN} \capt{no polyp} captions). The word suppresses false positives far from small lesions (Fig.~\ref{fig:steering}(a), first row), which no threshold can remove without also removing the lesion.

As a result, the text gain is concentrated on small lesions (Fig.~\ref{fig:axis}). Over the \nAxisN{} colonoscopy and BUSI test images with a lesion, it is \nBandSmall{} points \ci{\nBandSmallLo}{\nBandSmallHi} for lesions under \nBandCutLo\% of the image (the smallest 30\% of Kvasir-SEG training lesions), \nBandMid{} between \nBandCutLo{} and \nBandCutHi\%, and \nBandLarge{} \ci{\nBandLargeLo}{\nBandLargeHi} above \nBandCutHi\% (the largest 30\%). Much of the benefit of text is a rescue of small lesions, and it works because the caption says that they are small.

\subsection{Where the dependence comes from}
\label{sec:rq2}
The training captions create the dependence on extent. On colonoscopy, all five criteria are met (Table~\ref{tab:cf}; Table~\ref{tab:criteria} lists every registered criterion and its verdict). The Original models reproduce the published checkpoint to within \nRzeroGapAbs{} points (R0). Models trained with the size word follow a forced size word even when it is wrong, while the Size-blind and Text-free models ignore it (R1; Fig.~\ref{fig:retraining}). Deleting the size word from the training captions leaves \nRtwo\% of the text gain (R2). Keeping only the size word, which leaves just \nSizeOnlyDistinct{} distinct training captions such as \capt{small polyp, located in of the image} (only the template's fixed words remain), keeps \nRthree\% (R3). Captions unrelated to their images teach almost nothing (R4). The Size-blind models keep a small gain (\nTBNoSize{} points) because their remaining words still carry some extent information ($R^2_{\mathrm{cap}}=\nRcapCFSizeBlind$; Table~\ref{tab:score}).

These verdicts use the image bootstrap that the criteria specify, and they are less certain once seeds and video sequences are taken into account. R2 holds in all nine pairings of a Size-blind seed with an Original seed, but R3 holds in only \nw{\nRthreePairsMet} of the nine, because one Size-only seed gains just \nTBSizeOnlyMinSeed{} points. When the seeds are also resampled, both intervals cross their bars, and when video sequences are resampled, so does the upper bound of R2 (\ref{app:sens}, Table~\ref{tab:cluster}). Both point estimates stay on the required side.

BUSI reproduces the size response and the cost to segmentation from the image, but not every criterion (Table~\ref{tab:cf}). The Original models reproduce the published checkpoint (R0), and the Size-blind models ignore the size word (R1). On the 65 lesion images, the Size-blind models keep \nBRtwo\% of the text gain: below the R2 bar, but with an interval that reaches it (\emph{consistent}). On a second population specified in advance, the \nBusiLesionTVN{} lesion images of the test and validation sets, they keep \nBRtwoTV\% \ci{\nBRtwoTVLo}{\nBRtwoTVHi}, and R2 is met. R3 is not met, because BUSI captions carry more than the extent: the Size-only captions also lose the diagnosis word, and these models keep only \nBRthree\% of the gain. With the matched caption they score about as well as the Original models, but they score higher with the empty caption, so their share of the gain is small (Table~\ref{tab:perseed}). On BUSI, then, the size word is one of several cues that the Original models rely on. R4 is not met either, but only because of the lesion-free images: in \nBRfourEmptyPosN{} of \nBRfourEmptyEvalN{} evaluations of such images (\nBusiNoLesionTestN{} images, three seeds), the Mismatched models predict a lesion although the caption says \capt{No tumor in the breast ultrasound image.} This accounts for \nBRfourNormals{} of their \nBRfourGain{} points; on the 65 lesion images their gain is \nBRfourGainLes{} \ci{\nBRfourGainLesLo}{\nBRfourGainLesHi}, within the bar. With the seeds also resampled, the BUSI R2 and R3 intervals both cross their bars (\ref{app:sens}).

Chest X-ray meets every criterion that was run (Table~\ref{tab:cf}). The Original models reproduce the published checkpoint (R0). The Zone-blind models ignore the zone phrase (R1) and keep only \nQRtwo\% of the gain (R2), although their captions still state the side and number of regions ($R^2_{\mathrm{cap}}=\nQRcapZoneBlind$). The zone phrase alone teaches \nQRthree\% of the gain (R3). With the seeds also resampled, R2 stays well below its bar, but R3 holds in only \nw{\nQRthreePairsMet} of the nine seed pairings and its interval crosses the bar (\ref{app:sens}). Resampling subjects instead of images changes no conclusion.

On colonoscopy and chest X-ray, the published checkpoints depend on the caption even more than our retrained models. Their text gains are larger (\nPolypGain{} against \nTBOrig{} points on colonoscopy, \nTabTwoQataG{} against \nQTBOrig{} on chest X-ray), mainly because they score lower with the empty caption. Our retraining results therefore describe models trained with the published recipe; the stronger dependence of the published checkpoints themselves is shown only by the test-time edits of Section~\ref{sec:rq1}. Text gains also vary between checkpoints trained the same way, from \nTBOrigSeedMin{} to \nTBOrigSeedMax{} points across our three Original seeds on colonoscopy and from \nQTBOrigSeedMin{} to \nQTBOrigSeedMax{} on chest X-ray (Table~\ref{tab:perseed}), so they should be reported over several checkpoints. Every causal comparison in this paper is between models trained with the same recipe; on colonoscopy, these comparisons also hold when checkpoints are selected by validation Dice instead of taken at the last epoch (\ref{app:sens}).

\begin{figure*}[t]
\centering
\includegraphics[width=\textwidth]{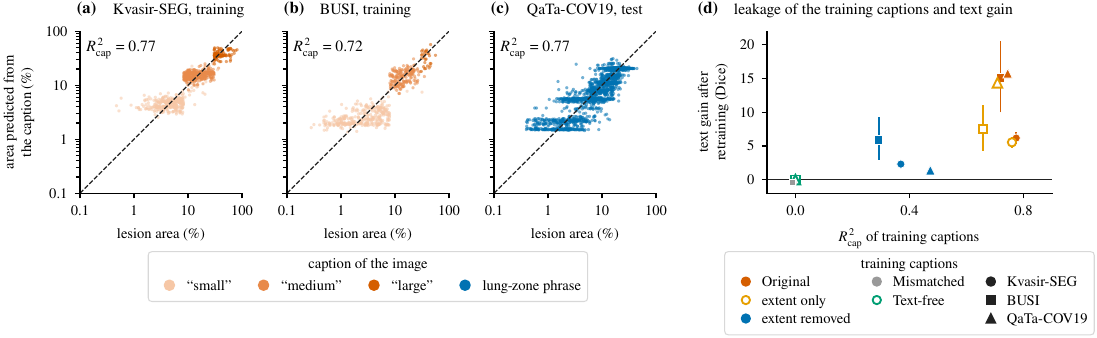}
\caption{Caption-only leakage score. (a)-(c) Lesion area predicted from the caption words alone (out-of-fold ridge regression) against the true area; dashed line: identity. (d) The score of each training-caption set against the text gain of the models retrained on it (three-seed means, 95\% CIs). \emph{Extent only}: Size-only and Zone-only; \emph{extent removed}: Size-blind and Zone-blind. The Mismatched and Text-free sets overlap near zero.}
\label{fig:score}
\end{figure*}

\subsection{The cost to segmentation from the image}
\label{sec:rq3}
Learning the leakage weakens segmentation from the image (Fig.~\ref{fig:retraining}(d)). We found this on colonoscopy and confirmed it on breast ultrasound and chest X-ray, where it was specified in advance. On colonoscopy, with the empty caption, every Size-blind model beats every Original model, by $\rho_{\text{Size-blind}}=\nRelyNoSize$ points on average (Table~\ref{tab:cf}), and the Text-free models lead by $\rho_{\text{Text-free}}=\nRelyTF$ \ci{\nRelyTFLo}{\nRelyTFHi} in all nine seed pairings. The Size-only models are no better than the Original models ($\rho_{\text{Size-only}}=\nRelySizeOnly$ \ci{\nRelySizeOnlyLo}{\nRelySizeOnlyHi}), so the size word alone produces the deficit. Both leads stay above zero when seeds or video sequences are also resampled (\ref{app:sens}).

On BUSI the deficit is larger. On the 65 lesion images, $\rho_{\text{Size-blind}}=\nBRelyNoSize$ points, $\rho_{\text{Text-free}}=\nBRelyTF$ \ci{\nBRelyTFLo}{\nBRelyTFHi} and $\rho_{\text{Size-only}}=\nBRelySizeOnly$ \ci{\nBRelySizeOnlyLo}{\nBRelySizeOnlyHi}. Even the Size-only models segment better from the image than the Original models, so here the size word produces only part of the deficit (and the Size-only interval includes zero when seeds are resampled). Trained and tested without any text, the BUSI models score \nBNullTF{} on these images, against \nBOracleOrig{} for the Original models with the matched caption.

On chest X-ray, the Zone-blind models lead the Original models by $\rho_{\text{Zone-blind}}=\nQRelyBlind$ points (Table~\ref{tab:cf}), by at least \nQPairsBlindMinAbs{} in every seed pairing, and match the Text-free models (\nQNullBlind{} against \nQNullTF{} Dice with the empty caption). The Zone-only models are no better than the Original models ($\rho_{\text{Zone-only}}=\nQRelyZoneOnly$ \ci{\nQRelyZoneOnlyLo}{\nQRelyZoneOnlyHi}). Again, the extent attribute alone produces the deficit.

The deficit is not caused by the caption simply being absent \citep{ma2022}. The Size-blind and Mismatched models never saw an empty caption in training either, yet they show no deficit: the Mismatched models lead the Original models by $\rho_{\text{Mismatched}}=\nRelyShuf$ points \ci{\nRelyShufLo}{\nRelyShufHi}, in all nine seed pairings. What sets the Original models apart is a caption that states the image's own extent. We read this as a label-derived form of modality imbalance \citep{du2023,huang2022,wang2020,wu2022}: when the caption supplies the extent, the model learns less about it from the image.

\begin{table}[!t]
\centering
\caption{Caption-only leakage of the released and edited caption sets.}
\label{tab:score}
\footnotesize
\setlength{\tabcolsep}{3.5pt}
\begin{tabular}{@{}lrrr@{}}
\toprule
& $R^2_{\mathrm{cap}}$ & \multicolumn{2}{c}{$I_{\mathrm{ext}}$}\\
\cmidrule(lr){2-2}\cmidrule(l){3-4}
Captions & train & train & test\\
\midrule
Kvasir-SEG & \nRcapTrKvasir & \nIextKvasirTr & \nIextKvasirTe\\
CVC-ClinicDB & \nRcapTrClinic & \nIextClinicTr & \nIextClinicTe\\
CVC-300 & - & - & \nIextCvcThreeTe$^\ast$\\
BKAI-IGH & \nRcapTrBkai & \nIextBkaiTr & \nIextBkaiTe\\
CVC-ColonDB & - & - & \nIextColonTe\\
BUSI & \nRcapTrBusi & \nIextBusiTr & \nIextBusiTe\\
QaTa-COV19$^\dagger$ & \nRcapTrQata & \nIextQataTr & \nIextQataTe\\
\addlinespace
\multicolumn{4}{@{}l}{\emph{Kvasir-SEG training-caption sets of Section~\ref{sec:m-cf}}$^\S$}\\
Size-only & \nRcapCFSizeOnly & \nIextCFSizeOnly & -\\
Size-blind$^\dagger$ & \nRcapCFSizeBlind & \nIextCFSizeBlind & -\\
Mismatched & \nRcapCFShuf & \sIextCFShuf & -\\
Text-free & \nRcapCFTF & \nIextCFTF & -\\
\addlinespace
\multicolumn{4}{@{}l}{\emph{BUSI training-caption sets}$^{\S\P}$}\\
Size-only & \nBRcapSizeOnly & \nBIextCFSizeOnly & -\\
Size-blind$^\dagger$ & \nBRcapSizeBlind & \nBIextCFSizeBlind & -\\
Mismatched & \nBRcapShuf & \nBIextCFShuf & -\\
\addlinespace
\multicolumn{4}{@{}l}{\emph{QaTa-COV19 training-caption sets}$^\S$}\\
Zone-only$^\dagger$ & \nQRcapZoneOnly & \nQIextCFZoneOnly & -\\
Zone-blind$^\dagger$ & \nQRcapZoneBlind & \nQIextCFZoneBlind & -\\
\bottomrule
\end{tabular}\\[2pt]
\begin{minipage}{\columnwidth}\footnotesize Images with a lesion; test-set $R^2_{\mathrm{cap}}$ in Table~\ref{tab:released}. With shuffled captions, $R^2_{\mathrm{cap}}\le\nRcapPermTrMax$. $I_{\mathrm{ext}}$: mutual information between extent token and area tercile, relative to the tercile entropy (Section~\ref{sec:m-score}); permutation $p=0.005$ except for CVC-300 ($^\ast p=\nIextCvcThreeP$), Mismatched and Text-free. $^\dagger$No size word: the whole caption is the extent token. $^\S$Training $R^2_{\mathrm{cap}}$: x-axis of Fig.~\ref{fig:score}(d). $^\P$Whole caption as extent token, so $I_{\mathrm{ext}}$ is not comparable with the BUSI row above. -: not scored.\end{minipage}
\end{table}

\subsection{Detecting the leakage from the captions alone}
\label{sec:res-score}
The leakage can be measured before any model is trained. From the caption alone, the regression predicts lesion area with $R^2_{\mathrm{cap}}$ between \nRcapTeMinNoCvc{} and \nRcapTeMaxNoCvc{} on six of the seven test sets (Table~\ref{tab:released}) and between \nRcapTrBkai{} and \nRcapTrKvasir{} on the five training-caption sets (Table~\ref{tab:score}); shuffled captions score at most zero. The predictions fall into one band per size word (Fig.~\ref{fig:score}(a)-(c)). The model-free measure $I_{\mathrm{ext}}$ agrees: it lies between \nIextBkaiTr{} and \nIextKvasirTe{} for every released caption set except CVC-300, each with the smallest possible $p$-value (Table~\ref{tab:score}).

CVC-300 scores low because \nCvcSmallN{} of its \nCvcN{} captions say \emph{small}, so the size word can explain at most \nCvcCeiling{} of the variance in log lesion area, and its $I_{\mathrm{ext}}$ (\nIextCvcThreeTe, $p=\nIextCvcThreeP$) cannot be told apart from chance. Yet the model still reads the size word ($S_{\mathrm{ext}}=\nTabTwoCvcThreeSuff\%$). A low score therefore does not certify that a benchmark is clean \citep{feng2019}.

Within each training dataset, the score ranks the edited caption sets of Section~\ref{sec:m-cf} in the same order as the text gains of the models trained on them, and puts the uninformative sets at zero (Fig.~\ref{fig:score}(d)): it predicts what a set of captions will teach. Across datasets, it does not. Kvasir-SEG's Size-blind captions score \nRcapCFSizeBlind{} and leave \nRtwo\% of the gain, while QaTa-COV19's Zone-blind captions score \nQRcapZoneBlind{} and leave only \nQRtwo\%. For template captions the score merely rediscovers Eq.~\eqref{eq:quantiser}; it is most useful for human-written captions, such as those of QaTa-COV19, and for benchmarks whose construction is not documented.

\subsection{Performance without the test mask}
\label{sec:rq4}
No annotation-free method we tested reaches the reported performance (Table~\ref{tab:nomask}). Each scores below the published model with its matched caption (\nRelLeakAll{} Dice), and the 95\% interval of each difference excludes zero; this also holds without the duplicated CVC-300 images, with video sequences resampled and, where scored, on the normalised surface Dice (\ref{app:sens}). On colonoscopy, the edge-aligned threshold, which trains nothing but whose rule was chosen on the test sets, beats the Text-free models, and predicting the caption from the image is no better than training without text. On QaTa-COV19, without retuning, the threshold reaches \nEdgeQata, between the published model's \nQataNull{} with the empty caption and \nQataOracle{} with the matched one; the Text-free models reach \nQNullTF.

ExtentSeg is the best annotation-free method (\nXSAll{} Dice), \nXSvsTFAllAbs{} points \ci{\nXSvsTFAllLo}{\nXSvsTFAllHi} above the Text-free models and \nXSgap{} below the published model. On the unseen centres, the area loss adds \nXSvsNoADGAbs{} points \ci{\nXSvsNoADGLo}{\nXSvsNoADGHi} (ExtentSeg against ExtentSeg without it), and the estimated extent adds \nXSvsNoEDGAbs{} points \ci{\nXSvsNoEDGLo}{\nXSvsNoEDGHi} (ExtentSeg against the extent-free control). The second gain is positive in \nw{\nXSvsNoEDGPairs} of nine seed pairings, but its interval includes zero, so it misses its registered bar (Table~\ref{tab:criteria}). Given the true extent, ExtentSeg matches the retrained Original models with the matched caption (\nXStrueAll{} against \nRetLeakAll; difference \nXStrueVsRetAll{} \ci{\nXStrueVsRetAllLo}{\nXStrueVsRetAllHi}). For models trained the same way, a twenty-level extent code with no language model therefore carries the benefit of the caption. About half of ExtentSeg's gap to the published result (\nXSgapEst{} of \nXSgap{} points) is the cost of estimating the extent from the image. The rest roughly equals the gap between the published checkpoint and our retrained Original models (\nRzeroGapAbs{} points), which is indistinguishable from zero when video sequences are resampled (\ci{\sClusterRetLeakAllDLo}{\sClusterRetLeakAllDHi}; Table~\ref{tab:cluster}).

\begin{table}[!t]
\centering
\caption{Methods with and without access to the test mask.}
\label{tab:nomask}
\footnotesize
\setlength{\tabcolsep}{2.2pt}
\begin{tabular}{@{}lrrl@{}}
\toprule
& \multicolumn{2}{c}{Dice} & \\
\cmidrule(lr){2-3}
Method & all & unseen & $\Delta$, all (95\% CI)\\
\midrule
\multicolumn{4}{@{}l}{\emph{Annotation-free}}\\
Published, empty caption & \nRelEmptyAll & \nRelEmptyDG & \nRelEmptyAllD{} \ci{\nRelEmptyAllLo}{\nRelEmptyAllHi}\\
Edge-aligned threshold & \nEdgeCutAll & \nEdgeCutDG & \nEdgeCutAllD{} \ci{\nEdgeCutAllLo}{\nEdgeCutAllHi}\\
Image-only describer & \nDescriberAll & \nDescriberDG & \nDescriberAllD{} \ci{\nDescriberAllLo}{\nDescriberAllHi}\\
Text-free (trained without text) & \nTextFreeAll & \nTextFreeDG & \nTextFreeAllD{} \ci{\nTextFreeAllLo}{\nTextFreeAllHi}\\
Extent-free control & \nXSnoEAll & \nXSnoEDG & \nXSnoEAllD{} \ci{\nXSnoEAllLo}{\nXSnoEAllHi}\\
ExtentSeg without area loss & \nXSnoAAll & \nXSnoADG & \nXSnoAAllD{} \ci{\nXSnoAAllLo}{\nXSnoAAllHi}\\
ExtentSeg & \textbf{\nXSAll} & \textbf{\nXSDG} & \nXSAllD{} \ci{\nXSAllLo}{\nXSAllHi}\\
\addlinespace
\multicolumn{4}{@{}l}{\emph{Uses the test mask}}\\
ExtentSeg, true extent$^\dagger$ & \nXStrueAll & \nXStrueDG & \nXStrueAllD{} \ci{\nXStrueAllLo}{\nXStrueAllHi}\\
Retrained, matched caption$^\dagger$ & \nRetLeakAll & \nRetLeakDG & \nRetLeakAllD{} \ci{\nRetLeakAllLo}{\nRetLeakAllHi}\\
Published, matched caption$^\dagger$ & \nRelLeakAll & \nRelLeakDG & -\\
\bottomrule
\end{tabular}\\[3pt]
\begin{minipage}{\columnwidth}\footnotesize All five colonoscopy test sets ($n=681$) and the unseen centres ($n=\nUnseenN$). $\Delta$: paired difference from the published model with its matched caption. Trained models: mean of three seeds. Best annotation-free result in bold. $^\dagger$Reads the test mask, directly or through its caption.\end{minipage}
\end{table}

\section{Discussion}
\label{sec:disc}

\subsection{What the reported numbers measure}
\label{sec:d-measure}
The scores reported with mask-derived captions mix three things: segmentation from the image, a dataset-level correction of how much the model segments, and the lesion extent read from the caption. Only the first is what a segmentation benchmark sets out to measure. The second needs no caption of the image itself, as the mismatched caption shows. The third is not available in deployment, where the mask is what the model must produce, yet it accounts for most of what the caption contributes on colonoscopy and chest X-ray; on BUSI, the captions also carry the diagnosis and whether a lesion is present.

Text can still help. A clinician's description could convey extent, ExtentSeg shows that a model can use such information, and segmentation guided by the findings of routine radiology reports is now benchmarked at scale \citep{cf2seg}. The problem is where these benchmarks get the extent from. A clinician's \capt{large mass} is an uncertain judgement that is often not recorded at all. The benchmark's size word is an exact threshold of the scoring mask, the zone phrase was written for the benchmark from the same images, and both are present for every test image. We did not test clinician-written descriptions; on colonoscopy and chest X-ray, the gains measured here are carried by the benchmark's extent attribute.

\subsection{Recommendations}
\label{sec:d-recs}
We make four recommendations. (1) Captions given to a model at test time should never be computed from the test annotation. (2) Results should be reported with the empty, mismatched and matched captions, together with the prior share and, for template captions, the sufficiency and necessity of each attribute, alongside the metric guidance of \citet{reinke2024} and \citet{maierhein2024}. (3) The caption-only score should be computed whenever a captioned benchmark is released, as \citet{tst2025} propose for multimodal question answering. (4) Text-guided models should be compared with a text-free model trained with the same recipe, over more than one checkpoint.

\subsection{Scope and threats to validity}
\label{sec:scope}
The published models cover two architectures and three modalities; the retraining covers MedCLIPSeg on three training sets. RecLMIS was tested only at test time and only on QaTa-COV19, and LViT's code release has no QaTa-COV19 checkpoint \citep{lvit_code}; retraining other architectures is the next step. We showed that models trained with the extent attribute segment worse from the image, but not where in the network the deficit arises. ExtentSeg was evaluated on colonoscopy only.

The main results hold without the duplicated or mislabelled test images and when video sequences or subjects are resampled (\ref{app:sens}); the quantities that become less certain are reported with their results in Section~\ref{sec:results}. The caption-only score was not recomputed with folds grouped by video or subject. On BUSI, the published model's extent shares rest on 65 images and change noticeably without the \sBusiDupTestLesion{} near-duplicate lesion images ($S_{\mathrm{ext}}=\sBusiDedupSuff\%$, $N_{\mathrm{ext}}=\sBusiDedupNec\%$), while the retraining verdicts stay the same. Demographics and acquisition devices are not reported for most of the datasets, so we could not analyse those biases. Every model reads $224\times224$ inputs, the size fixed by the frozen encoders of the audited checkpoints, so higher input resolutions were not tested.

\section{Conclusion}
The benchmarks audited here hand text-guided segmenters, at test time, a caption that was computed from the scoring mask or written by the benchmark's annotators, and the models have learned to read the lesion's extent from it. On colonoscopy and chest X-ray, the benefit attributed to language is largely the benefit of knowing the size of the answer in advance; on breast ultrasound, the diagnosis word shares it. The habit is not free: on all three training sets, models trained on the full captions segment worse from the image alone than models trained without the extent attribute, and without the test mask no method we tested reaches the reported scores. Telling language from leakage takes three checks, which we recommend as standard practice: caption edits at test time, a caption-only score, and a text-free model trained with the same recipe.

\enlargethispage{5\baselineskip}
\section*{CRediT authorship contribution statement}
\textbf{Niran Nataraj}: Conceptualization, Methodology, Software, Validation, Formal analysis, Investigation, Data curation, Visualization, Writing - original draft, Writing - review \& editing. \textbf{Ankita Das}: Methodology, Validation, Investigation, Writing - original draft, Writing - review \& editing. \textbf{C. Krishna Mohan}: Resources, Supervision, Project administration, Writing - review \& editing. \textbf{Kenji Kawashima}: Resources, Supervision, Project administration, Writing - review \& editing.

\section*{Declaration of competing interest}
Kenji Kawashima holds stock in RiverField Inc., a surgical robotics company. Niran Nataraj, Ankita Das and C. Krishna Mohan declare that they have no known competing financial interests or personal relationships that could have appeared to influence the work reported in this paper.

\section*{Funding}
This research did not receive any specific grant from funding agencies in the public, commercial, or not-for-profit sectors.

\section*{Data availability}
All seven datasets are public (Kvasir-SEG, CVC-ClinicDB, CVC-300, CVC-ColonDB, BKAI-IGH, BUSI, QaTa-COV19), as are the captions and checkpoints released with MedCLIPSeg \citep{medclipseg_data,medclipseg_ckpt} and RecLMIS \citep{reclmis_code}. The edited caption sets, per-image score tables, the analysis code that generates every number and figure in this article, and the criteria of \ref{app:registration} will be made public at \url{https://github.com/felirox/ExtentProbe} on acceptance; until then the repository is available to the editors and reviewers on request. The authors' code is released under the PolyForm Noncommercial License 1.0.0; the patch to the MedCLIPSeg code keeps that code's Apache License 2.0, and the derived caption sets keep the licences of the captions they are built from (MedVLSM and LViT, MIT License). The dataset images are not redistributed; they are obtained from their original sources.

\section*{Declaration of generative AI and AI-assisted technologies in the manuscript preparation process}
During the preparation of this work, the authors used Grammarly to improve the clarity, readability, and academic presentation of the English text. After using this service, the authors reviewed and edited the manuscript as needed and take full responsibility for the content of the published article.

\onecolumn
\appendix
\counterwithin{table}{section}
\renewcommand{\thetable}{\Alph{section}.\arabic{table}}
\section{Registration record}
\label{app:registration}
Here, \emph{registration} means that a criterion was written in the project's dated research record before the results it decides were seen, and was committed to version control; no external registry was used. For each family of registered criteria reported in this paper, Table~\ref{tab:registration} gives where the criteria were written, the first commit that contains them, and when the first and the deciding results were seen. Commit times are those recorded by version control, in India Standard Time (IST, UTC+5:30, where UTC is Coordinated Universal Time) for September 2026 and Japan Standard Time (JST, UTC+9) for October 2026; other times come from the dated research record (the project's wiki and run logs). The text of every registered criterion, with its commit time, is included in the code release (see Data availability).

\begin{table*}[!tb]
\centering
\caption{Registration record of the criteria in Table~\ref{tab:criteria}.}
\label{tab:registration}
\footnotesize
\setlength{\tabcolsep}{3pt}
\begin{tabular}{@{}>{\raggedright\arraybackslash}p{2.6cm}>{\raggedright\arraybackslash}p{3.2cm}>{\raggedright\arraybackslash}p{3.0cm}>{\raggedright\arraybackslash}p{2.7cm}>{\raggedright\arraybackslash}p{2.9cm}>{\raggedright\arraybackslash}p{2.5cm}@{}}
\toprule
Criteria & Written (dated record) & First commit containing them & First run & First result seen & Deciding result\\
\midrule
Retraining, Kvasir-SEG (R0-R4) & research plan, dated by its author 22 Sep about 16:30 IST, while the first Original run was training & \texttt{e19ee2a}, 22 Sep 23:06 IST (audit script with the bars) & 22 Sep 16:16 IST (first Original run, by the project log; its log was last written at 17:20 IST) & seed 1: R0 22 Sep 18:40, R1-R2 21:12, R3 23:35 IST; R4 23 Sep 02:10 IST & three seeds: R0-R3 23 Sep 13:02-15:03 IST; R4 7 Oct\\
Retraining, BUSI and QaTa-COV19 (R0-R4, $\rho$) & declaration, 5 Oct; Amendment~1 (R4 intervals, $\rho$, \emph{consistent}) the same day & \texttt{9f31c22}/\texttt{e2c2051}, 5 Oct 16:14-16:15 JST; amendment \texttt{0e94675}, 16:40 JST & queue started 5 Oct 16:20 JST with two Kvasir-SEG Mismatched seeds; first BUSI run after 17:15 JST; first QaTa-COV19 run 6 Oct & BUSI 7 Oct; QaTa-COV19 10 Oct & BUSI 7 Oct; QaTa-COV19 10 Oct\\
ExtentSeg against the published model, and its area loss & architecture plan, 23 Sep 01:28 IST, before any output of the model & \texttt{e484275}, 23 Sep 15:27 IST & first run queued 23 Sep 01:31 IST & seed 1: 23 Sep 11:40 IST & three seeds: 24 Sep 01:40 IST\\
ExtentSeg against its extent-free control & research plan, 23 Sep, before any output of the control & \texttt{d7160da}, 23 Sep 21:13 IST (audit script with the bar) & control built 23 Sep 12:50 IST & seed 1: 23 Sep 21:20 IST & three seeds: after 24 Sep 01:25 IST, when seeds 2-3 of the control were launched\\
RecLMIS (reproduction, extent necessity) & research plan, 22 Sep, before either was run & \texttt{e19ee2a}, 22 Sep 23:06 IST & - (published checkpoint) & 22 Sep & 22 Sep\\
Caption style & research plan, 22 Sep, before any rewritten caption was generated & \texttt{e19ee2a}, 22 Sep 23:06 IST & - (published checkpoints) & 22 Sep & 22 Sep\\
\bottomrule
\end{tabular}
\end{table*}

For BUSI and QaTa-COV19 the record is complete in version control: the criteria and their amendment were committed before the first of these models began training. For the other families, the dated research record places the criteria before the results they decide (for Kvasir-SEG, after the first Original run had started and before it finished), but the first commit that contains them came later. The Kvasir-SEG criteria were first committed after the single-seed results of R0-R2 had been seen and before those of R3 and R4 and every three-seed result; the criteria for ExtentSeg against the published model, after the seed-1 result and before the three-seed result; and those for RecLMIS and the caption-style test, after the results themselves. The bar for ExtentSeg against its extent-free control was committed before its first result. For RecLMIS, the matched caption in MedCLIPSeg's wording had been partly run by mistake, and its running mean seen, before the criteria were written; that condition is therefore reported as an observation (\ref{app:sens}), not as a test.

\section{Registered criteria}
\label{app:criteria}
Table~\ref{tab:criteria} lists every registered criterion whose analysis appears in this paper, with its bar, its result and its verdict. Criteria registered while ExtentSeg was being developed, each tested on one seed, are not in the table and are summarised here. Two concerned caption-based variants that are not reported; neither was met. Six concerned ExtentSeg itself, its decoding or its controls. One of them was met: ExtentSeg without the area loss outscores the best caption-based variant. The other five were not met. ExtentSeg did not beat a fixed offset of the output logits for each extent bin by two points, so most of what its extent input provides acts like such an offset. Re-estimating the extent from the predicted mask, mixing the outputs of the conditioned and unconditioned passes, and an extent head that also reads the area of its own map each failed to improve Dice. Finally, the area loss did not improve the extent-free control on its first seed.

\begin{table*}[!tb]
\centering
\caption{Registered criteria reported in this paper.}
\label{tab:criteria}
\footnotesize
\setlength{\tabcolsep}{3pt}
\begin{tabular}{@{}>{\raggedright\arraybackslash}p{3.3cm}>{\raggedright\arraybackslash}p{5.6cm}>{\raggedright\arraybackslash}p{4.3cm}>{\raggedright\arraybackslash}p{3.6cm}@{}}
\toprule
Criterion & Bar & Result & Verdict\\
\midrule
R0-R4 and $\rho$, retraining on three training sets & Section~\ref{sec:m-cf} & Table~\ref{tab:cf} & Kvasir-SEG: R0-R4 met ($\rho$ not registered). QaTa-COV19: every criterion run was met. BUSI: R0, R1 and $\rho$ met; R2 consistent, and met on the second population specified in advance (Section~\ref{sec:rq2}); R3 and R4 not met\\
ExtentSeg against the published model & ExtentSeg without area loss (the registered model) minus the published model with its matched caption, unseen centres: interval above zero & \nXSnoADGD{} \ci{\nXSnoADGLo}{\nXSnoADGHi}; with the area loss, \nXSDGD{} \ci{\nXSDGLo}{\nXSDGHi} & not met\\
Area loss of ExtentSeg & ExtentSeg minus ExtentSeg without area loss, unseen centres: at least $+0.5$, interval excluding zero & \nXSvsNoADG{} \ci{\nXSvsNoADGLo}{\nXSvsNoADGHi} & met\\
Extent input of ExtentSeg & ExtentSeg minus its extent-free control, unseen centres: at least $+0.5$, interval excluding zero & \nXSvsNoEDG{} \ci{\nXSvsNoEDGLo}{\nXSvsNoEDGHi} & not met: positive in \nw{\nXSvsNoEDGPairs} of nine seed pairings, interval includes zero (Section~\ref{sec:rq4})\\
RecLMIS reproduction & Dice with its own captions within 1.0 of the reported \nRecPublished & \nRecRepro & met\\
RecLMIS extent necessity & deleting the zone phrase costs at least 50\% of the text gain, deletion written in MedCLIPSeg's wording & \nRecNecReg\% \ci{\nRecNecRegLo}{\nRecNecRegHi} & not met; in RecLMIS's own wording the necessity is \nTabTwoRecNec\% (Table~\ref{tab:released})\\
Caption style & the four restyles of the same content (terse, verbose, clinical, list) each within 1.0 point of the terse one, pooled over the seven test sets, each paired interval within $\pm1.5$ & largest difference \nParaMaxShiftTerse{} points (list style; Table~\ref{tab:para}) & not met\\
\bottomrule
\end{tabular}\\[2pt]
\begin{minipage}{\textwidth}\footnotesize Retrained models and ExtentSeg: three-seed means; RecLMIS and caption style: published checkpoints; image bootstrap throughout. The RecLMIS necessity criterion deleted the zone phrase in MedCLIPSeg's wording, not RecLMIS's own (Table~\ref{tab:released} gives the matched value). The caption-style criterion missed only through the list style, which is not a sentence.\end{minipage}
\end{table*}

\section{Implementation and reproducibility}
\label{app:impl}
All retrained models start from the released UniMed-CLIP ViT-B/16 weights (SHA-256 prefix \texttt{bfdce18ca2473c56}, where SHA-256 is the 256-bit Secure Hash Algorithm) and are trained with the MedCLIPSeg code \citep{medclipseg_code} at commit \texttt{31d2220} with a local patch that adds ExtentSeg (patch digest \texttt{4bdb08c88d52534b}, SHA-256 prefix). The released version of the patch (digest \texttt{7de4a5d0d2762fb6}) keeps only the code these runs use: with ExtentSeg off it gives the outputs of the published models bit for bit (32 test images each of Kvasir-SEG and BUSI), and on the same inputs and seeds it reproduces the probability maps of the original patch and, after one training epoch, its weights bit for bit; the training-caption sets are passed through the released configuration options. Models are trained and run with PyTorch \citep{pytorch}; the regressions and classifiers of Sections~\ref{sec:m-score} and~\ref{sec:m-free} and of \ref{app:qata} use scikit-learn \citep{sklearn}. Inference with a fixed seed is deterministic on a given machine: on eight CVC-300 images, two runs of the 30-pass stochastic inference with the same seed give bitwise identical probability maps, and a different seed changes them; across machines the maps can differ slightly. The SHA-256 digests of the published checkpoints and of the last-epoch checkpoint of every retrained model are part of the code release (\texttt{paper/data/ckpt\_digests.json}; see Data availability); the digests of the four published checkpoints equal those of the files distributed by their authors (checked on 22 October 2026).

The provenance of the colonoscopy and BUSI captions (Section~\ref{sec:quantiser}) was checked by comparing the caption files released with MedCLIPSeg with the prompt files of \citet{poudel2024} \citep[commit \texttt{dc36a3a};][]{medvlsm_code}: all \nPoudelRows{} rows of the original training, validation and test caption files of the six colonoscopy and BUSI sets equal a P6 prompt of the same image (for BUSI, one of the two P6 variants stored for most images).

The BUSI and QaTa-COV19 retraining and the Kvasir-SEG Mismatched seeds 2 and 3 ran on an NVIDIA RTX 5000 Ada graphics processing unit; one-epoch timings on that card put a 100-epoch run at about 1.0-1.2 h for Kvasir-SEG, 0.8-0.9 h for BUSI and 6.5 h for QaTa-COV19, with a peak memory of about 9.6 GB (project run log of 5 October 2026). The other Kvasir-SEG retrained models, ExtentSeg and its controls were trained on one NVIDIA Quadro RTX 6000 (24 GB). All runs used Python 3.13.9, PyTorch 2.12.0 built for CUDA 12.6, Transformers 4.56.0 and NumPy 2.5.3.

RecLMIS is run from its released code \citep{reclmis_code} at commit \texttt{d3265c8} with the released QaTa-COV19 checkpoint and test script (Section~\ref{sec:setup}). Two changes, neither on the segmentation path, were needed: the attention module imports a softmax from \texttt{fairseq}, which is not among the released requirements and does not build here, so that one call is inlined as \texttt{torch.nn.functional.softmax(x, dim=-1, dtype=torch.float32)}; and the image file names differ between the RecLMIS and MedCLIPSeg distributions of the dataset, which is handled with symbolic links rather than by editing the data. A third change makes the test script write an optional per-image score file. The code changes and the script that creates the links are part of the code release (see Data availability).

\emph{Further method details.} The released MedCLIPSeg evaluation configuration loads the latest checkpoint file (\texttt{TEST.USE\_LATEST}); the epoch of the published checkpoints is not stated. The ridge penalty of the caption-only score (Section~\ref{sec:m-score}) is chosen by leave-one-out cross-validation inside each training fold, and its chance level averages five shuffles. On QaTa-COV19 the edge-aligned threshold averages seven fixed captions: \capt{Bilateral pulmonary infection involving two regions with involvement of \ldots} completed with each of \capt{the lower right lung}, \capt{the lower left lung and the lower right lung}, \capt{the middle lower left lung and the middle lower right lung} and \capt{all left lung and all right lung}, and \capt{a small portion of the lungs} with each of its three size words. In ExtentSeg, $\mathbf c$ holds the geometric centres of the bins, with the outer bins closed at half the first edge and at 1.5 times the last; the area term excludes the \sXSBinZeroN{} smallest of the \sXSBinZeroTrainN{} training lesions (those below the first bin edge, whatever rotation the training augmentation applies).

\section{Sensitivity analyses}
\label{app:sens}
Each analysis in this appendix recomputes quantities of Sections~\ref{sec:rq1}-\ref{sec:rq4} from the same per-image scores, with the bootstrap of Section~\ref{sec:stats}; none trains a new model.

\emph{Video-sequence clusters.} Colonoscopy clusters are formed from the images alone, without reference to any score. CVC-ColonDB and CVC-ClinicDB test frames are chained in file order, and a new cluster starts at a gap of more than three file numbers or where the correlation of adjacent $32\times32$ grey-level images falls below 0.6 (0.5 and 0.7 in Table~\ref{tab:cluster}). CVC-300 images that repeat a CVC-ColonDB image join its cluster, the other CVC-300 images are chained in the same way, and clusters that share a byte-identical image are merged. Kvasir-SEG and BKAI-IGH have no sequence information, so each of their images is its own cluster. Clusters are resampled within each test set, and a cluster that spans CVC-300 and CVC-ColonDB is drawn with CVC-ColonDB.

\emph{Duplicated and mislabelled test images.} Table~\ref{tab:dedup} removes the duplicated, near-duplicated and mislabelled test images of Section~\ref{sec:data}. Without the \sDedupDropN{} duplicated CVC-300 images, no $\Delta$ of Table~\ref{tab:nomask} moves by more than \sDedupTabFiveMaxShift{} points, and ExtentSeg remains the best annotation-free method (\sDedupXSAll{} Dice, $\Delta=\sDedupXSAllD$ \ci{\sDedupXSAllDLo}{\sDedupXSAllDHi}).

\emph{Video sequences and subjects.} Table~\ref{tab:cluster} resamples colonoscopy video sequences instead of images. Resampling the \sQSubjN{} QaTa-COV19 test subjects instead of the \sQSubjImagesN{} images moves no interval bound of the retrained or published MedCLIPSeg models by more than 1.1 points: R2 is \ci{\sQClusterRtwoLo}{\sQClusterRtwoHi}\%, R3 \ci{\sQClusterRthreeLo}{\sQClusterRthreeHi}\%, $\rho_{\text{Zone-blind}}$ \ci{\sQClusterRelyBlindLo}{\sQClusterRelyBlindHi}, $\rho_{\text{Zone-only}}$ \ci{\sQClusterRelyZoneOnlyLo}{\sQClusterRelyZoneOnlyHiThree} (its upper bound now just above zero) and $\pi$ \ci{\sQClusterPiLo}{\sQClusterPiHi}\%. For RecLMIS, the prior-share interval widens most, from \ci{\sRecPiMixedLo}{\sRecPiMixedHi}\% to \ci{\sQRecClusterPiLo}{\sQRecClusterPiHi}\%, and $S_{\mathrm{ext}}$ is \ci{\sQRecClusterSuffLo}{\sQRecClusterSuffHi}\%.

\emph{Training seeds.} When the three seeds of each model are resampled together with the images, the colonoscopy empty-caption differences stay above zero: $\rho_{\text{Size-blind}}$ is \ci{\nRelyNoSizeHierLo}{\nRelyNoSizeHierHi} and $\rho_{\text{Text-free}}$ \ci{\nRelyTFHierLo}{\nRelyTFHierHi}, while R2 is \ci{\nRtwoHierLo}{\nRtwoHierHi}\% and R3 \ci{\nRthreeHierLo}{\nRthreeHierHi}\%, so both intervals cross their bars. On BUSI, R2 is \ci{\sBRtwoHierLo}{\sBRtwoHierHi}\% and R3 \ci{\sBRthreeHierLo}{\sBRthreeHierHi}\% on the 65 lesion images, so both intervals cross their bars (R2 meets its bar in \nw{\sBRtwoPairsMet} of the nine seed pairings, R3 in \nw{\sBRthreePairsMet}). $\rho_{\text{Size-blind}}$, $\rho_{\text{Mismatched}}$ and $\rho_{\text{Text-free}}$ stay above zero (\ci{\sBRelyNoSizeHierLo}{\sBRelyNoSizeHierHi}, \ci{\sBRelyShufHierLo}{\sBRelyShufHierHi} and \ci{\sBRelyTFHierLo}{\sBRelyTFHierHi}), while $\rho_{\text{Size-only}}$ includes zero (\ci{\sBRelySizeOnlyHierLo}{\sBRelySizeOnlyHierHi}). On QaTa-COV19, R2 is \ci{\nQRtwoHierLo}{\nQRtwoHierHi}\% and R3 \ci{\nQRthreeHierLo}{\nQRthreeHierHi}\%, so only R3 crosses its bar; $\rho_{\text{Zone-blind}}$ is \ci{\nQRelyBlindHierLo}{\nQRelyBlindHierHi} and $\rho_{\text{Zone-only}}$ \ci{\nQRelyZoneOnlyHierLo}{\nQRelyZoneOnlyHierHi}. Table~\ref{tab:perseed} gives the retrained models seed by seed.

\emph{Checkpoint selection.} When each Kvasir-SEG run's checkpoint is selected by validation Dice instead of taken at the last epoch, $G_{\text{Original}}=\nGbest$ \ci{\nGbestLo}{\nGbestHi}, R2 is \nRtwoBest\% \ci{\nRtwoBestLo}{\nRtwoBestHi}, R3 \nRthreeBest\% \ci{\nRthreeBestLo}{\nRthreeBestHi} and $\rho_{\text{Text-free}}=\nRelyTFBest$ \ci{\nRelyTFBestLo}{\nRelyTFBestHi}.

\emph{Other scores and tests.} Table~\ref{tab:nsd} repeats Table~\ref{tab:released} on the normalised surface Dice. On that score ExtentSeg is again the best of the annotation-free methods of Table~\ref{tab:nomask} that were scored on it (\sNsdXSAll{}, against \sNsdRelLeakAll{} for the published model with its matched caption). Table~\ref{tab:para} gives the caption-style test, and Table~\ref{tab:rules} the label-free rules from which the edge-aligned threshold was chosen.

\emph{RecLMIS.} On QaTa-COV19 ($n=\sSplitQataTest$), RecLMIS scores \nRecDiceOracle{} Dice with the matched caption in MedCLIPSeg's wording, against \nTabTwoRecMatch{} in its own. In the LViT wording, it scores \nRecDiceZoneOnlyLvit{} with the zone phrase alone ($S_{\mathrm{ext}}$ \ci{\nRecSuffLo}{\nRecSuffHi}\%) and \nRecDiceDelLvit{} with the zone phrase deleted ($N_{\mathrm{ext}}$ \ci{\nRecNecLo}{\nRecNecHi}\%); the registered deletion in MedCLIPSeg's wording gives \nRecDiceDel. Forcing the zone phrase to one or to six zones gives \nRecDiceOneZone{} and \nRecDiceSixZones. With matched and mismatched captions both taken from MedCLIPSeg's set, its prior share is \sRecPiSameRegister\% \ci{\sRecPiSameRegisterLo}{\sRecPiSameRegisterHi}.

\emph{Mismatched captions identical to the image's own.} Outside QaTa-COV19 such captions are rare (at most \sSameTextColonN{} test images per test set). Removing them changes $\pi$ by at most \sSameTextColonMaxShift{} points on any colonoscopy test set (pooled: \sSameTextPolypPiExcl\% \ci{\sSameTextPolypPiExclLo}{\sSameTextPolypPiExclHi}) and moves BUSI further below zero (\sSameTextBusiPiExcl\%). QaTa-COV19 is reported in Section~\ref{sec:rq1}. The excluded images are those whose caption is shared with other test images, not a random subset, so these values show where the prior gain comes from; they do not estimate it for a permutation of caption texts.

\begin{table*}[!tb]
\centering
\caption{Main quantities without duplicated or mislabelled test images.}
\label{tab:dedup}
\footnotesize
\setlength{\tabcolsep}{3.5pt}
\begin{tabular}{@{}lrrrrrrrr@{}}
\toprule
& \multicolumn{2}{c}{Colonoscopy, all ($n=\sDedupN{}$)} & \multicolumn{2}{c}{Unseen centres ($n=\sDedupNDG{}$)} & \multicolumn{2}{c}{Without \capt{no polyp} ($n=\sNoPolypN{}$)} & \multicolumn{2}{c}{BUSI lesions ($n=\sBusiDedupN{}$)}\\
\cmidrule(lr){2-3}\cmidrule(lr){4-5}\cmidrule(lr){6-7}\cmidrule(l){8-9}
Quantity & value (95\% CI) & change & value (95\% CI) & change & value (95\% CI) & change & value (95\% CI) & change\\
\midrule

$G$ (points) & \sDedupG{}~\ci{\sDedupGLo{}}{\sDedupGHi{}} & \sDedupGDelta{} & \sDedupGDG{}~\ci{\sDedupGDGLo{}}{\sDedupGDGHi{}} & \sDedupGDGDelta{} & \sNoPolypG{}~\ci{\sNoPolypGLo{}}{\sNoPolypGHi{}} & \sNoPolypGDelta{} & \sBusiDedupG{}~\ci{\sBusiDedupGLo{}}{\sBusiDedupGHi{}} & \sBusiDedupGDelta{}\\
$P$ (points) & \sDedupP{}~\ci{\sDedupPLo{}}{\sDedupPHi{}} & \sDedupPDelta{} & \sDedupPDG{}~\ci{\sDedupPDGLo{}}{\sDedupPDGHi{}} & \sDedupPDGDelta{} & \sNoPolypP{}~\ci{\sNoPolypPLo{}}{\sNoPolypPHi{}} & \sNoPolypPDelta{} & \sBusiDedupP{}~\ci{\sBusiDedupPLo{}}{\sBusiDedupPHi{}} & \sBusiDedupPDelta{}\\
$\pi$ (\%) & \sDedupPi{}~\ci{\sDedupPiLo{}}{\sDedupPiHi{}} & \sDedupPiDelta{} & \sDedupPiDG{}~\ci{\sDedupPiDGLo{}}{\sDedupPiDGHi{}} & \sDedupPiDGDelta{} & \sNoPolypPi{}~\ci{\sNoPolypPiLo{}}{\sNoPolypPiHi{}} & \sNoPolypPiDelta{} & \sBusiDedupPi{}~\ci{\sBusiDedupPiLo{}}{\sBusiDedupPiHi{}} & \sBusiDedupPiDelta{}\\
$S_{\mathrm{ext}}$ (\%) & \sDedupSuff{}~\ci{\sDedupSuffLo{}}{\sDedupSuffHi{}} & \sDedupSuffDelta{} & \sDedupSuffDG{}~\ci{\sDedupSuffDGLo{}}{\sDedupSuffDGHi{}} & \sDedupSuffDGDelta{} & \sNoPolypSuff{}~\ci{\sNoPolypSuffLo{}}{\sNoPolypSuffHi{}} & \sNoPolypSuffDelta{} & \sBusiDedupSuff{}~\ci{\sBusiDedupSuffLo{}}{\sBusiDedupSuffHi{}} & \sBusiDedupSuffDelta{}\\
$N_{\mathrm{ext}}$ (\%) & \sDedupNec{}~\ci{\sDedupNecLo{}}{\sDedupNecHi{}} & \sDedupNecDelta{} & \sDedupNecDG{}~\ci{\sDedupNecDGLo{}}{\sDedupNecDGHi{}} & \sDedupNecDGDelta{} & \sNoPolypNec{}~\ci{\sNoPolypNecLo{}}{\sNoPolypNecHi{}} & \sNoPolypNecDelta{} & \sBusiDedupNec{}~\ci{\sBusiDedupNecLo{}}{\sBusiDedupNecHi{}} & \sBusiDedupNecDelta{}\\
R2 (\%) & \sDedupRtwo{}~\ci{\sDedupRtwoLo{}}{\sDedupRtwoHi{}} & \sDedupRtwoDelta{} & \sDedupRtwoDG{}~\ci{\sDedupRtwoDGLo{}}{\sDedupRtwoDGHi{}} & \sDedupRtwoDGDelta{} & - & - & \sBusiDedupRtwo{}~\ci{\sBusiDedupRtwoLo{}}{\sBusiDedupRtwoHi{}} & \sBusiDedupRtwoDelta{}\\
R3 (\%) & \sDedupRthree{}~\ci{\sDedupRthreeLo{}}{\sDedupRthreeHi{}} & \sDedupRthreeDelta{} & \sDedupRthreeDG{}~\ci{\sDedupRthreeDGLo{}}{\sDedupRthreeDGHi{}} & \sDedupRthreeDGDelta{} & - & - & \sBusiDedupRthree{}~\ci{\sBusiDedupRthreeLo{}}{\sBusiDedupRthreeHi{}} & \sBusiDedupRthreeDelta{}\\
$\rho_{\text{Size-blind}}$ & \sDedupRelyNoSize{}~\ci{\sDedupRelyNoSizeLo{}}{\sDedupRelyNoSizeHi{}} & \sDedupRelyNoSizeDelta{} & \sDedupRelyNoSizeDG{}~\ci{\sDedupRelyNoSizeDGLo{}}{\sDedupRelyNoSizeDGHi{}} & \sDedupRelyNoSizeDGDelta{} & - & - & \sBusiDedupRelyNoSize{}~\ci{\sBusiDedupRelyNoSizeLo{}}{\sBusiDedupRelyNoSizeHi{}} & \sBusiDedupRelyNoSizeDelta{}\\
$\rho_{\text{Size-only}}$ & \sDedupRelySizeOnly{}~\ci{\sDedupRelySizeOnlyLo{}}{\sDedupRelySizeOnlyHi{}} & \sDedupRelySizeOnlyDelta{} & \sDedupRelySizeOnlyDG{}~\ci{\sDedupRelySizeOnlyDGLo{}}{\sDedupRelySizeOnlyDGHi{}} & \sDedupRelySizeOnlyDGDelta{} & - & - & \sBusiDedupRelySizeOnly{}~\ci{\sBusiDedupRelySizeOnlyLo{}}{\sBusiDedupRelySizeOnlyHi{}} & \sBusiDedupRelySizeOnlyDelta{}\\
$\rho_{\text{Mismatched}}$ & \sDedupRelyShuf{}~\ci{\sDedupRelyShufLo{}}{\sDedupRelyShufHi{}} & \sDedupRelyShufDelta{} & \sDedupRelyShufDG{}~\ci{\sDedupRelyShufDGLo{}}{\sDedupRelyShufDGHi{}} & \sDedupRelyShufDGDelta{} & - & - & \sBusiDedupRelyShuf{}~\ci{\sBusiDedupRelyShufLo{}}{\sBusiDedupRelyShufHi{}} & \sBusiDedupRelyShufDelta{}\\
$\rho_{\text{Text-free}}$ & \sDedupRelyTF{}~\ci{\sDedupRelyTFLo{}}{\sDedupRelyTFHi{}} & \sDedupRelyTFDelta{} & \sDedupRelyTFDG{}~\ci{\sDedupRelyTFDGLo{}}{\sDedupRelyTFDGHi{}} & \sDedupRelyTFDGDelta{} & - & - & \sBusiDedupRelyTF{}~\ci{\sBusiDedupRelyTFLo{}}{\sBusiDedupRelyTFHi{}} & \sBusiDedupRelyTFDelta{}\\
\bottomrule
\end{tabular}\\[2pt]
\begin{minipage}{\textwidth}\footnotesize Recomputed on reduced populations with the image bootstrap. Colonoscopy: without the \sDedupDropN{} CVC-300 images that are also CVC-ColonDB images. \capt{no polyp}: without the \nw{\sBkaiNoPolypN} BKAI-IGH images so captioned despite a polyp in the mask. BUSI: lesion images without the \sBusiDupTestLesion{} near-duplicates of training or validation images. Change: value minus the full-population value (percentage points for $\pi$, shares and ratios).\end{minipage}
\end{table*}

\begin{table*}[!tb]
\centering
\caption{Colonoscopy intervals when video sequences are resampled instead of images.}
\label{tab:cluster}
\footnotesize
\setlength{\tabcolsep}{4pt}
\begin{tabular}{@{}lrrrrr@{}}
\toprule
Quantity ($n=681$) & Estimate & Images & Sequences, 0.6 & Sequences, 0.5 & Sequences, 0.7\\
\midrule

$G$ (points) & \sClusterG{} & \ci{\sClusterGImgLo{}}{\sClusterGImgHi{}} & \ci{\sClusterGLo{}}{\sClusterGHi{}} & \ci{\sClusterFiveGLo{}}{\sClusterFiveGHi{}} & \ci{\sClusterSevenGLo{}}{\sClusterSevenGHi{}}\\
$P$ (points) & \sClusterP{} & \ci{\sClusterPImgLo{}}{\sClusterPImgHi{}} & \ci{\sClusterPLo{}}{\sClusterPHi{}} & \ci{\sClusterFivePLo{}}{\sClusterFivePHi{}} & \ci{\sClusterSevenPLo{}}{\sClusterSevenPHi{}}\\
$\pi$ (\%) & \sClusterPi{} & \ci{\sClusterPiImgLo{}}{\sClusterPiImgHi{}} & \ci{\sClusterPiLo{}}{\sClusterPiHi{}} & \ci{\sClusterFivePiLo{}}{\sClusterFivePiHi{}} & \ci{\sClusterSevenPiLo{}}{\sClusterSevenPiHi{}}\\
$S_{\mathrm{ext}}$ (\%) & \sClusterSuff{} & \ci{\sClusterSuffImgLo{}}{\sClusterSuffImgHi{}} & \ci{\sClusterSuffLo{}}{\sClusterSuffHi{}} & \ci{\sClusterFiveSuffLo{}}{\sClusterFiveSuffHi{}} & \ci{\sClusterSevenSuffLo{}}{\sClusterSevenSuffHi{}}\\
$N_{\mathrm{ext}}$ (\%) & \sClusterNec{} & \ci{\sClusterNecImgLo{}}{\sClusterNecImgHi{}} & \ci{\sClusterNecLo{}}{\sClusterNecHi{}} & \ci{\sClusterFiveNecLo{}}{\sClusterFiveNecHi{}} & \ci{\sClusterSevenNecLo{}}{\sClusterSevenNecHi{}}\\
R2 (\%) & \sClusterRtwo{} & \ci{\sClusterRtwoImgLo{}}{\sClusterRtwoImgHi{}} & \ci{\sClusterRtwoLo{}}{\sClusterRtwoHi{}} & \ci{\sClusterFiveRtwoLo{}}{\sClusterFiveRtwoHi{}} & \ci{\sClusterSevenRtwoLo{}}{\sClusterSevenRtwoHi{}}\\
R3 (\%) & \sClusterRthree{} & \ci{\nRthreeLo{}}{\nRthreeHi{}} & \ci{\sClusterRthreeLo{}}{\sClusterRthreeHi{}} & \ci{\sClusterFiveRthreeLo{}}{\sClusterFiveRthreeHi{}} & \ci{\sClusterSevenRthreeLo{}}{\sClusterSevenRthreeHi{}}\\
$\rho_{\text{Size-blind}}$ & \sClusterRelyNoSize{} & \ci{\sClusterRelyNoSizeImgLo{}}{\sClusterRelyNoSizeImgHi{}} & \ci{\sClusterRelyNoSizeLo{}}{\sClusterRelyNoSizeHi{}} & \ci{\sClusterFiveRelyNoSizeLo{}}{\sClusterFiveRelyNoSizeHi{}} & \ci{\sClusterSevenRelyNoSizeLo{}}{\sClusterSevenRelyNoSizeHi{}}\\
$\rho_{\text{Size-only}}$ & \sClusterRelySizeOnly{} & \ci{\sClusterRelySizeOnlyImgLo{}}{\sClusterRelySizeOnlyImgHi{}} & \ci{\sClusterRelySizeOnlyLo{}}{\sClusterRelySizeOnlyHi{}} & \ci{\sClusterFiveRelySizeOnlyLo{}}{\sClusterFiveRelySizeOnlyHi{}} & \ci{\sClusterSevenRelySizeOnlyLo{}}{\sClusterSevenRelySizeOnlyHi{}}\\
$\rho_{\text{Mismatched}}$ & \sClusterRelyShuf{} & \ci{\sClusterRelyShufImgLo{}}{\sClusterRelyShufImgHi{}} & \ci{\sClusterRelyShufLo{}}{\sClusterRelyShufHi{}} & \ci{\sClusterFiveRelyShufLo{}}{\sClusterFiveRelyShufHi{}} & \ci{\sClusterSevenRelyShufLo{}}{\sClusterSevenRelyShufHi{}}\\
$\rho_{\text{Text-free}}$ & \sClusterRelyTF{} & \ci{\sClusterRelyTFImgLo{}}{\sClusterRelyTFImgHi{}} & \ci{\sClusterRelyTFLo{}}{\sClusterRelyTFHi{}} & \ci{\sClusterFiveRelyTFLo{}}{\sClusterFiveRelyTFHi{}} & \ci{\sClusterSevenRelyTFLo{}}{\sClusterSevenRelyTFHi{}}\\
\addlinespace
\multicolumn{6}{@{}l}{\emph{$\Delta$ of Table~\ref{tab:nomask}, all five test sets}}\\
Published, empty caption & \nRelEmptyAllD{} & \ci{\nRelEmptyAllLo{}}{\nRelEmptyAllHi{}} & \ci{\sClusterRelEmptyAllDLo{}}{\sClusterRelEmptyAllDHi{}} & \ci{\sClusterFiveRelEmptyAllDLo{}}{\sClusterFiveRelEmptyAllDHi{}} & \ci{\sClusterSevenRelEmptyAllDLo{}}{\sClusterSevenRelEmptyAllDHi{}}\\
Edge-aligned threshold & \nEdgeCutAllD{} & \ci{\nEdgeCutAllLo{}}{\nEdgeCutAllHi{}} & \ci{\sClusterEdgeCutAllDLo{}}{\sClusterEdgeCutAllDHi{}} & \ci{\sClusterFiveEdgeCutAllDLo{}}{\sClusterFiveEdgeCutAllDHi{}} & \ci{\sClusterSevenEdgeCutAllDLo{}}{\sClusterSevenEdgeCutAllDHi{}}\\
Image-only describer & \nDescriberAllD{} & \ci{\nDescriberAllLo{}}{\nDescriberAllHi{}} & \ci{\sClusterDescriberAllDLo{}}{\sClusterDescriberAllDHi{}} & \ci{\sClusterFiveDescriberAllDLo{}}{\sClusterFiveDescriberAllDHi{}} & \ci{\sClusterSevenDescriberAllDLo{}}{\sClusterSevenDescriberAllDHi{}}\\
Text-free (trained without text) & \nTextFreeAllD{} & \ci{\nTextFreeAllLo{}}{\nTextFreeAllHi{}} & \ci{\sClusterTextFreeAllDLo{}}{\sClusterTextFreeAllDHi{}} & \ci{\sClusterFiveTextFreeAllDLo{}}{\sClusterFiveTextFreeAllDHi{}} & \ci{\sClusterSevenTextFreeAllDLo{}}{\sClusterSevenTextFreeAllDHi{}}\\
Extent-free control & \nXSnoEAllD{} & \ci{\nXSnoEAllLo{}}{\nXSnoEAllHi{}} & \ci{\sClusterXSnoEAllDLo{}}{\sClusterXSnoEAllDHi{}} & \ci{\sClusterFiveXSnoEAllDLo{}}{\sClusterFiveXSnoEAllDHi{}} & \ci{\sClusterSevenXSnoEAllDLo{}}{\sClusterSevenXSnoEAllDHi{}}\\
ExtentSeg without area loss & \nXSnoAAllD{} & \ci{\nXSnoAAllLo{}}{\nXSnoAAllHi{}} & \ci{\sClusterXSnoAAllDLo{}}{\sClusterXSnoAAllDHi{}} & \ci{\sClusterFiveXSnoAAllDLo{}}{\sClusterFiveXSnoAAllDHi{}} & \ci{\sClusterSevenXSnoAAllDLo{}}{\sClusterSevenXSnoAAllDHi{}}\\
ExtentSeg & \nXSAllD{} & \ci{\nXSAllLo{}}{\nXSAllHi{}} & \ci{\sClusterXSAllDLo{}}{\sClusterXSAllDHi{}} & \ci{\sClusterFiveXSAllDLo{}}{\sClusterFiveXSAllDHi{}} & \ci{\sClusterSevenXSAllDLo{}}{\sClusterSevenXSAllDHi{}}\\
ExtentSeg, true extent$^\dagger$ & \nXStrueAllD{} & \ci{\nXStrueAllLo{}}{\nXStrueAllHi{}} & \ci{\sClusterXStrueAllDLo{}}{\sClusterXStrueAllDHi{}} & \ci{\sClusterFiveXStrueAllDLo{}}{\sClusterFiveXStrueAllDHi{}} & \ci{\sClusterSevenXStrueAllDLo{}}{\sClusterSevenXStrueAllDHi{}}\\
Retrained, matched caption$^\dagger$ & \nRetLeakAllD{} & \ci{\nRetLeakAllLo{}}{\nRetLeakAllHi{}} & \ci{\sClusterRetLeakAllDLo{}}{\sClusterRetLeakAllDHi{}} & \ci{\sClusterFiveRetLeakAllDLo{}}{\sClusterFiveRetLeakAllDHi{}} & \ci{\sClusterSevenRetLeakAllDLo{}}{\sClusterSevenRetLeakAllDHi{}}\\
\bottomrule
\end{tabular}\\[2pt]
\begin{minipage}{\textwidth}\footnotesize 95\% intervals. Sequences: clusters at a correlation threshold of 0.6 (primary; \sClusterNClusters{} clusters), 0.5 (\sClusterFiveNClusters) and 0.7 (\sClusterSevenNClusters), resampled within each test set; the estimates do not change. $^\dagger$Reads the test mask.\end{minipage}
\end{table*}

\begin{table*}[!tb]
\centering
\caption{Table~\ref{tab:released} on the normalised surface Dice, and RecLMIS on the intersection over union.}
\label{tab:nsd}
\footnotesize
\setlength{\tabcolsep}{4pt}
\begin{tabular}{@{}lrrrrrrr@{}}
\toprule
Test set & Empty & Mism. & Match. & $G$ (95\% CI) & $\pi$ (95\% CI) & $S_{\mathrm{ext}}$ (95\% CI) & $N_{\mathrm{ext}}$ (95\% CI)\\
\midrule
\multicolumn{8}{@{}l}{\emph{MedCLIPSeg, normalised surface Dice}}\\

Kvasir-SEG & \sNsdKvasirEmpty{} & \sNsdKvasirMism{} & \sNsdKvasirMatch{} & \sNsdKvasirG{}~\ci{\sNsdKvasirGLo{}}{\sNsdKvasirGHi{}} & \sNsdKvasirPi{}~\ci{\sNsdKvasirPiLo{}}{\sNsdKvasirPiHi{}} & \sNsdKvasirSuff{}~\ci{\sNsdKvasirSuffLo{}}{\sNsdKvasirSuffHi{}} & \sNsdKvasirNec{}~\ci{\sNsdKvasirNecLo{}}{\sNsdKvasirNecHi{}}\\
CVC-ClinicDB & \sNsdClinicEmpty{} & \sNsdClinicMism{} & \sNsdClinicMatch{} & \sNsdClinicG{}~\ci{\sNsdClinicGLo{}}{\sNsdClinicGHi{}} & \sNsdClinicPi{}~\ci{\sNsdClinicPiLo{}}{\sNsdClinicPiHi{}} & \sNsdClinicSuff{}~\ci{\sNsdClinicSuffLo{}}{\sNsdClinicSuffHi{}} & \sNsdClinicNec{}~\ci{\sNsdClinicNecLo{}}{\sNsdClinicNecHi{}}\\
CVC-300 & \sNsdCvcThreeEmpty{} & \sNsdCvcThreeMism{} & \sNsdCvcThreeMatch{} & \sNsdCvcThreeG{}~\ci{\sNsdCvcThreeGLo{}}{\sNsdCvcThreeGHi{}} & \sNsdCvcThreePi{}~\ci{\sNsdCvcThreePiLo{}}{\sNsdCvcThreePiHi{}} & \sNsdCvcThreeSuff{}~\ci{\sNsdCvcThreeSuffLo{}}{\sNsdCvcThreeSuffHi{}} & \sNsdCvcThreeNec{}~\ci{\sNsdCvcThreeNecLo{}}{\sNsdCvcThreeNecHi{}}\\
BKAI-IGH & \sNsdBkaiEmpty{} & \sNsdBkaiMism{} & \sNsdBkaiMatch{} & \sNsdBkaiG{}~\ci{\sNsdBkaiGLo{}}{\sNsdBkaiGHi{}} & \sNsdBkaiPi{}~\ci{\sNsdBkaiPiLo{}}{\sNsdBkaiPiHi{}} & \sNsdBkaiSuff{}~\ci{\sNsdBkaiSuffLo{}}{\sNsdBkaiSuffHi{}} & \sNsdBkaiNec{}~\ci{\sNsdBkaiNecLo{}}{\sNsdBkaiNecHi{}}\\
CVC-ColonDB & \sNsdColonEmpty{} & \sNsdColonMism{} & \sNsdColonMatch{} & \sNsdColonG{}~\ci{\sNsdColonGLo{}}{\sNsdColonGHi{}} & \sNsdColonPi{}~\ci{\sNsdColonPiLo{}}{\sNsdColonPiHi{}} & \sNsdColonSuff{}~\ci{\sNsdColonSuffLo{}}{\sNsdColonSuffHi{}} & \sNsdColonNec{}~\ci{\sNsdColonNecLo{}}{\sNsdColonNecHi{}}\\
BUSI$^\ddagger$ & \sNsdBusiEmpty{} & \sNsdBusiMism{} & \sNsdBusiMatch{} & \sNsdBusiG{}~\ci{\sNsdBusiGLo{}}{\sNsdBusiGHi{}} & \sNsdBusiPi{}~\ci{\sNsdBusiPiLo{}}{\sNsdBusiPiHi{}} & \sNsdBusiSuff{}~\ci{\sNsdBusiSuffLo{}}{\sNsdBusiSuffHi{}} & \sNsdBusiNec{}~\ci{\sNsdBusiNecLo{}}{\sNsdBusiNecHi{}}\\
QaTa-COV19 & \sNsdQataEmpty{} & \sNsdQataMism{} & \sNsdQataMatch{} & \sNsdQataG{}~\ci{\sNsdQataGLo{}}{\sNsdQataGHi{}} & \sNsdQataPi{}~\ci{\sNsdQataPiLo{}}{\sNsdQataPiHi{}} & \sNsdQataSuff{}~\ci{\sNsdQataSuffLo{}}{\sNsdQataSuffHi{}} & \sNsdQataNec{}~\ci{\sNsdQataNecLo{}}{\sNsdQataNecHi{}}\\
Colonoscopy, all & \sNsdPolypEmpty{} & \sNsdPolypMism{} & \sNsdPolypMatch{} & \sNsdPolypG{}~\ci{\sNsdPolypGLo{}}{\sNsdPolypGHi{}} & \sNsdPolypPi{}~\ci{\sNsdPolypPiLo{}}{\sNsdPolypPiHi{}} & \sNsdPolypSuff{}~\ci{\sNsdPolypSuffLo{}}{\sNsdPolypSuffHi{}} & \sNsdPolypNec{}~\ci{\sNsdPolypNecLo{}}{\sNsdPolypNecHi{}}\\
Unseen centres & \sNsdPolypDGEmpty{} & \sNsdPolypDGMism{} & \sNsdPolypDGMatch{} & \sNsdPolypDGG{}~\ci{\sNsdPolypDGGLo{}}{\sNsdPolypDGGHi{}} & \sNsdPolypDGPi{}~\ci{\sNsdPolypDGPiLo{}}{\sNsdPolypDGPiHi{}} & \sNsdPolypDGSuff{}~\ci{\sNsdPolypDGSuffLo{}}{\sNsdPolypDGSuffHi{}} & \sNsdPolypDGNec{}~\ci{\sNsdPolypDGNecLo{}}{\sNsdPolypDGNecHi{}}\\
\addlinespace
\multicolumn{8}{@{}l}{\emph{RecLMIS, intersection over union}}\\
QaTa-COV19 & \sRecIouEmpty{} & \sRecIouMism{} & \sRecIouMatch{} & \sRecIouG{}~\ci{\sRecIouGLo{}}{\sRecIouGHi{}} & \sRecIouPi{}~\ci{\sRecIouPiLo{}}{\sRecIouPiHi{}} & \sRecIouSuff{}~\ci{\sRecIouSuffLo{}}{\sRecIouSuffHi{}} & \sRecIouNec{}~\ci{\sRecIouNecLo{}}{\sRecIouNecHi{}}\\
\bottomrule
\end{tabular}\\[2pt]
\begin{minipage}{\textwidth}\footnotesize Normalised surface Dice at a tolerance of 2 pixels; populations and captions as in Table~\ref{tab:released}. $^\ddagger$$G$ and $\pi$ on all 78 images, $S_{\mathrm{ext}}$ and $N_{\mathrm{ext}}$ on the 65 with a lesion. RecLMIS has no per-image surface score, so its intersection over union is shown.\end{minipage}
\end{table*}

\begin{table*}[!tb]
\centering
\caption{MedCLIPSeg under captions rewritten in other styles (Dice).}
\label{tab:para}
\footnotesize
\setlength{\tabcolsep}{2.5pt}
\begin{tabular}{@{}lrrrrrrrrr@{}}
\toprule
& & & \multicolumn{4}{c}{Same content, other style} & \multicolumn{2}{c}{Content slots emptied} & \\
\cmidrule(lr){4-7}\cmidrule(lr){8-9}
Test set & $n$ & Released & Terse & Verbose & Clinical & List & Clinical & List & Empty\\
\midrule

Kvasir-SEG & \sParaKvasirN{} & \sParaKvasirMatch{} & \sParaKvasirTerse{} & \sParaKvasirVerbose{} & \sParaKvasirClinical{} & \sParaKvasirList{} & \sParaKvasirClinicalFrame{} & \sParaKvasirListFrame{} & \sParaKvasirEmpty{}\\
CVC-ClinicDB & \sParaClinicN{} & \sParaClinicMatch{} & \sParaClinicTerse{} & \sParaClinicVerbose{} & \sParaClinicClinical{} & \sParaClinicList{} & \sParaClinicClinicalFrame{} & \sParaClinicListFrame{} & \sParaClinicEmpty{}\\
CVC-300 & \sParaCvcThreeN{} & \sParaCvcThreeMatch{} & \sParaCvcThreeTerse{} & \sParaCvcThreeVerbose{} & \sParaCvcThreeClinical{} & \sParaCvcThreeList{} & \sParaCvcThreeClinicalFrame{} & \sParaCvcThreeListFrame{} & \sParaCvcThreeEmpty{}\\
BKAI-IGH & \sParaBkaiN{} & \sParaBkaiMatch{} & \sParaBkaiTerse{} & \sParaBkaiVerbose{} & \sParaBkaiClinical{} & \sParaBkaiList{} & \sParaBkaiClinicalFrame{} & \sParaBkaiListFrame{} & \sParaBkaiEmpty{}\\
CVC-ColonDB & \sParaColonN{} & \sParaColonMatch{} & \sParaColonTerse{} & \sParaColonVerbose{} & \sParaColonClinical{} & \sParaColonList{} & \sParaColonClinicalFrame{} & \sParaColonListFrame{} & \sParaColonEmpty{}\\
BUSI & \sParaBusiN{} & \sParaBusiMatch{} & \sParaBusiTerse{} & \sParaBusiVerbose{} & \sParaBusiClinical{} & \sParaBusiList{} & \sParaBusiClinicalFrame{} & \sParaBusiListFrame{} & \sParaBusiEmpty{}\\
QaTa-COV19 & \sParaQataN{} & \sParaQataMatch{} & \sParaQataTerse{} & \sParaQataVerbose{} & \sParaQataClinical{} & \sParaQataList{} & \sParaQataClinicalFrame{} & \sParaQataListFrame{} & \sParaQataEmpty{}\\
Pooled & \sParaPooledN{} & \sParaPooledMatch{} & \sParaPooledTerse{} & \sParaPooledVerbose{} & \sParaPooledClinical{} & \sParaPooledList{} & \sParaPooledClinicalFrame{} & \sParaPooledListFrame{} & \sParaPooledEmpty{}\\
\addlinespace
Pooled $-$ released &  & - & \sParaPooledTerseD{}~\ci{\sParaPooledTerseDLo{}}{\sParaPooledTerseDHi{}} & \sParaPooledVerboseD{}~\ci{\sParaPooledVerboseDLo{}}{\sParaPooledVerboseDHi{}} & \sParaPooledClinicalD{}~\ci{\sParaPooledClinicalDLo{}}{\sParaPooledClinicalDHi{}} & \sParaPooledListD{}~\ci{\sParaPooledListDLo{}}{\sParaPooledListDHi{}} & \sParaPooledClinicalFrameD{} & \sParaPooledListFrameD{} & \sParaPooledEmptyD{}\\
Pooled $-$ terse &  &  & - & \sParaPooledVerboseDT{}~\ci{\sParaPooledVerboseDTLo{}}{\sParaPooledVerboseDTHi{}} & \sParaPooledClinicalDT{}~\ci{\sParaPooledClinicalDTLo{}}{\sParaPooledClinicalDTHi{}} & \sParaPooledListDT{}~\ci{\sParaPooledListDTLo{}}{\sParaPooledListDTHi{}} &  &  & \\
\bottomrule
\end{tabular}\\[2pt]
\begin{minipage}{\textwidth}\footnotesize Each caption is parsed into its content slots and rendered again in four styles: terse (LViT style), verbose (MedCLIPSeg style), clinical (a findings sentence) and list (no sentence). Terse is the reference of the registered criterion, because the parser drops text outside the slots. \emph{Content slots emptied}: the wording is kept and the slots are emptied. Last rows: paired differences of pooled means (95\% CI).\end{minipage}
\end{table*}

\begin{table*}[!tb]
\centering
\caption{Label-free rules for cutting the averaged score map of the edge-aligned threshold (Dice).}
\label{tab:rules}
\footnotesize
\setlength{\tabcolsep}{4pt}
\begin{tabular}{@{}lrrrrrrrrr@{}}
\toprule
Rule & Kvasir-SEG & CVC-ClinicDB & CVC-300 & BKAI-IGH & CVC-ColonDB & Colonoscopy, all & Unseen & BUSI & QaTa-COV19\\
\midrule

half & \sRuleKvasirHalf{} & \sRuleClinicHalf{} & \sRuleCvcThreeHalf{} & \sRuleBkaiHalf{} & \sRuleColonHalf{} & \sRulePolypHalfAll{} & \sRulePolypHalfDG{} & \sRuleBusiHalf{} & \sRuleQataHalf{}\\
Otsu & \sRuleKvasirOtsu{} & \sRuleClinicOtsu{} & \sRuleCvcThreeOtsu{} & \sRuleBkaiOtsu{} & \sRuleColonOtsu{} & \sRulePolypOtsuAll{} & \sRulePolypOtsuDG{} & \sRuleBusiOtsu{} & \sRuleQataOtsu{}\\
drop & \sRuleKvasirDrop{} & \sRuleClinicDrop{} & \sRuleCvcThreeDrop{} & \sRuleBkaiDrop{} & \sRuleColonDrop{} & \sRulePolypDropAll{} & \sRulePolypDropDG{} & \sRuleBusiDrop{} & \sRuleQataDrop{}\\
plateau & \sRuleKvasirPlateau{} & \sRuleClinicPlateau{} & \sRuleCvcThreePlateau{} & \sRuleBkaiPlateau{} & \sRuleColonPlateau{} & \sRulePolypPlateauAll{} & \sRulePolypPlateauDG{} & \sRuleBusiPlateau{} & \sRuleQataPlateau{}\\
agree & \sRuleKvasirAgree{} & \sRuleClinicAgree{} & \sRuleCvcThreeAgree{} & \sRuleBkaiAgree{} & \sRuleColonAgree{} & \sRulePolypAgreeAll{} & \sRulePolypAgreeDG{} & \sRuleBusiAgree{} & \sRuleQataAgree{}\\
grad$^\ast$ & \sRuleKvasirGrad{} & \sRuleClinicGrad{} & \sRuleCvcThreeGrad{} & \sRuleBkaiGrad{} & \sRuleColonGrad{} & \sRulePolypGradAll{} & \sRulePolypGradDG{} & \sRuleBusiGrad{} & \sRuleQataGrad{}\\
compact & \sRuleKvasirCompact{} & \sRuleClinicCompact{} & \sRuleCvcThreeCompact{} & \sRuleBkaiCompact{} & \sRuleColonCompact{} & \sRulePolypCompactAll{} & \sRulePolypCompactDG{} & \sRuleBusiCompact{} & -\\
gradimg & \sRuleKvasirGradImg{} & \sRuleClinicGradImg{} & \sRuleCvcThreeGradImg{} & \sRuleBkaiGradImg{} & \sRuleColonGradImg{} & \sRulePolypGradImgAll{} & \sRulePolypGradImgDG{} & \sRuleBusiGradImg{} & -\\
gradcomp & \sRuleKvasirGradComp{} & \sRuleClinicGradComp{} & \sRuleCvcThreeGradComp{} & \sRuleBkaiGradComp{} & \sRuleColonGradComp{} & \sRulePolypGradCompAll{} & \sRulePolypGradCompDG{} & \sRuleBusiGradComp{} & -\\
contrast & \sRuleKvasirContrast{} & \sRuleClinicContrast{} & \sRuleCvcThreeContrast{} & \sRuleBkaiContrast{} & \sRuleColonContrast{} & \sRulePolypContrastAll{} & \sRulePolypContrastDG{} & \sRuleBusiContrast{} & -\\
gradwin & \sRuleKvasirGradWin{} & \sRuleClinicGradWin{} & \sRuleCvcThreeGradWin{} & \sRuleBkaiGradWin{} & \sRuleColonGradWin{} & \sRulePolypGradWinAll{} & \sRulePolypGradWinDG{} & \sRuleBusiGradWin{} & -\\
gdrop & \sRuleKvasirGDrop{} & \sRuleClinicGDrop{} & \sRuleCvcThreeGDrop{} & \sRuleBkaiGDrop{} & \sRuleColonGDrop{} & \sRulePolypGDropAll{} & \sRulePolypGDropDG{} & \sRuleBusiGDrop{} & \sRuleQataGDrop{}\\
\addlinespace
Published, empty caption & \sRuleKvasirEmpty{} & \sRuleClinicEmpty{} & \sRuleCvcThreeEmpty{} & \sRuleBkaiEmpty{} & \sRuleColonEmpty{} & \sRulePolypEmptyAll{} & \sRulePolypEmptyDG{} & \sRuleBusiEmpty{} & \nTabTwoQataEmpty{}\\
Published, matched caption$^\dagger$ & \sRuleKvasirMatch{} & \sRuleClinicMatch{} & \sRuleCvcThreeMatch{} & \sRuleBkaiMatch{} & \sRuleColonMatch{} & \nRelLeakAll{} & \sRulePolypMatchDG{} & \sRuleBusiMatch{} & \sRuleQataMatch{}\\
Cut at the true area$^\dagger$ & \sRuleKvasirGtAreaCut{} & \sRuleClinicGtAreaCut{} & \sRuleCvcThreeGtAreaCut{} & \sRuleBkaiGtAreaCut{} & \sRuleColonGtAreaCut{} & \sRulePolypGtAreaCutAll{} & \sRulePolypGtAreaCutDG{} & \sRuleBusiGtAreaCut{} & \sRuleQataGtAreaCut{}\\
Best threshold$^\dagger$ & \sRuleKvasirBest{} & \sRuleClinicBest{} & \sRuleCvcThreeBest{} & \sRuleBkaiBest{} & \sRuleColonBest{} & \sRulePolypBestAll{} & \sRulePolypBestDG{} & \sRuleBusiBest{} & \sRuleQataBest{}\\
\bottomrule
\end{tabular}\\[2pt]
\begin{minipage}{\textwidth}\footnotesize Each rule sets how many top-scoring pixels to keep. half: $s>0.5$; Otsu: \citet{otsu1979}; drop: steepest fall of the sorted scores; plateau: area most stable to the threshold; agree: best agreement with the single-caption masks; grad: steepest mean boundary gradient, Eq.~\eqref{eq:edgecut}; compact: compactness times mean score; gradimg: image gradient on the boundary; gradcomp: grad times compactness; contrast: mean score inside minus outside; gradwin: grad within a factor of four of the empty-caption area; gdrop: geometric mean of the grad and drop areas. $^\ast$Rule used in the paper. -: not evaluated. $^\dagger$Uses the test mask.\end{minipage}
\end{table*}

\begin{table*}[!tb]
\centering
\caption{Per-seed results of the retrained models.}
\label{tab:perseed}
\footnotesize
\setlength{\tabcolsep}{4pt}
\begin{tabular}{@{}lrrrrrrrrr@{}}
\toprule
& \multicolumn{3}{c}{Dice, empty caption} & \multicolumn{3}{c}{Dice, matched caption} & \multicolumn{3}{c}{Text gain $G$}\\
\cmidrule(lr){2-4}\cmidrule(lr){5-7}\cmidrule(l){8-10}
Model & seed 1 & seed 2 & seed 3 & seed 1 & seed 2 & seed 3 & seed 1 & seed 2 & seed 3\\
\midrule

\multicolumn{10}{@{}l}{\emph{Kvasir-SEG, five colonoscopy test sets ($n=681$)}}\\
Original & \sSeedKvasirOrigAllEmptyOne{} & \sSeedKvasirOrigAllEmptyTwo{} & \sSeedKvasirOrigAllEmptyThree{} & \sSeedKvasirOrigAllMatchOne{} & \sSeedKvasirOrigAllMatchTwo{} & \sSeedKvasirOrigAllMatchThree{} & \sSeedKvasirOrigAllGOne{} & \sSeedKvasirOrigAllGTwo{} & \sSeedKvasirOrigAllGThree{}\\
Size-blind & \sSeedKvasirNoSizeAllEmptyOne{} & \sSeedKvasirNoSizeAllEmptyTwo{} & \sSeedKvasirNoSizeAllEmptyThree{} & \sSeedKvasirNoSizeAllMatchOne{} & \sSeedKvasirNoSizeAllMatchTwo{} & \sSeedKvasirNoSizeAllMatchThree{} & \sSeedKvasirNoSizeAllGOne{} & \sSeedKvasirNoSizeAllGTwo{} & \sSeedKvasirNoSizeAllGThree{}\\
Size-only & \sSeedKvasirSizeOnlyAllEmptyOne{} & \sSeedKvasirSizeOnlyAllEmptyTwo{} & \sSeedKvasirSizeOnlyAllEmptyThree{} & \sSeedKvasirSizeOnlyAllMatchOne{} & \sSeedKvasirSizeOnlyAllMatchTwo{} & \sSeedKvasirSizeOnlyAllMatchThree{} & \sSeedKvasirSizeOnlyAllGOne{} & \sSeedKvasirSizeOnlyAllGTwo{} & \sSeedKvasirSizeOnlyAllGThree{}\\
Mismatched & \sSeedKvasirShufAllEmptyOne{} & \sSeedKvasirShufAllEmptyTwo{} & \sSeedKvasirShufAllEmptyThree{} & \sSeedKvasirShufAllMatchOne{} & \sSeedKvasirShufAllMatchTwo{} & \sSeedKvasirShufAllMatchThree{} & \sSeedKvasirShufAllGOne{} & \sSeedKvasirShufAllGTwo{} & \sSeedKvasirShufAllGThree{}\\
Text-free & \sSeedKvasirTFAllEmptyOne{} & \sSeedKvasirTFAllEmptyTwo{} & \sSeedKvasirTFAllEmptyThree{} & \sSeedKvasirTFAllMatchOne{} & \sSeedKvasirTFAllMatchTwo{} & \sSeedKvasirTFAllMatchThree{} & \sSeedKvasirTFAllGOne{} & \sSeedKvasirTFAllGTwo{} & \sSeedKvasirTFAllGThree{}\\
\multicolumn{10}{@{}l}{\emph{BUSI, 65 test images with a lesion}}\\
Original & \sSeedBusiOrigLesEmptyOne{} & \sSeedBusiOrigLesEmptyTwo{} & \sSeedBusiOrigLesEmptyThree{} & \sSeedBusiOrigLesMatchOne{} & \sSeedBusiOrigLesMatchTwo{} & \sSeedBusiOrigLesMatchThree{} & \sSeedBusiOrigLesGOne{} & \sSeedBusiOrigLesGTwo{} & \sSeedBusiOrigLesGThree{}\\
Size-blind & \sSeedBusiNoSizeLesEmptyOne{} & \sSeedBusiNoSizeLesEmptyTwo{} & \sSeedBusiNoSizeLesEmptyThree{} & \sSeedBusiNoSizeLesMatchOne{} & \sSeedBusiNoSizeLesMatchTwo{} & \sSeedBusiNoSizeLesMatchThree{} & \sSeedBusiNoSizeLesGOne{} & \sSeedBusiNoSizeLesGTwo{} & \sSeedBusiNoSizeLesGThree{}\\
Size-only & \sSeedBusiSizeOnlyLesEmptyOne{} & \sSeedBusiSizeOnlyLesEmptyTwo{} & \sSeedBusiSizeOnlyLesEmptyThree{} & \sSeedBusiSizeOnlyLesMatchOne{} & \sSeedBusiSizeOnlyLesMatchTwo{} & \sSeedBusiSizeOnlyLesMatchThree{} & \sSeedBusiSizeOnlyLesGOne{} & \sSeedBusiSizeOnlyLesGTwo{} & \sSeedBusiSizeOnlyLesGThree{}\\
Mismatched & \sSeedBusiShufLesEmptyOne{} & \sSeedBusiShufLesEmptyTwo{} & \sSeedBusiShufLesEmptyThree{} & \sSeedBusiShufLesMatchOne{} & \sSeedBusiShufLesMatchTwo{} & \sSeedBusiShufLesMatchThree{} & \sSeedBusiShufLesGOne{} & \sSeedBusiShufLesGTwo{} & \sSeedBusiShufLesGThree{}\\
Text-free & \sSeedBusiTFLesEmptyOne{} & \sSeedBusiTFLesEmptyTwo{} & \sSeedBusiTFLesEmptyThree{} & \sSeedBusiTFLesMatchOne{} & \sSeedBusiTFLesMatchTwo{} & \sSeedBusiTFLesMatchThree{} & \sSeedBusiTFLesGOne{} & \sSeedBusiTFLesGTwo{} & \sSeedBusiTFLesGThree{}\\
\multicolumn{10}{@{}l}{\emph{QaTa-COV19 ($n=2113$)}}\\
Original & \sSeedQataOrigEmptyOne{} & \sSeedQataOrigEmptyTwo{} & \sSeedQataOrigEmptyThree{} & \sSeedQataOrigMatchOne{} & \sSeedQataOrigMatchTwo{} & \sSeedQataOrigMatchThree{} & \sSeedQataOrigGOne{} & \sSeedQataOrigGTwo{} & \sSeedQataOrigGThree{}\\
Zone-blind & \sSeedQataBlindEmptyOne{} & \sSeedQataBlindEmptyTwo{} & \sSeedQataBlindEmptyThree{} & \sSeedQataBlindMatchOne{} & \sSeedQataBlindMatchTwo{} & \sSeedQataBlindMatchThree{} & \sSeedQataBlindGOne{} & \sSeedQataBlindGTwo{} & \sSeedQataBlindGThree{}\\
Zone-only & \sSeedQataOnlyEmptyOne{} & \sSeedQataOnlyEmptyTwo{} & \sSeedQataOnlyEmptyThree{} & \sSeedQataOnlyMatchOne{} & \sSeedQataOnlyMatchTwo{} & \sSeedQataOnlyMatchThree{} & \sSeedQataOnlyGOne{} & \sSeedQataOnlyGTwo{} & \sSeedQataOnlyGThree{}\\
Text-free & \sSeedQataTFEmptyOne{} & \sSeedQataTFEmptyTwo{} & \sSeedQataTFEmptyThree{} & \sSeedQataTFMatchOne{} & \sSeedQataTFMatchTwo{} & \sSeedQataTFMatchThree{} & \sSeedQataTFGOne{} & \sSeedQataTFGTwo{} & \sSeedQataTFGThree{}\\
\multicolumn{10}{@{}l}{\emph{ExtentSeg and its controls, five colonoscopy test sets: Dice under the conditioning named}}\\
ExtentSeg, own estimate & \sSeedXSAllPredOne{} & \sSeedXSAllPredTwo{} & \sSeedXSAllPredThree{} &  &  &  &  &  & \\
ExtentSeg without area loss, own estimate & \sSeedXSnoAAllPredOne{} & \sSeedXSnoAAllPredTwo{} & \sSeedXSnoAAllPredThree{} &  &  &  &  &  & \\
Extent-free control & \sSeedXSnoEAllPredOne{} & \sSeedXSnoEAllPredTwo{} & \sSeedXSnoEAllPredThree{} &  &  &  &  &  & \\
ExtentSeg, true extent$^\dagger$ & \sSeedXSAllTrueOne{} & \sSeedXSAllTrueTwo{} & \sSeedXSAllTrueThree{} &  &  &  &  &  & \\
\bottomrule
\end{tabular}\\[2pt]
\begin{minipage}{\textwidth}\footnotesize Last-epoch checkpoints, each scored on the full test population. $^\dagger$Reads the test mask.\end{minipage}
\end{table*}

\section{Provenance of the QaTa-COV19 extent phrase}
\label{app:qata}
The QaTa-COV19 captions were written for the benchmark by LViT's annotators \citep{lvit}; the dataset contains no radiology reports. To test whether the zone phrase is nonetheless a fixed function of the mask, we compared it with zone sets computed from the mask and a lung field. The lung field comes from the ChestX-Det lung segmentation network of TorchXRayVision \citep{torchxrayvision,chestxdet}, with image thirds as a fallback. The analysis plan was fixed before any rule or classifier was compared with a test caption.

\emph{Sides.} In training captions that name one side, \capt{left} captions place \nQataSideLeft\% of the mask pixels in the image-right half, and \capt{right} captions \nQataSideRight\%. The annotations therefore name sides by image position, the opposite of the radiological convention; test images with a visible cardiac silhouette and side markers confirm this.

\emph{Zones.} The overlap rule selected on the training split (a zone counts as involved when the mask covers at least \nQataTau\% of it) reproduces the full zone set on \nQataZoneRule\% \ci{\nQataZoneRuleLo}{\nQataZoneRuleHi} of the \sSplitQataTest{} test images, against \nQataZoneMajority\% for the most frequent set (intervals resample test subjects). Its mean Jaccard overlap with the caption is \nQataJaccard, and the per-zone Cohen's $\kappa$ ranges from \nQataKappaMin{} to \nQataKappaMax: the upper zones agree almost perfectly, and where the two disagree, the rule mostly calls a middle or lower zone involved that the caption does not name. A gradient-boosting classifier on \nQataClfFeatures{} mask-shape features reaches \nQataClfAcc\% \ci{\nQataClfLo}{\nQataClfHi}. The side (\nQataLaterality\%) and the stated number of regions (\nQataCountCC\%, against connected components) can almost always be derived from the mask; the zone set cannot. LViT's own captions, which RecLMIS reads, parse to the same zones as MedCLIPSeg's on all \sSplitQataTest{} test images.

\emph{Interpretation.} In an analysis not planned in advance, the number of named zones correlates with mask area with a Spearman rank correlation of $r_s=\nQataZoneAreaSpearman$ on the \sSplitQataTest{} test images. The zone phrase is therefore a human annotation of extent that carries about as much extent information as the computed size word ($R^2_{\mathrm{cap}}$ \nRcapTeQata{} against \nRcapTeKvasir{} on Kvasir-SEG; Table~\ref{tab:released}). Our analysis plan reserved \emph{leakage by construction} for captions computed from the mask, and the paper keeps that distinction: only the colonoscopy and BUSI size words are functions of the mask. We still use \emph{extent leakage} for QaTa-COV19, because its zone phrase, written for the benchmark and checked against the masks, gives the model at test time extent information that does not exist at deployment.

\twocolumn
\bibliography{references}

\begin{thebibliography}{101}
\expandafter\ifx\csname natexlab\endcsname\relax\def\natexlab#1{#1}\fi
\providecommand{\url}[1]{\texttt{#1}}
\providecommand{\href}[2]{#2}
\providecommand{\path}[1]{#1}
\providecommand{\DOIprefix}{doi:}
\providecommand{\ArXivprefix}{arXiv:}
\providecommand{\URLprefix}{URL: }
\providecommand{\Pubmedprefix}{pmid:}
\providecommand{\doi}[1]{\href{http://dx.doi.org/#1}{\path{#1}}}
\providecommand{\Pubmed}[1]{\href{pmid:#1}{\path{#1}}}
\providecommand{\bibinfo}[2]{#2}
\ifx\xfnm\relax \def\xfnm[#1]{\unskip,\space#1}\fi
\bibitem[{Agrawal et~al.(2018)Agrawal, Batra, Parikh and
  Kembhavi}]{agrawal2018}
\bibinfo{author}{Agrawal, A.}, \bibinfo{author}{Batra, D.},
  \bibinfo{author}{Parikh, D.}, \bibinfo{author}{Kembhavi, A.},
  \bibinfo{year}{2018}.
\newblock \bibinfo{title}{Don't just assume; look and answer: Overcoming priors
  for visual question answering}, in: \bibinfo{booktitle}{Proceedings of the
  IEEE/CVF Conference on Computer Vision and Pattern Recognition (CVPR)}, pp.
  \bibinfo{pages}{4971-4980}.
\newblock \DOIprefix\doi{10.1109/CVPR.2018.00522}.
\bibitem[{Al-Dhabyani et~al.(2020)Al-Dhabyani, Gomaa, Khaled and Fahmy}]{busi}
\bibinfo{author}{Al-Dhabyani, W.}, \bibinfo{author}{Gomaa, M.},
  \bibinfo{author}{Khaled, H.}, \bibinfo{author}{Fahmy, A.},
  \bibinfo{year}{2020}.
\newblock \bibinfo{title}{Dataset of breast ultrasound images}.
\newblock \bibinfo{journal}{Data Brief} \bibinfo{volume}{28},
  \bibinfo{pages}{104863}.
\newblock \DOIprefix\doi{10.1016/j.dib.2019.104863}.
\bibitem[{Ba et~al.(2016)Ba, Kiros and Hinton}]{ba2016}
\bibinfo{author}{Ba, J.L.}, \bibinfo{author}{Kiros, J.R.},
  \bibinfo{author}{Hinton, G.E.}, \bibinfo{year}{2016}.
\newblock \bibinfo{title}{Layer normalization}.
\newblock \bibinfo{howpublished}{arXiv preprint arXiv:1607.06450}.
\newblock \DOIprefix\doi{10.48550/arXiv.1607.06450}.
\bibitem[{Bateson et~al.(2021)Bateson, Dolz, Kervadec, Lombaert and {Ben
  Ayed}}]{bateson2021}
\bibinfo{author}{Bateson, M.}, \bibinfo{author}{Dolz, J.},
  \bibinfo{author}{Kervadec, H.}, \bibinfo{author}{Lombaert, H.},
  \bibinfo{author}{{Ben Ayed}, I.}, \bibinfo{year}{2021}.
\newblock \bibinfo{title}{Constrained domain adaptation for image
  segmentation}.
\newblock \bibinfo{journal}{IEEE Trans. Med. Imaging} \bibinfo{volume}{40},
  \bibinfo{pages}{1875-1887}.
\newblock \DOIprefix\doi{10.1109/TMI.2021.3067688}.
\bibitem[{Bateson et~al.(2022)Bateson, Kervadec, Dolz, Lombaert and {Ben
  Ayed}}]{bateson2022}
\bibinfo{author}{Bateson, M.}, \bibinfo{author}{Kervadec, H.},
  \bibinfo{author}{Dolz, J.}, \bibinfo{author}{Lombaert, H.},
  \bibinfo{author}{{Ben Ayed}, I.}, \bibinfo{year}{2022}.
\newblock \bibinfo{title}{Source-free domain adaptation for image
  segmentation}.
\newblock \bibinfo{journal}{Med. Image Anal.} \bibinfo{volume}{82},
  \bibinfo{pages}{102617}.
\newblock \DOIprefix\doi{10.1016/j.media.2022.102617}.
\bibitem[{Bernal et~al.(2015)Bernal, S{\'a}nchez, Fern{\'a}ndez-Esparrach, Gil,
  Rodr{\'i}guez and Vilari{\~n}o}]{clinicdb}
\bibinfo{author}{Bernal, J.}, \bibinfo{author}{S{\'a}nchez, F.J.},
  \bibinfo{author}{Fern{\'a}ndez-Esparrach, G.}, \bibinfo{author}{Gil, D.},
  \bibinfo{author}{Rodr{\'i}guez, C.}, \bibinfo{author}{Vilari{\~n}o, F.},
  \bibinfo{year}{2015}.
\newblock \bibinfo{title}{{WM-DOVA} maps for accurate polyp highlighting in
  colonoscopy: Validation vs. saliency maps from physicians}.
\newblock \bibinfo{journal}{Comput. Med. Imaging Graph.} \bibinfo{volume}{43},
  \bibinfo{pages}{99-111}.
\newblock \DOIprefix\doi{10.1016/j.compmedimag.2015.02.007}.
\bibitem[{Bernal et~al.(2012)Bernal, S{\'a}nchez and Vilari{\~n}o}]{bernal2012}
\bibinfo{author}{Bernal, J.}, \bibinfo{author}{S{\'a}nchez, J.},
  \bibinfo{author}{Vilari{\~n}o, F.}, \bibinfo{year}{2012}.
\newblock \bibinfo{title}{Towards automatic polyp detection with a polyp
  appearance model}.
\newblock \bibinfo{journal}{Pattern Recognit.} \bibinfo{volume}{45},
  \bibinfo{pages}{3166-3182}.
\newblock \DOIprefix\doi{10.1016/j.patcog.2012.03.002}.
\bibitem[{Bhardwaj et~al.(2025)Bhardwaj, Tambe and Neog}]{vitexnet}
\bibinfo{author}{Bhardwaj, R.}, \bibinfo{author}{Tambe, U.Y.},
  \bibinfo{author}{Neog, D.R.}, \bibinfo{year}{2025}.
\newblock \bibinfo{title}{{ViTexNet}: Vision-text guided dynamic convolution
  network for medical image segmentation}, in: \bibinfo{booktitle}{Medical
  Image Computing and Computer Assisted Intervention - {MICCAI} 2025},
  \bibinfo{publisher}{Springer Nature Switzerland}, \bibinfo{address}{Cham}.
  pp. \bibinfo{pages}{690-699}.
\newblock \DOIprefix\doi{10.1007/978-3-032-04981-0_65}.
\bibitem[{{Brown II} et~al.(2026){Brown II}, Yang, Yang, Fergus and
  Xie}]{tst2025}
\bibinfo{author}{{Brown II}, E.L.}, \bibinfo{author}{Yang, J.},
  \bibinfo{author}{Yang, S.}, \bibinfo{author}{Fergus, R.},
  \bibinfo{author}{Xie, S.}, \bibinfo{year}{2026}.
\newblock \bibinfo{title}{Benchmark designers should ``train on the test set''
  to expose exploitable non-visual shortcuts}.
\newblock \bibinfo{howpublished}{Conference on Language Modeling (COLM)}.
\newblock \URLprefix \url{https://openreview.net/forum?id=ggtdDv1l9s}.
  \bibinfo{note}{arXiv:2511.04655v2}.
\bibitem[{Cao et~al.(2026)Cao, Pham and Pham}]{cao2026}
\bibinfo{author}{Cao, D.P.M.}, \bibinfo{author}{Pham, H.D.},
  \bibinfo{author}{Pham, H.}, \bibinfo{year}{2026}.
\newblock \bibinfo{title}{When oracle conditioning misleads deployment:
  Conditioning-availability bias in echocardiographic segmentation}.
\newblock \bibinfo{howpublished}{arXiv preprint arXiv:2608.03342}.
\newblock \DOIprefix\doi{10.48550/arXiv.2608.03342}. \bibinfo{note}{{Accepted
  at the MICCAI 2026 Workshop on Fairness of {AI} in Medical Imaging (FAIMI)}}.
\bibitem[{Chen et~al.(2026)Chen, Pan, Tian, Tian and Wu}]{tmc}
\bibinfo{author}{Chen, G.}, \bibinfo{author}{Pan, H.}, \bibinfo{author}{Tian,
  Y.}, \bibinfo{author}{Tian, Z.}, \bibinfo{author}{Wu, X.},
  \bibinfo{year}{2026}.
\newblock \bibinfo{title}{Text-guided multi-stage cross-perception network for
  medical image segmentation}.
\newblock \bibinfo{journal}{J. Imaging Inform. Med.}
  \DOIprefix\doi{10.1007/s10278-025-01811-0}. \bibinfo{note}{{Online ahead of
  print}}.
\bibitem[{Chen et~al.(2024a)Chen, Li, Dong, Zhang, Zang, Chen, Duan, Wang,
  Qiao, Lin and Zhao}]{mmstar}
\bibinfo{author}{Chen, L.}, \bibinfo{author}{Li, J.}, \bibinfo{author}{Dong,
  X.}, \bibinfo{author}{Zhang, P.}, \bibinfo{author}{Zang, Y.},
  \bibinfo{author}{Chen, Z.}, \bibinfo{author}{Duan, H.},
  \bibinfo{author}{Wang, J.}, \bibinfo{author}{Qiao, Y.}, \bibinfo{author}{Lin,
  D.}, \bibinfo{author}{Zhao, F.}, \bibinfo{year}{2024}a.
\newblock \bibinfo{title}{Are we on the right way for evaluating large
  vision-language models?}, in: \bibinfo{booktitle}{Advances in Neural
  Information Processing Systems}, pp. \bibinfo{pages}{27056-27087}.
\newblock \DOIprefix\doi{10.52202/079017-0850}.
\bibitem[{Chen et~al.(2024b)Chen, Wei, Zheng, Hu, Shi, Xiong, Zhu and
  Mou}]{causalclipseg}
\bibinfo{author}{Chen, Y.}, \bibinfo{author}{Wei, M.}, \bibinfo{author}{Zheng,
  Z.}, \bibinfo{author}{Hu, J.}, \bibinfo{author}{Shi, Y.},
  \bibinfo{author}{Xiong, S.}, \bibinfo{author}{Zhu, X.X.},
  \bibinfo{author}{Mou, L.}, \bibinfo{year}{2024}b.
\newblock \bibinfo{title}{{CausalCLIPSeg}: Unlocking {CLIP}'s potential in
  referring medical image segmentation with causal intervention}, in:
  \bibinfo{booktitle}{Medical Image Computing and Computer Assisted
  Intervention - MICCAI 2024}, \bibinfo{publisher}{Springer}. pp.
  \bibinfo{pages}{77-87}.
\newblock \DOIprefix\doi{10.1007/978-3-031-72384-1_8}.
\bibitem[{Cirik et~al.(2018)Cirik, Morency and Berg-Kirkpatrick}]{cirik2018}
\bibinfo{author}{Cirik, V.}, \bibinfo{author}{Morency, L.P.},
  \bibinfo{author}{Berg-Kirkpatrick, T.}, \bibinfo{year}{2018}.
\newblock \bibinfo{title}{Visual referring expression recognition: What do
  systems actually learn?}, in: \bibinfo{booktitle}{Proceedings of the 2018
  Conference of the North American Chapter of the Association for Computational
  Linguistics: Human Language Technologies (NAACL-HLT), Volume 2 (Short
  Papers)}, pp. \bibinfo{pages}{781-787}.
\newblock \DOIprefix\doi{10.18653/v1/N18-2123}.
\bibitem[{Cohen et~al.(2022)Cohen, Viviano, Bertin, Morrison, Torabian,
  Guarrera, Lungren, Chaudhari, Brooks, Hashir and Bertrand}]{torchxrayvision}
\bibinfo{author}{Cohen, J.P.}, \bibinfo{author}{Viviano, J.D.},
  \bibinfo{author}{Bertin, P.}, \bibinfo{author}{Morrison, P.},
  \bibinfo{author}{Torabian, P.}, \bibinfo{author}{Guarrera, M.},
  \bibinfo{author}{Lungren, M.P.}, \bibinfo{author}{Chaudhari, A.},
  \bibinfo{author}{Brooks, R.}, \bibinfo{author}{Hashir, M.},
  \bibinfo{author}{Bertrand, H.}, \bibinfo{year}{2022}.
\newblock \bibinfo{title}{{TorchXRayVision}: A library of chest {X}-ray
  datasets and models}, in: \bibinfo{booktitle}{Proceedings of the
  International Conference on Medical Imaging with Deep Learning (MIDL)}, pp.
  \bibinfo{pages}{231-249}.
\newblock \URLprefix \url{https://proceedings.mlr.press/v172/cohen22a.html}.
\bibitem[{Degerli et~al.(2021)Degerli, Ahishali, Yamac, Kiranyaz, Chowdhury,
  Hameed, Hamid, Mazhar and Gabbouj}]{degerli2021qata}
\bibinfo{author}{Degerli, A.}, \bibinfo{author}{Ahishali, M.},
  \bibinfo{author}{Yamac, M.}, \bibinfo{author}{Kiranyaz, S.},
  \bibinfo{author}{Chowdhury, M.E.H.}, \bibinfo{author}{Hameed, K.},
  \bibinfo{author}{Hamid, T.}, \bibinfo{author}{Mazhar, R.},
  \bibinfo{author}{Gabbouj, M.}, \bibinfo{year}{2021}.
\newblock \bibinfo{title}{{COVID-19} infection map generation and detection
  from chest {X}-ray images}.
\newblock \bibinfo{journal}{Health Inf. Sci. Syst.} \bibinfo{volume}{9},
  \bibinfo{pages}{15}.
\newblock \DOIprefix\doi{10.1007/s13755-021-00146-8}.
\bibitem[{Degerli et~al.(2022)Degerli, Kiranyaz, Chowdhury and Gabbouj}]{qata}
\bibinfo{author}{Degerli, A.}, \bibinfo{author}{Kiranyaz, S.},
  \bibinfo{author}{Chowdhury, M.E.H.}, \bibinfo{author}{Gabbouj, M.},
  \bibinfo{year}{2022}.
\newblock \bibinfo{title}{{OSegNet}: Operational segmentation network for
  {COVID-19} detection using chest {X}-ray images}, in:
  \bibinfo{booktitle}{Proceedings of the IEEE International Conference on Image
  Processing (ICIP)}, pp. \bibinfo{pages}{2306-2310}.
\newblock \DOIprefix\doi{10.1109/ICIP46576.2022.9897412}.
\bibitem[{DeGrave et~al.(2021)DeGrave, Janizek and Lee}]{degrave2021}
\bibinfo{author}{DeGrave, A.J.}, \bibinfo{author}{Janizek, J.D.},
  \bibinfo{author}{Lee, S.I.}, \bibinfo{year}{2021}.
\newblock \bibinfo{title}{{AI} for radiographic {COVID-19} detection selects
  shortcuts over signal}.
\newblock \bibinfo{journal}{Nat. Mach. Intell.} \bibinfo{volume}{3},
  \bibinfo{pages}{610-619}.
\newblock \DOIprefix\doi{10.1038/s42256-021-00338-7}.
\bibitem[{Dhakal et~al.(2024)Dhakal, Adhikari, Thapaliya and
  Khanal}]{vlsmadapter}
\bibinfo{author}{Dhakal, M.}, \bibinfo{author}{Adhikari, R.},
  \bibinfo{author}{Thapaliya, S.}, \bibinfo{author}{Khanal, B.},
  \bibinfo{year}{2024}.
\newblock \bibinfo{title}{{VLSM-Adapter}: Finetuning vision-language
  segmentation efficiently with lightweight blocks}, in:
  \bibinfo{booktitle}{Medical Image Computing and Computer Assisted
  Intervention - {MICCAI} 2024}, \bibinfo{publisher}{Springer Nature
  Switzerland}, \bibinfo{address}{Cham}. pp. \bibinfo{pages}{712-722}.
\newblock \DOIprefix\doi{10.1007/978-3-031-72114-4_68}.
\bibitem[{Dice(1945)}]{dice1945}
\bibinfo{author}{Dice, L.R.}, \bibinfo{year}{1945}.
\newblock \bibinfo{title}{Measures of the amount of ecologic association
  between species}.
\newblock \bibinfo{journal}{Ecology} \bibinfo{volume}{26},
  \bibinfo{pages}{297-302}.
\newblock \DOIprefix\doi{10.2307/1932409}.
\bibitem[{Dosovitskiy et~al.(2021)Dosovitskiy, Beyer, Kolesnikov, Weissenborn,
  Zhai, Unterthiner, Dehghani, Minderer, Heigold, Gelly, Uszkoreit and
  Houlsby}]{vit}
\bibinfo{author}{Dosovitskiy, A.}, \bibinfo{author}{Beyer, L.},
  \bibinfo{author}{Kolesnikov, A.}, \bibinfo{author}{Weissenborn, D.},
  \bibinfo{author}{Zhai, X.}, \bibinfo{author}{Unterthiner, T.},
  \bibinfo{author}{Dehghani, M.}, \bibinfo{author}{Minderer, M.},
  \bibinfo{author}{Heigold, G.}, \bibinfo{author}{Gelly, S.},
  \bibinfo{author}{Uszkoreit, J.}, \bibinfo{author}{Houlsby, N.},
  \bibinfo{year}{2021}.
\newblock \bibinfo{title}{An image is worth 16x16 words: Transformers for image
  recognition at scale}.
\newblock \bibinfo{howpublished}{International Conference on Learning
  Representations (ICLR)}.
\newblock \URLprefix \url{https://openreview.net/forum?id=YicbFdNTTy}.
\bibitem[{Du et~al.(2023)Du, Teng, Li, Liu, Yuan, Wang, Yuan and Zhao}]{du2023}
\bibinfo{author}{Du, C.}, \bibinfo{author}{Teng, J.}, \bibinfo{author}{Li, T.},
  \bibinfo{author}{Liu, Y.}, \bibinfo{author}{Yuan, T.}, \bibinfo{author}{Wang,
  Y.}, \bibinfo{author}{Yuan, Y.}, \bibinfo{author}{Zhao, H.},
  \bibinfo{year}{2023}.
\newblock \bibinfo{title}{On uni-modal feature learning in supervised
  multi-modal learning}, in: \bibinfo{booktitle}{Proceedings of the 40th
  International Conference on Machine Learning (ICML)}, pp.
  \bibinfo{pages}{8632-8656}.
\newblock \URLprefix \url{https://proceedings.mlr.press/v202/du23e.html}.
\bibitem[{Efron and Tibshirani(1993)}]{efron1994}
\bibinfo{author}{Efron, B.}, \bibinfo{author}{Tibshirani, R.J.},
  \bibinfo{year}{1993}.
\newblock \bibinfo{title}{An Introduction to the Bootstrap}.
\newblock Monographs on Statistics and Applied Probability,
  \bibinfo{publisher}{Chapman \& Hall}, \bibinfo{address}{New York, NY, USA}.
\newblock \DOIprefix\doi{10.1007/978-1-4899-4541-9}.
\bibitem[{Feng et~al.(2019)Feng, Wallace and Boyd-Graber}]{feng2019}
\bibinfo{author}{Feng, S.}, \bibinfo{author}{Wallace, E.},
  \bibinfo{author}{Boyd-Graber, J.}, \bibinfo{year}{2019}.
\newblock \bibinfo{title}{Misleading failures of partial-input baselines}, in:
  \bibinfo{booktitle}{Proceedings of the 57th Annual Meeting of the Association
  for Computational Linguistics (ACL)}, pp. \bibinfo{pages}{5533-5538}.
\newblock \DOIprefix\doi{10.18653/v1/P19-1554}.
\bibitem[{Geirhos et~al.(2020)Geirhos, Jacobsen, Michaelis, Zemel, Brendel,
  Bethge and Wichmann}]{geirhos2020}
\bibinfo{author}{Geirhos, R.}, \bibinfo{author}{Jacobsen, J.H.},
  \bibinfo{author}{Michaelis, C.}, \bibinfo{author}{Zemel, R.},
  \bibinfo{author}{Brendel, W.}, \bibinfo{author}{Bethge, M.},
  \bibinfo{author}{Wichmann, F.A.}, \bibinfo{year}{2020}.
\newblock \bibinfo{title}{Shortcut learning in deep neural networks}.
\newblock \bibinfo{journal}{Nat. Mach. Intell.} \bibinfo{volume}{2},
  \bibinfo{pages}{665-673}.
\newblock \DOIprefix\doi{10.1038/s42256-020-00257-z}.
\bibitem[{Goyal et~al.(2017)Goyal, Khot, Summers-Stay, Batra and
  Parikh}]{goyal2017}
\bibinfo{author}{Goyal, Y.}, \bibinfo{author}{Khot, T.},
  \bibinfo{author}{Summers-Stay, D.}, \bibinfo{author}{Batra, D.},
  \bibinfo{author}{Parikh, D.}, \bibinfo{year}{2017}.
\newblock \bibinfo{title}{Making the {V} in {VQA} matter: Elevating the role of
  image understanding in visual question answering}, in:
  \bibinfo{booktitle}{Proceedings of the IEEE Conference on Computer Vision and
  Pattern Recognition (CVPR)}, pp. \bibinfo{pages}{6325-6334}.
\newblock \DOIprefix\doi{10.1109/CVPR.2017.670}.
\bibitem[{Gu et~al.(2022)Gu, Tinn, Cheng, Lucas, Usuyama, Liu, Naumann, Gao and
  Poon}]{pubmedbert}
\bibinfo{author}{Gu, Y.}, \bibinfo{author}{Tinn, R.}, \bibinfo{author}{Cheng,
  H.}, \bibinfo{author}{Lucas, M.}, \bibinfo{author}{Usuyama, N.},
  \bibinfo{author}{Liu, X.}, \bibinfo{author}{Naumann, T.},
  \bibinfo{author}{Gao, J.}, \bibinfo{author}{Poon, H.}, \bibinfo{year}{2022}.
\newblock \bibinfo{title}{Domain-specific language model pretraining for
  biomedical natural language processing}.
\newblock \bibinfo{journal}{ACM Trans. Comput. Healthc.} \bibinfo{volume}{3},
  \bibinfo{pages}{1-23}.
\newblock \DOIprefix\doi{10.1145/3458754}.
\bibitem[{Gururangan et~al.(2018)Gururangan, Swayamdipta, Levy, Schwartz,
  Bowman and Smith}]{gururangan2018}
\bibinfo{author}{Gururangan, S.}, \bibinfo{author}{Swayamdipta, S.},
  \bibinfo{author}{Levy, O.}, \bibinfo{author}{Schwartz, R.},
  \bibinfo{author}{Bowman, S.}, \bibinfo{author}{Smith, N.A.},
  \bibinfo{year}{2018}.
\newblock \bibinfo{title}{Annotation artifacts in natural language inference
  data}, in: \bibinfo{booktitle}{Proceedings of the 2018 Conference of the
  North American Chapter of the Association for Computational Linguistics:
  Human Language Technologies (NAACL-HLT), Volume 2 (Short Papers)}, pp.
  \bibinfo{pages}{107-112}.
\newblock \DOIprefix\doi{10.18653/v1/N18-2017}.
\bibitem[{Hessel and Lee(2020)}]{hessel2020}
\bibinfo{author}{Hessel, J.}, \bibinfo{author}{Lee, L.}, \bibinfo{year}{2020}.
\newblock \bibinfo{title}{Does my multimodal model learn cross-modal
  interactions? {It's} harder to tell than you might think!}, in:
  \bibinfo{booktitle}{Proceedings of the 2020 Conference on Empirical Methods
  in Natural Language Processing (EMNLP)}, pp. \bibinfo{pages}{861-877}.
\newblock \DOIprefix\doi{10.18653/v1/2020.emnlp-main.62}.
\bibitem[{Hoerl and Kennard(1970)}]{hoerl1970}
\bibinfo{author}{Hoerl, A.E.}, \bibinfo{author}{Kennard, R.W.},
  \bibinfo{year}{1970}.
\newblock \bibinfo{title}{Ridge regression: Biased estimation for nonorthogonal
  problems}.
\newblock \bibinfo{journal}{Technometrics} \bibinfo{volume}{12},
  \bibinfo{pages}{55-67}.
\newblock \DOIprefix\doi{10.1080/00401706.1970.10488634}.
\bibitem[{Huang(2025)}]{reclmis_code}
\bibinfo{author}{Huang, X.}, \bibinfo{year}{2025}.
\newblock \bibinfo{title}{{RecLMIS}: Code repository}.
\newblock \bibinfo{howpublished}{GitHub repository
  \mbox{ShawnHuang497/RecLMIS}, commit d3265c8}.
\newblock \URLprefix \url{https://github.com/ShawnHuang497/RecLMIS}.
  \bibinfo{note}{{[software] (accessed 21 October 2026)}}.
\bibitem[{Huang et~al.(2025)Huang, Li, Cao, Chen, You and An}]{reclmis}
\bibinfo{author}{Huang, X.}, \bibinfo{author}{Li, H.}, \bibinfo{author}{Cao,
  M.}, \bibinfo{author}{Chen, L.}, \bibinfo{author}{You, C.},
  \bibinfo{author}{An, D.}, \bibinfo{year}{2025}.
\newblock \bibinfo{title}{Cross-modal conditioned reconstruction for
  language-guided medical image segmentation}.
\newblock \bibinfo{journal}{IEEE Trans. Med. Imaging} \bibinfo{volume}{44},
  \bibinfo{pages}{1821-1835}.
\newblock \DOIprefix\doi{10.1109/TMI.2024.3523333}.
\bibitem[{Huang et~al.(2022)Huang, Lin, Zhou, Yang and Huang}]{huang2022}
\bibinfo{author}{Huang, Y.}, \bibinfo{author}{Lin, J.}, \bibinfo{author}{Zhou,
  C.}, \bibinfo{author}{Yang, H.}, \bibinfo{author}{Huang, L.},
  \bibinfo{year}{2022}.
\newblock \bibinfo{title}{Modality competition: What makes joint training of
  multi-modal network fail in deep learning? ({Provably})}, in:
  \bibinfo{booktitle}{Proceedings of the 39th International Conference on
  Machine Learning (ICML)}, pp. \bibinfo{pages}{9226-9259}.
\newblock \URLprefix \url{https://proceedings.mlr.press/v162/huang22e.html}.
\bibitem[{Jha et~al.(2020)Jha, Smedsrud, Riegler, Halvorsen, de~Lange, Johansen
  and Johansen}]{kvasirseg}
\bibinfo{author}{Jha, D.}, \bibinfo{author}{Smedsrud, P.H.},
  \bibinfo{author}{Riegler, M.A.}, \bibinfo{author}{Halvorsen, P.},
  \bibinfo{author}{de~Lange, T.}, \bibinfo{author}{Johansen, D.},
  \bibinfo{author}{Johansen, H.D.}, \bibinfo{year}{2020}.
\newblock \bibinfo{title}{{Kvasir-SEG}: A segmented polyp dataset}, in:
  \bibinfo{booktitle}{MultiMedia Modeling - {MMM} 2020},
  \bibinfo{publisher}{Springer International Publishing},
  \bibinfo{address}{Cham}. pp. \bibinfo{pages}{451-462}.
\newblock \DOIprefix\doi{10.1007/978-3-030-37734-2_37}.
\bibitem[{Kapoor and Narayanan(2023)}]{kapoor2023}
\bibinfo{author}{Kapoor, S.}, \bibinfo{author}{Narayanan, A.},
  \bibinfo{year}{2023}.
\newblock \bibinfo{title}{Leakage and the reproducibility crisis in
  machine-learning-based science}.
\newblock \bibinfo{journal}{Patterns} \bibinfo{volume}{4},
  \bibinfo{pages}{100804}.
\newblock \DOIprefix\doi{10.1016/j.patter.2023.100804}.
\bibitem[{Kaufman et~al.(2012)Kaufman, Rosset, Perlich and
  Stitelman}]{kaufman2012}
\bibinfo{author}{Kaufman, S.}, \bibinfo{author}{Rosset, S.},
  \bibinfo{author}{Perlich, C.}, \bibinfo{author}{Stitelman, O.},
  \bibinfo{year}{2012}.
\newblock \bibinfo{title}{Leakage in data mining: Formulation, detection, and
  avoidance}.
\newblock \bibinfo{journal}{ACM Trans. Knowl. Discov. Data}
  \bibinfo{volume}{6}, \bibinfo{pages}{1-21}.
\newblock \DOIprefix\doi{10.1145/2382577.2382579}.
\bibitem[{Kervadec et~al.(2019a)Kervadec, Dolz, Granger and {Ben
  Ayed}}]{kervadec2019curr}
\bibinfo{author}{Kervadec, H.}, \bibinfo{author}{Dolz, J.},
  \bibinfo{author}{Granger, {\'E}.}, \bibinfo{author}{{Ben Ayed}, I.},
  \bibinfo{year}{2019}a.
\newblock \bibinfo{title}{Curriculum semi-supervised segmentation}, in:
  \bibinfo{booktitle}{Medical Image Computing and Computer Assisted
  Intervention - {MICCAI} 2019}, \bibinfo{publisher}{Springer International
  Publishing}, \bibinfo{address}{Cham}. pp. \bibinfo{pages}{568-576}.
\newblock \DOIprefix\doi{10.1007/978-3-030-32245-8_63}.
\bibitem[{Kervadec et~al.(2019b)Kervadec, Dolz, Tang, Granger, Boykov and {Ben
  Ayed}}]{kervadec2019}
\bibinfo{author}{Kervadec, H.}, \bibinfo{author}{Dolz, J.},
  \bibinfo{author}{Tang, M.}, \bibinfo{author}{Granger, {\'E}.},
  \bibinfo{author}{Boykov, Y.}, \bibinfo{author}{{Ben Ayed}, I.},
  \bibinfo{year}{2019}b.
\newblock \bibinfo{title}{Constrained-{CNN} losses for weakly supervised
  segmentation}.
\newblock \bibinfo{journal}{Med. Image Anal.} \bibinfo{volume}{54},
  \bibinfo{pages}{88-99}.
\newblock \DOIprefix\doi{10.1016/j.media.2019.02.009}.
\bibitem[{Khattak et~al.(2024)Khattak, Kunhimon, Naseer, Khan and
  Khan}]{unimedclip}
\bibinfo{author}{Khattak, M.U.}, \bibinfo{author}{Kunhimon, S.},
  \bibinfo{author}{Naseer, M.}, \bibinfo{author}{Khan, S.},
  \bibinfo{author}{Khan, F.S.}, \bibinfo{year}{2024}.
\newblock \bibinfo{title}{{UniMed-CLIP}: Towards a unified image-text
  pretraining paradigm for diverse medical imaging modalities}.
\newblock \bibinfo{howpublished}{arXiv preprint arXiv:2412.10372}.
\newblock \DOIprefix\doi{10.48550/arXiv.2412.10372}.
\bibitem[{Kingma and Ba(2015)}]{kingma2015}
\bibinfo{author}{Kingma, D.P.}, \bibinfo{author}{Ba, J.}, \bibinfo{year}{2015}.
\newblock \bibinfo{title}{{Adam}: A method for stochastic optimization}.
\newblock \bibinfo{howpublished}{International Conference on Learning
  Representations (ICLR)}.
\newblock \URLprefix \url{https://arxiv.org/abs/1412.6980}.
\bibitem[{Kirillov et~al.(2023)Kirillov, Mintun, Ravi, Mao, Rolland, Gustafson,
  Xiao, Whitehead, Berg, Lo, Doll{\'a}r and Girshick}]{sam}
\bibinfo{author}{Kirillov, A.}, \bibinfo{author}{Mintun, E.},
  \bibinfo{author}{Ravi, N.}, \bibinfo{author}{Mao, H.},
  \bibinfo{author}{Rolland, C.}, \bibinfo{author}{Gustafson, L.},
  \bibinfo{author}{Xiao, T.}, \bibinfo{author}{Whitehead, S.},
  \bibinfo{author}{Berg, A.C.}, \bibinfo{author}{Lo, W.Y.},
  \bibinfo{author}{Doll{\'a}r, P.}, \bibinfo{author}{Girshick, R.},
  \bibinfo{year}{2023}.
\newblock \bibinfo{title}{Segment anything}, in:
  \bibinfo{booktitle}{Proceedings of the IEEE/CVF International Conference on
  Computer Vision (ICCV)}, pp. \bibinfo{pages}{3992-4003}.
\newblock \DOIprefix\doi{10.1109/ICCV51070.2023.00371}.
\bibitem[{Koleilat(2026a)}]{medclipseg_data}
\bibinfo{author}{Koleilat, T.}, \bibinfo{year}{2026}a.
\newblock \bibinfo{title}{{MedCLIPSeg} datasets: Images, masks and text
  prompts}.
\newblock \bibinfo{howpublished}{Hugging Face dataset repository
  TahaKoleilat/MedCLIPSeg, revision b52c8f5}.
\newblock \URLprefix
  \url{https://huggingface.co/datasets/TahaKoleilat/MedCLIPSeg}.
  \bibinfo{note}{{[dataset] (accessed 21 October 2026)}}.
\bibitem[{Koleilat(2026b)}]{medclipseg_ckpt}
\bibinfo{author}{Koleilat, T.}, \bibinfo{year}{2026}b.
\newblock \bibinfo{title}{{MedCLIPSeg} model checkpoints}.
\newblock \bibinfo{howpublished}{Hugging Face model repository
  TahaKoleilat/MedCLIPSeg, revision 69bfb8f}.
\newblock \URLprefix \url{https://huggingface.co/TahaKoleilat/MedCLIPSeg}.
  \bibinfo{note}{{[software] (accessed 21 October 2026)}}.
\bibitem[{Koleilat(2026c)}]{medclipseg_code}
\bibinfo{author}{Koleilat, T.}, \bibinfo{year}{2026}c.
\newblock \bibinfo{title}{{MedCLIPSeg}: Official code repository}.
\newblock \bibinfo{howpublished}{GitHub repository
  \mbox{HealthX-Lab/MedCLIPSeg}, commit 31d2220}.
\newblock \URLprefix \url{https://github.com/HealthX-Lab/MedCLIPSeg}.
  \bibinfo{note}{{[software] (accessed 21 October 2026)}}.
\bibitem[{Koleilat et~al.(2026)Koleilat, Asgariandehkordi, Nejatimanzari,
  Barile, Xiao and Rivaz}]{medclipseg}
\bibinfo{author}{Koleilat, T.}, \bibinfo{author}{Asgariandehkordi, H.},
  \bibinfo{author}{Nejatimanzari, O.}, \bibinfo{author}{Barile, B.},
  \bibinfo{author}{Xiao, Y.}, \bibinfo{author}{Rivaz, H.},
  \bibinfo{year}{2026}.
\newblock \bibinfo{title}{{MedCLIPSeg}: Probabilistic vision-language
  adaptation for data-efficient and generalizable medical image segmentation},
  in: \bibinfo{booktitle}{Proceedings of the IEEE/CVF Conference on Computer
  Vision and Pattern Recognition (CVPR)}, pp. \bibinfo{pages}{1406-1417}.
\newblock \URLprefix
  \url{https://openaccess.thecvf.com/content/CVPR2026/html/Koleilat_MedCLIPSeg_Probabilistic_Vision-Language_Adaptation_for_Data-Efficient_and_Generalizable_Medical_Image_CVPR_2026_paper.html}.
  \bibinfo{note}{{arXiv:2602.20423}}.
\bibitem[{Koleilat et~al.(2025)Koleilat, Asgariandehkordi, Rivaz and
  Xiao}]{medclipsamv2}
\bibinfo{author}{Koleilat, T.}, \bibinfo{author}{Asgariandehkordi, H.},
  \bibinfo{author}{Rivaz, H.}, \bibinfo{author}{Xiao, Y.},
  \bibinfo{year}{2025}.
\newblock \bibinfo{title}{{MedCLIP-SAMv2}: Towards universal text-driven
  medical image segmentation}.
\newblock \bibinfo{journal}{Med. Image Anal.} \bibinfo{volume}{106},
  \bibinfo{pages}{103749}.
\newblock \DOIprefix\doi{10.1016/j.media.2025.103749}.
\bibitem[{Lapuschkin et~al.(2019)Lapuschkin, W{\"a}ldchen, Binder, Montavon,
  Samek and M{\"u}ller}]{lapuschkin2019}
\bibinfo{author}{Lapuschkin, S.}, \bibinfo{author}{W{\"a}ldchen, S.},
  \bibinfo{author}{Binder, A.}, \bibinfo{author}{Montavon, G.},
  \bibinfo{author}{Samek, W.}, \bibinfo{author}{M{\"u}ller, K.R.},
  \bibinfo{year}{2019}.
\newblock \bibinfo{title}{Unmasking {Clever Hans} predictors and assessing what
  machines really learn}.
\newblock \bibinfo{journal}{Nat. Commun.} \bibinfo{volume}{10},
  \bibinfo{pages}{1096}.
\newblock \DOIprefix\doi{10.1038/s41467-019-08987-4}.
\bibitem[{Lemay et~al.(2021)Lemay, Gros, Vincent, Liu, Cohen and
  Cohen-Adad}]{lemay2021}
\bibinfo{author}{Lemay, A.}, \bibinfo{author}{Gros, C.},
  \bibinfo{author}{Vincent, O.}, \bibinfo{author}{Liu, Y.},
  \bibinfo{author}{Cohen, J.P.}, \bibinfo{author}{Cohen-Adad, J.},
  \bibinfo{year}{2021}.
\newblock \bibinfo{title}{Benefits of linear conditioning for segmentation
  using metadata}, in: \bibinfo{booktitle}{Proceedings of the Fourth Conference
  on Medical Imaging with Deep Learning (MIDL)}, pp. \bibinfo{pages}{416-430}.
\newblock \URLprefix \url{https://proceedings.mlr.press/v143/lemay21a.html}.
\bibitem[{Li et~al.(2026)Li, Juvekar, Zhang, Liu, Wahed, Nguyen, Shen, Yu and
  Lourentzou}]{hallusegbench}
\bibinfo{author}{Li, X.}, \bibinfo{author}{Juvekar, A.},
  \bibinfo{author}{Zhang, J.}, \bibinfo{author}{Liu, X.},
  \bibinfo{author}{Wahed, M.}, \bibinfo{author}{Nguyen, K.A.},
  \bibinfo{author}{Shen, Y.}, \bibinfo{author}{Yu, T.},
  \bibinfo{author}{Lourentzou, I.}, \bibinfo{year}{2026}.
\newblock \bibinfo{title}{Counterfactual segmentation reasoning: Diagnosing and
  mitigating pixel-grounding hallucination}, in:
  \bibinfo{booktitle}{Proceedings of the IEEE/CVF Conference on Computer Vision
  and Pattern Recognition Findings Track (CVPRF)}, pp.
  \bibinfo{pages}{7453-7463}.
\newblock \DOIprefix\doi{10.1109/CVPRF71909.2026.00715}.
\bibitem[{Li(2026)}]{lvit_code}
\bibinfo{author}{Li, Z.}, \bibinfo{year}{2026}.
\newblock \bibinfo{title}{{LViT}: Code repository and {QaTa-COV19} and
  {MosMedData+} text annotations}.
\newblock \bibinfo{howpublished}{GitHub repository \mbox{HUANGLIZI/LViT},
  commit ba90775}.
\newblock \URLprefix \url{https://github.com/HUANGLIZI/LViT}.
  \bibinfo{note}{{[dataset] (accessed 21 October 2026)}}.
\bibitem[{Li et~al.(2024)Li, Li, Li, Wang, Guo, Lu, Jin, Zhang and Hong}]{lvit}
\bibinfo{author}{Li, Z.}, \bibinfo{author}{Li, Y.}, \bibinfo{author}{Li, Q.},
  \bibinfo{author}{Wang, P.}, \bibinfo{author}{Guo, D.}, \bibinfo{author}{Lu,
  L.}, \bibinfo{author}{Jin, D.}, \bibinfo{author}{Zhang, Y.},
  \bibinfo{author}{Hong, Q.}, \bibinfo{year}{2024}.
\newblock \bibinfo{title}{{LViT}: Language meets vision transformer in medical
  image segmentation}.
\newblock \bibinfo{journal}{IEEE Trans. Med. Imaging} \bibinfo{volume}{43},
  \bibinfo{pages}{96-107}.
\newblock \DOIprefix\doi{10.1109/TMI.2023.3291719}.
\bibitem[{Lian et~al.(2021)Lian, Liu, Zhang, Gao, Liu, Zhang and
  Yu}]{chestxdet}
\bibinfo{author}{Lian, J.}, \bibinfo{author}{Liu, J.}, \bibinfo{author}{Zhang,
  S.}, \bibinfo{author}{Gao, K.}, \bibinfo{author}{Liu, X.},
  \bibinfo{author}{Zhang, D.}, \bibinfo{author}{Yu, Y.}, \bibinfo{year}{2021}.
\newblock \bibinfo{title}{A structure-aware relation network for thoracic
  diseases detection and segmentation}.
\newblock \bibinfo{journal}{IEEE Trans. Med. Imaging} \bibinfo{volume}{40},
  \bibinfo{pages}{2042-2052}.
\newblock \DOIprefix\doi{10.1109/TMI.2021.3070847}.
\bibitem[{Liang et~al.(2026)Liang, Wang, Li, Liu and Wang}]{cafe2026}
\bibinfo{author}{Liang, S.}, \bibinfo{author}{Wang, Z.}, \bibinfo{author}{Li,
  Y.}, \bibinfo{author}{Liu, X.}, \bibinfo{author}{Wang, H.},
  \bibinfo{year}{2026}.
\newblock \bibinfo{title}{From pixels to concepts: Do segmentation models
  understand what they segment?}
\newblock \bibinfo{howpublished}{arXiv preprint arXiv:2605.09591}.
\newblock \DOIprefix\doi{10.48550/arXiv.2605.09591}.
\bibitem[{Lin et~al.(2026)Lin, Zhao, Xiong, Zhang, Du and Zhu}]{tgcnet}
\bibinfo{author}{Lin, G.}, \bibinfo{author}{Zhao, H.}, \bibinfo{author}{Xiong,
  Y.}, \bibinfo{author}{Zhang, L.}, \bibinfo{author}{Du, B.},
  \bibinfo{author}{Zhu, W.}, \bibinfo{year}{2026}.
\newblock \bibinfo{title}{{TGC-Net}: A structure-aware and semantically-aligned
  framework for text-guided medical image segmentation}.
\newblock \bibinfo{journal}{Inf. Fusion} \bibinfo{volume}{135},
  \bibinfo{pages}{104438}.
\newblock \DOIprefix\doi{10.1016/j.inffus.2026.104438}.
\bibitem[{Liu et~al.(2026)Liu, Zhu and Shi}]{textsens2609}
\bibinfo{author}{Liu, Z.}, \bibinfo{author}{Zhu, Z.}, \bibinfo{author}{Shi,
  X.}, \bibinfo{year}{2026}.
\newblock \bibinfo{title}{Characterizing text branch sensitivity in medical
  vision-language segmentation via evidence decoupling}.
\newblock \bibinfo{howpublished}{arXiv preprint arXiv:2609.02663}.
\newblock \DOIprefix\doi{10.48550/arXiv.2609.02663}.
\bibitem[{Loshchilov and Hutter(2017)}]{loshchilov2017}
\bibinfo{author}{Loshchilov, I.}, \bibinfo{author}{Hutter, F.},
  \bibinfo{year}{2017}.
\newblock \bibinfo{title}{{SGDR}: Stochastic gradient descent with warm
  restarts}.
\newblock \bibinfo{howpublished}{International Conference on Learning
  Representations (ICLR)}.
\newblock \URLprefix \url{https://arxiv.org/abs/1608.03983}.
\bibitem[{Ma et~al.(2024)Ma, He, Li, Han, You and Wang}]{medsam}
\bibinfo{author}{Ma, J.}, \bibinfo{author}{He, Y.}, \bibinfo{author}{Li, F.},
  \bibinfo{author}{Han, L.}, \bibinfo{author}{You, C.}, \bibinfo{author}{Wang,
  B.}, \bibinfo{year}{2024}.
\newblock \bibinfo{title}{Segment anything in medical images}.
\newblock \bibinfo{journal}{Nat. Commun.} \bibinfo{volume}{15},
  \bibinfo{pages}{654}.
\newblock \DOIprefix\doi{10.1038/s41467-024-44824-z}.
\bibitem[{Ma et~al.(2022)Ma, Ren, Zhao, Testuggine and Peng}]{ma2022}
\bibinfo{author}{Ma, M.}, \bibinfo{author}{Ren, J.}, \bibinfo{author}{Zhao,
  L.}, \bibinfo{author}{Testuggine, D.}, \bibinfo{author}{Peng, X.},
  \bibinfo{year}{2022}.
\newblock \bibinfo{title}{Are multimodal transformers robust to missing
  modality?}, in: \bibinfo{booktitle}{Proceedings of the IEEE/CVF Conference on
  Computer Vision and Pattern Recognition (CVPR)}, pp.
  \bibinfo{pages}{18156-18165}.
\newblock \DOIprefix\doi{10.1109/CVPR52688.2022.01764}.
\bibitem[{Maier-Hein et~al.(2018)Maier-Hein, Eisenmann, Reinke, Onogur,
  Stankovic, Scholz, Arbel, Bogunovic, Bradley, Carass et~al.}]{maierhein2018}
\bibinfo{author}{Maier-Hein, L.}, \bibinfo{author}{Eisenmann, M.},
  \bibinfo{author}{Reinke, A.}, \bibinfo{author}{Onogur, S.},
  \bibinfo{author}{Stankovic, M.}, \bibinfo{author}{Scholz, P.},
  \bibinfo{author}{Arbel, T.}, \bibinfo{author}{Bogunovic, H.},
  \bibinfo{author}{Bradley, A.P.}, \bibinfo{author}{Carass, A.}, et~al.,
  \bibinfo{year}{2018}.
\newblock \bibinfo{title}{Why rankings of biomedical image analysis
  competitions should be interpreted with care}.
\newblock \bibinfo{journal}{Nat. Commun.} \bibinfo{volume}{9},
  \bibinfo{pages}{5217}.
\newblock \DOIprefix\doi{10.1038/s41467-018-07619-7}.
\bibitem[{Maier-Hein et~al.(2024)Maier-Hein, Reinke, Godau, Tizabi, Buettner,
  Christodoulou, Glocker, Isensee, Kleesiek, Kozubek et~al.}]{maierhein2024}
\bibinfo{author}{Maier-Hein, L.}, \bibinfo{author}{Reinke, A.},
  \bibinfo{author}{Godau, P.}, \bibinfo{author}{Tizabi, M.D.},
  \bibinfo{author}{Buettner, F.}, \bibinfo{author}{Christodoulou, E.},
  \bibinfo{author}{Glocker, B.}, \bibinfo{author}{Isensee, F.},
  \bibinfo{author}{Kleesiek, J.}, \bibinfo{author}{Kozubek, M.}, et~al.,
  \bibinfo{year}{2024}.
\newblock \bibinfo{title}{Metrics reloaded: Recommendations for image analysis
  validation}.
\newblock \bibinfo{journal}{Nat. Methods} \bibinfo{volume}{21},
  \bibinfo{pages}{195-212}.
\newblock \DOIprefix\doi{10.1038/s41592-023-02151-z}.
\bibitem[{Mazurowski et~al.(2023)Mazurowski, Dong, Gu, Yang, Konz and
  Zhang}]{mazurowski2023}
\bibinfo{author}{Mazurowski, M.A.}, \bibinfo{author}{Dong, H.},
  \bibinfo{author}{Gu, H.}, \bibinfo{author}{Yang, J.}, \bibinfo{author}{Konz,
  N.}, \bibinfo{author}{Zhang, Y.}, \bibinfo{year}{2023}.
\newblock \bibinfo{title}{Segment anything model for medical image analysis: An
  experimental study}.
\newblock \bibinfo{journal}{Med. Image Anal.} \bibinfo{volume}{89},
  \bibinfo{pages}{102918}.
\newblock \DOIprefix\doi{10.1016/j.media.2023.102918}.
\bibitem[{Miller(1955)}]{miller1955}
\bibinfo{author}{Miller, G.A.}, \bibinfo{year}{1955}.
\newblock \bibinfo{title}{Note on the bias of information estimates}, in:
  \bibinfo{editor}{Quastler, H.} (Ed.), \bibinfo{booktitle}{Information Theory
  in Psychology: Problems and Methods}. \bibinfo{publisher}{Free Press},
  \bibinfo{address}{Glencoe, IL, USA}, pp. \bibinfo{pages}{95-100}.
\bibitem[{Morozov et~al.(2020)Morozov, Andreychenko, Blokhin, Gelezhe, Gonchar,
  Nikolaev, Pavlov, Chernina and Gombolevskiy}]{morozov2020}
\bibinfo{author}{Morozov, S.P.}, \bibinfo{author}{Andreychenko, A.E.},
  \bibinfo{author}{Blokhin, I.A.}, \bibinfo{author}{Gelezhe, P.B.},
  \bibinfo{author}{Gonchar, A.P.}, \bibinfo{author}{Nikolaev, A.E.},
  \bibinfo{author}{Pavlov, N.A.}, \bibinfo{author}{Chernina, V.Y.},
  \bibinfo{author}{Gombolevskiy, V.A.}, \bibinfo{year}{2020}.
\newblock \bibinfo{title}{{MosMedData}: Data set of 1110 chest {CT} scans
  performed during the {COVID-19} epidemic}.
\newblock \bibinfo{journal}{Digit. Diagn.} \bibinfo{volume}{1},
  \bibinfo{pages}{49-59}.
\newblock \DOIprefix\doi{10.17816/dd46826}.
\bibitem[{Neverova et~al.(2016)Neverova, Wolf, Taylor and Nebout}]{moddrop}
\bibinfo{author}{Neverova, N.}, \bibinfo{author}{Wolf, C.},
  \bibinfo{author}{Taylor, G.}, \bibinfo{author}{Nebout, F.},
  \bibinfo{year}{2016}.
\newblock \bibinfo{title}{{ModDrop}: Adaptive multi-modal gesture recognition}.
\newblock \bibinfo{journal}{IEEE Trans. Pattern Anal. Mach. Intell.}
  \bibinfo{volume}{38}, \bibinfo{pages}{1692-1706}.
\newblock \DOIprefix\doi{10.1109/TPAMI.2015.2461544}.
\bibitem[{{Ngoc Lan} et~al.(2021){Ngoc Lan}, An, Hang, Long, Trung, Thuy and
  Sang}]{bkai}
\bibinfo{author}{{Ngoc Lan}, P.}, \bibinfo{author}{An, N.S.},
  \bibinfo{author}{Hang, D.V.}, \bibinfo{author}{Long, D.V.},
  \bibinfo{author}{Trung, T.Q.}, \bibinfo{author}{Thuy, N.T.},
  \bibinfo{author}{Sang, D.V.}, \bibinfo{year}{2021}.
\newblock \bibinfo{title}{{NeoUNet}: Towards accurate colon polyp segmentation
  and neoplasm detection}, in: \bibinfo{booktitle}{Advances in Visual Computing
  - {ISVC} 2021}, \bibinfo{publisher}{Springer International Publishing},
  \bibinfo{address}{Cham}. pp. \bibinfo{pages}{15-28}.
\newblock \DOIprefix\doi{10.1007/978-3-030-90436-4_2}.
\bibitem[{Nikolov et~al.(2021)Nikolov, Blackwell, Zverovitch, Mendes, Livne,
  {De Fauw}, Patel, Meyer, Askham, Romera-Paredes et~al.}]{nsd}
\bibinfo{author}{Nikolov, S.}, \bibinfo{author}{Blackwell, S.},
  \bibinfo{author}{Zverovitch, A.}, \bibinfo{author}{Mendes, R.},
  \bibinfo{author}{Livne, M.}, \bibinfo{author}{{De Fauw}, J.},
  \bibinfo{author}{Patel, Y.}, \bibinfo{author}{Meyer, C.},
  \bibinfo{author}{Askham, H.}, \bibinfo{author}{Romera-Paredes, B.}, et~al.,
  \bibinfo{year}{2021}.
\newblock \bibinfo{title}{Clinically applicable segmentation of head and neck
  anatomy for radiotherapy: Deep learning algorithm development and validation
  study}.
\newblock \bibinfo{journal}{J. Med. Internet Res.} \bibinfo{volume}{23},
  \bibinfo{pages}{e26151}.
\newblock \DOIprefix\doi{10.2196/26151}.
\bibitem[{{Ong Ly} et~al.(2024){Ong Ly}, Unnikrishnan, Tadic, Patel, Duhamel,
  Kandel, Moayedi, Brudno, Hope, Ross and McIntosh}]{ongly2024}
\bibinfo{author}{{Ong Ly}, C.}, \bibinfo{author}{Unnikrishnan, B.},
  \bibinfo{author}{Tadic, T.}, \bibinfo{author}{Patel, T.},
  \bibinfo{author}{Duhamel, J.}, \bibinfo{author}{Kandel, S.},
  \bibinfo{author}{Moayedi, Y.}, \bibinfo{author}{Brudno, M.},
  \bibinfo{author}{Hope, A.}, \bibinfo{author}{Ross, H.},
  \bibinfo{author}{McIntosh, C.}, \bibinfo{year}{2024}.
\newblock \bibinfo{title}{Shortcut learning in medical {AI} hinders
  generalization: Method for estimating {AI} model generalization without
  external data}.
\newblock \bibinfo{journal}{npj Digit. Med.} \bibinfo{volume}{7},
  \bibinfo{pages}{124}.
\newblock \DOIprefix\doi{10.1038/s41746-024-01118-4}.
\bibitem[{Otsu(1979)}]{otsu1979}
\bibinfo{author}{Otsu, N.}, \bibinfo{year}{1979}.
\newblock \bibinfo{title}{A threshold selection method from gray-level
  histograms}.
\newblock \bibinfo{journal}{IEEE Trans. Syst. Man Cybern.} \bibinfo{volume}{9},
  \bibinfo{pages}{62-66}.
\newblock \DOIprefix\doi{10.1109/TSMC.1979.4310076}.
\bibitem[{Paszke et~al.(2019)Paszke, Gross, Massa, Lerer, Bradbury, Chanan,
  Killeen, Lin, Gimelshein, Antiga, Desmaison, K{\"o}pf, Yang, DeVito, Raison,
  Tejani, Chilamkurthy, Steiner, Fang, Bai and Chintala}]{pytorch}
\bibinfo{author}{Paszke, A.}, \bibinfo{author}{Gross, S.},
  \bibinfo{author}{Massa, F.}, \bibinfo{author}{Lerer, A.},
  \bibinfo{author}{Bradbury, J.}, \bibinfo{author}{Chanan, G.},
  \bibinfo{author}{Killeen, T.}, \bibinfo{author}{Lin, Z.},
  \bibinfo{author}{Gimelshein, N.}, \bibinfo{author}{Antiga, L.},
  \bibinfo{author}{Desmaison, A.}, \bibinfo{author}{K{\"o}pf, A.},
  \bibinfo{author}{Yang, E.}, \bibinfo{author}{DeVito, Z.},
  \bibinfo{author}{Raison, M.}, \bibinfo{author}{Tejani, A.},
  \bibinfo{author}{Chilamkurthy, S.}, \bibinfo{author}{Steiner, B.},
  \bibinfo{author}{Fang, L.}, \bibinfo{author}{Bai, J.},
  \bibinfo{author}{Chintala, S.}, \bibinfo{year}{2019}.
\newblock \bibinfo{title}{{PyTorch}: An imperative style, high-performance deep
  learning library}, in: \bibinfo{booktitle}{Advances in Neural Information
  Processing Systems}, pp. \bibinfo{pages}{8026-8037}.
\bibitem[{Pedregosa et~al.(2011)Pedregosa, Varoquaux, Gramfort, Michel,
  Thirion, Grisel, Blondel, Prettenhofer, Weiss, Dubourg, Vanderplas, Passos,
  Cournapeau, Brucher, Perrot and Duchesnay}]{sklearn}
\bibinfo{author}{Pedregosa, F.}, \bibinfo{author}{Varoquaux, G.},
  \bibinfo{author}{Gramfort, A.}, \bibinfo{author}{Michel, V.},
  \bibinfo{author}{Thirion, B.}, \bibinfo{author}{Grisel, O.},
  \bibinfo{author}{Blondel, M.}, \bibinfo{author}{Prettenhofer, P.},
  \bibinfo{author}{Weiss, R.}, \bibinfo{author}{Dubourg, V.},
  \bibinfo{author}{Vanderplas, J.}, \bibinfo{author}{Passos, A.},
  \bibinfo{author}{Cournapeau, D.}, \bibinfo{author}{Brucher, M.},
  \bibinfo{author}{Perrot, M.}, \bibinfo{author}{Duchesnay, {\'E}.},
  \bibinfo{year}{2011}.
\newblock \bibinfo{title}{Scikit-learn: Machine learning in {Python}}.
\newblock \bibinfo{journal}{J. Mach. Learn. Res.} \bibinfo{volume}{12},
  \bibinfo{pages}{2825-2830}.
\newblock \URLprefix \url{https://jmlr.org/papers/v12/pedregosa11a.html}.
\bibitem[{Peng et~al.(2022)Peng, Wei, Deng, Wang and Hu}]{peng2022}
\bibinfo{author}{Peng, X.}, \bibinfo{author}{Wei, Y.}, \bibinfo{author}{Deng,
  A.}, \bibinfo{author}{Wang, D.}, \bibinfo{author}{Hu, D.},
  \bibinfo{year}{2022}.
\newblock \bibinfo{title}{Balanced multimodal learning via on-the-fly gradient
  modulation}, in: \bibinfo{booktitle}{Proceedings of the IEEE/CVF Conference
  on Computer Vision and Pattern Recognition (CVPR)}, pp.
  \bibinfo{pages}{8228-8237}.
\newblock \DOIprefix\doi{10.1109/CVPR52688.2022.00806}.
\bibitem[{Pfungst(1911)}]{pfungst1911}
\bibinfo{author}{Pfungst, O.}, \bibinfo{year}{1911}.
\newblock \bibinfo{title}{Clever {Hans} (The Horse of {Mr. von Osten}): A
  Contribution to Experimental Animal and Human Psychology}.
\newblock \bibinfo{publisher}{Henry Holt and Company}, \bibinfo{address}{New
  York, NY, USA}.
\newblock \DOIprefix\doi{10.5962/bhl.title.56164}. \bibinfo{note}{{Translated
  from the German by C. L. Rahn}}.
\bibitem[{Poliak et~al.(2018)Poliak, Naradowsky, Haldar, Rudinger and {Van
  Durme}}]{poliak2018}
\bibinfo{author}{Poliak, A.}, \bibinfo{author}{Naradowsky, J.},
  \bibinfo{author}{Haldar, A.}, \bibinfo{author}{Rudinger, R.},
  \bibinfo{author}{{Van Durme}, B.}, \bibinfo{year}{2018}.
\newblock \bibinfo{title}{Hypothesis only baselines in natural language
  inference}, in: \bibinfo{booktitle}{Proceedings of the Seventh Joint
  Conference on Lexical and Computational Semantics (*SEM)}, pp.
  \bibinfo{pages}{180-191}.
\newblock \DOIprefix\doi{10.18653/v1/S18-2023}.
\bibitem[{Poudel(2024)}]{medvlsm_code}
\bibinfo{author}{Poudel, K.}, \bibinfo{year}{2024}.
\newblock \bibinfo{title}{{MedVLSM}: Exploring transfer learning in medical
  image segmentation using vision-language models, official code and prompt
  files}.
\newblock \bibinfo{howpublished}{GitHub repository \mbox{naamiinepal/medvlsm},
  commit dc36a3a}.
\newblock \URLprefix \url{https://github.com/naamiinepal/medvlsm}.
  \bibinfo{note}{{[software and data] (accessed 21 October 2026)}}.
\bibitem[{Poudel et~al.(2024)Poudel, Dhakal, Bhandari, Adhikari, Thapaliya and
  Khanal}]{poudel2024}
\bibinfo{author}{Poudel, K.}, \bibinfo{author}{Dhakal, M.},
  \bibinfo{author}{Bhandari, P.}, \bibinfo{author}{Adhikari, R.},
  \bibinfo{author}{Thapaliya, S.}, \bibinfo{author}{Khanal, B.},
  \bibinfo{year}{2024}.
\newblock \bibinfo{title}{Exploring transfer learning in medical image
  segmentation using vision-language models}, in:
  \bibinfo{booktitle}{Proceedings of the International Conference on Medical
  Imaging with Deep Learning (MIDL)}, pp. \bibinfo{pages}{1142-1165}.
\newblock \URLprefix \url{https://proceedings.mlr.press/v250/poudel24a.html}.
\bibitem[{Radford et~al.(2021)Radford, Kim, Hallacy, Ramesh, Goh, Agarwal,
  Sastry, Askell, Mishkin, Clark, Krueger and Sutskever}]{clip}
\bibinfo{author}{Radford, A.}, \bibinfo{author}{Kim, J.W.},
  \bibinfo{author}{Hallacy, C.}, \bibinfo{author}{Ramesh, A.},
  \bibinfo{author}{Goh, G.}, \bibinfo{author}{Agarwal, S.},
  \bibinfo{author}{Sastry, G.}, \bibinfo{author}{Askell, A.},
  \bibinfo{author}{Mishkin, P.}, \bibinfo{author}{Clark, J.},
  \bibinfo{author}{Krueger, G.}, \bibinfo{author}{Sutskever, I.},
  \bibinfo{year}{2021}.
\newblock \bibinfo{title}{Learning transferable visual models from natural
  language supervision}, in: \bibinfo{booktitle}{Proceedings of the 38th
  International Conference on Machine Learning (ICML)}, pp.
  \bibinfo{pages}{8748-8763}.
\newblock \URLprefix \url{https://proceedings.mlr.press/v139/radford21a.html}.
\bibitem[{Rahman and Hammond(2026)}]{ddcmd}
\bibinfo{author}{Rahman, M.M.}, \bibinfo{author}{Hammond, T.},
  \bibinfo{year}{2026}.
\newblock \bibinfo{title}{Dual-domain cross-modal decoding for clinical
  text-guided medical image segmentation}, in: \bibinfo{booktitle}{Medical
  Image Computing and Computer Assisted Intervention - {MICCAI} 2026},
  \bibinfo{publisher}{Springer Nature Switzerland}, \bibinfo{address}{Cham}.
  pp. \bibinfo{pages}{135-145}.
\newblock \DOIprefix\doi{10.1007/978-3-032-38082-1_13}.
\bibitem[{Reinke et~al.(2024)Reinke, Tizabi, Baumgartner, Eisenmann,
  Heckmann-N{\"o}tzel, Kavur, R{\"a}dsch, Sudre, Acion, Antonelli
  et~al.}]{reinke2024}
\bibinfo{author}{Reinke, A.}, \bibinfo{author}{Tizabi, M.D.},
  \bibinfo{author}{Baumgartner, M.}, \bibinfo{author}{Eisenmann, M.},
  \bibinfo{author}{Heckmann-N{\"o}tzel, D.}, \bibinfo{author}{Kavur, A.E.},
  \bibinfo{author}{R{\"a}dsch, T.}, \bibinfo{author}{Sudre, C.H.},
  \bibinfo{author}{Acion, L.}, \bibinfo{author}{Antonelli, M.}, et~al.,
  \bibinfo{year}{2024}.
\newblock \bibinfo{title}{Understanding metric-related pitfalls in image
  analysis validation}.
\newblock \bibinfo{journal}{Nat. Methods} \bibinfo{volume}{21},
  \bibinfo{pages}{182-194}.
\newblock \DOIprefix\doi{10.1038/s41592-023-02150-0}.
\bibitem[{Shan et~al.(2025)Shan, Li, Li, Li, Tian and Hong}]{stpnet}
\bibinfo{author}{Shan, D.}, \bibinfo{author}{Li, Z.}, \bibinfo{author}{Li, Y.},
  \bibinfo{author}{Li, Q.}, \bibinfo{author}{Tian, J.}, \bibinfo{author}{Hong,
  Q.}, \bibinfo{year}{2025}.
\newblock \bibinfo{title}{{STPNet}: Scale-aware text prompt network for medical
  image segmentation}.
\newblock \bibinfo{journal}{IEEE Trans. Image Process.} \bibinfo{volume}{34},
  \bibinfo{pages}{3169-3180}.
\newblock \DOIprefix\doi{10.1109/TIP.2025.3571672}.
\bibitem[{Srikanth and Rudinger(2022)}]{srikanth2022}
\bibinfo{author}{Srikanth, N.}, \bibinfo{author}{Rudinger, R.},
  \bibinfo{year}{2022}.
\newblock \bibinfo{title}{Partial-input baselines show that {NLI} models can
  ignore context, but they don't}, in: \bibinfo{booktitle}{Proceedings of the
  2022 Conference of the North American Chapter of the Association for
  Computational Linguistics: Human Language Technologies (NAACL-HLT)}, pp.
  \bibinfo{pages}{4753-4763}.
\newblock \DOIprefix\doi{10.18653/v1/2022.naacl-main.350}.
\bibitem[{Tajbakhsh et~al.(2016)Tajbakhsh, Gurudu and Liang}]{colondb}
\bibinfo{author}{Tajbakhsh, N.}, \bibinfo{author}{Gurudu, S.R.},
  \bibinfo{author}{Liang, J.}, \bibinfo{year}{2016}.
\newblock \bibinfo{title}{Automated polyp detection in colonoscopy videos using
  shape and context information}.
\newblock \bibinfo{journal}{IEEE Trans. Med. Imaging} \bibinfo{volume}{35},
  \bibinfo{pages}{630-644}.
\newblock \DOIprefix\doi{10.1109/TMI.2015.2487997}.
\bibitem[{Tomar et~al.(2022)Tomar, Jha, Bagci and Ali}]{tganet}
\bibinfo{author}{Tomar, N.K.}, \bibinfo{author}{Jha, D.},
  \bibinfo{author}{Bagci, U.}, \bibinfo{author}{Ali, S.}, \bibinfo{year}{2022}.
\newblock \bibinfo{title}{{TGANet}: Text-guided attention for improved polyp
  segmentation}, in: \bibinfo{booktitle}{Medical Image Computing and Computer
  Assisted Intervention - {MICCAI} 2022}, \bibinfo{publisher}{Springer Nature
  Switzerland}, \bibinfo{address}{Cham}. pp. \bibinfo{pages}{151-160}.
\newblock \DOIprefix\doi{10.1007/978-3-031-16437-8_15}.
\bibitem[{Tomar et~al.(2025)Tomar, Jha, Biswas and Bagci}]{fanetv2}
\bibinfo{author}{Tomar, N.K.}, \bibinfo{author}{Jha, D.},
  \bibinfo{author}{Biswas, K.}, \bibinfo{author}{Bagci, U.},
  \bibinfo{year}{2025}.
\newblock \bibinfo{title}{Transformer-enhanced iterative feedback mechanism for
  polyp segmentation}, in: \bibinfo{booktitle}{Proceedings of the IEEE
  International Conference on Acoustics, Speech and Signal Processing
  (ICASSP)}, pp. \bibinfo{pages}{1-5}.
\newblock \DOIprefix\doi{10.1109/ICASSP49660.2025.10890567}.
\bibitem[{Valcarcel et~al.(2020)Valcarcel, Muschelli, Pham, Martin, Yushkevich,
  Brandstadter, Patterson, Schindler, Calabresi, Bakshi and Shinohara}]{tapas}
\bibinfo{author}{Valcarcel, A.M.}, \bibinfo{author}{Muschelli, J.},
  \bibinfo{author}{Pham, D.L.}, \bibinfo{author}{Martin, M.L.},
  \bibinfo{author}{Yushkevich, P.}, \bibinfo{author}{Brandstadter, R.},
  \bibinfo{author}{Patterson, K.R.}, \bibinfo{author}{Schindler, M.K.},
  \bibinfo{author}{Calabresi, P.A.}, \bibinfo{author}{Bakshi, R.},
  \bibinfo{author}{Shinohara, R.T.}, \bibinfo{year}{2020}.
\newblock \bibinfo{title}{{TAPAS}: A thresholding approach for probability map
  automatic segmentation in multiple sclerosis}.
\newblock \bibinfo{journal}{Neuroimage Clin.} \bibinfo{volume}{27},
  \bibinfo{pages}{102256}.
\newblock \DOIprefix\doi{10.1016/j.nicl.2020.102256}.
\bibitem[{Vapnik and Izmailov(2015)}]{vapnik2015}
\bibinfo{author}{Vapnik, V.}, \bibinfo{author}{Izmailov, R.},
  \bibinfo{year}{2015}.
\newblock \bibinfo{title}{Learning using privileged information: Similarity
  control and knowledge transfer}.
\newblock \bibinfo{journal}{J. Mach. Learn. Res.} \bibinfo{volume}{16},
  \bibinfo{pages}{2023-2049}.
\newblock \URLprefix \url{https://jmlr.org/papers/v16/vapnik15b.html}.
\bibitem[{V{\'a}squez-Venegas et~al.(2025)V{\'a}squez-Venegas, Wu, Sundar,
  Pr{\^o}a, Beloy, Medina, McNichol, Parvataneni, Kurtzman, Mirshawka,
  Aguirre-Jerez, Ebner and Celi}]{vasquez2024}
\bibinfo{author}{V{\'a}squez-Venegas, C.}, \bibinfo{author}{Wu, C.},
  \bibinfo{author}{Sundar, S.}, \bibinfo{author}{Pr{\^o}a, R.},
  \bibinfo{author}{Beloy, F.J.}, \bibinfo{author}{Medina, J.R.},
  \bibinfo{author}{McNichol, M.}, \bibinfo{author}{Parvataneni, K.},
  \bibinfo{author}{Kurtzman, N.}, \bibinfo{author}{Mirshawka, F.},
  \bibinfo{author}{Aguirre-Jerez, M.}, \bibinfo{author}{Ebner, D.K.},
  \bibinfo{author}{Celi, L.A.}, \bibinfo{year}{2025}.
\newblock \bibinfo{title}{Detecting and mitigating the {Clever Hans} effect in
  medical imaging: A scoping review}.
\newblock \bibinfo{journal}{J. Imaging Inform. Med.} \bibinfo{volume}{38},
  \bibinfo{pages}{2563-2579}.
\newblock \DOIprefix\doi{10.1007/s10278-024-01335-z}.
\bibitem[{V{\'a}zquez et~al.(2017)V{\'a}zquez, Bernal, S{\'a}nchez,
  Fern{\'a}ndez-Esparrach, L{\'o}pez, Romero, Drozdzal and Courville}]{cvc300}
\bibinfo{author}{V{\'a}zquez, D.}, \bibinfo{author}{Bernal, J.},
  \bibinfo{author}{S{\'a}nchez, F.J.},
  \bibinfo{author}{Fern{\'a}ndez-Esparrach, G.}, \bibinfo{author}{L{\'o}pez,
  A.M.}, \bibinfo{author}{Romero, A.}, \bibinfo{author}{Drozdzal, M.},
  \bibinfo{author}{Courville, A.}, \bibinfo{year}{2017}.
\newblock \bibinfo{title}{A benchmark for endoluminal scene segmentation of
  colonoscopy images}.
\newblock \bibinfo{journal}{J. Healthc. Eng.} \bibinfo{volume}{2017},
  \bibinfo{pages}{4037190}.
\newblock \DOIprefix\doi{10.1155/2017/4037190}.
\bibitem[{Wang et~al.(2026)Wang, Zhang, Yu, Ong, Huang, Li, Zhang, Cohan, Yu
  and Yao}]{medqamm}
\bibinfo{author}{Wang, B.}, \bibinfo{author}{Zhang, Y.}, \bibinfo{author}{Yu,
  J.}, \bibinfo{author}{Ong, C.S.}, \bibinfo{author}{Huang, J.},
  \bibinfo{author}{Li, Z.}, \bibinfo{author}{Zhang, Z.},
  \bibinfo{author}{Cohan, A.}, \bibinfo{author}{Yu, H.}, \bibinfo{author}{Yao,
  Z.}, \bibinfo{year}{2026}.
\newblock \bibinfo{title}{{MedQA-MM}: Shortcuts behind medical visual
  reasoning}, in: \bibinfo{booktitle}{Proceedings of the 2026 Conference on
  Empirical Methods in Natural Language Processing},
  \bibinfo{publisher}{Association for Computational Linguistics},
  \bibinfo{address}{Budapest, Hungary}. pp. \bibinfo{pages}{10708-10758}.
\newblock \URLprefix \url{https://aclanthology.org/2026.emnlp-main.467/}.
\bibitem[{Wang et~al.(2025)Wang, Lin and Yan}]{arseg}
\bibinfo{author}{Wang, Q.}, \bibinfo{author}{Lin, X.}, \bibinfo{author}{Yan,
  Z.}, \bibinfo{year}{2025}.
\newblock \bibinfo{title}{Towards robust medical image referring segmentation
  with incomplete textual prompts}, in: \bibinfo{booktitle}{Medical Image
  Computing and Computer Assisted Intervention - {MICCAI} 2025},
  \bibinfo{publisher}{Springer Nature Switzerland}, \bibinfo{address}{Cham}.
  pp. \bibinfo{pages}{636-646}.
\newblock \DOIprefix\doi{10.1007/978-3-032-04981-0_60}.
\bibitem[{Wang et~al.(2020)Wang, Tran and Feiszli}]{wang2020}
\bibinfo{author}{Wang, W.}, \bibinfo{author}{Tran, D.},
  \bibinfo{author}{Feiszli, M.}, \bibinfo{year}{2020}.
\newblock \bibinfo{title}{What makes training multi-modal classification
  networks hard?}, in: \bibinfo{booktitle}{Proceedings of the IEEE/CVF
  Conference on Computer Vision and Pattern Recognition (CVPR)}, pp.
  \bibinfo{pages}{12692-12702}.
\newblock \DOIprefix\doi{10.1109/CVPR42600.2020.01271}.
\bibitem[{Wu et~al.(2022)Wu, Jastrzebski, Cho and Geras}]{wu2022}
\bibinfo{author}{Wu, N.}, \bibinfo{author}{Jastrzebski, S.},
  \bibinfo{author}{Cho, K.}, \bibinfo{author}{Geras, K.J.},
  \bibinfo{year}{2022}.
\newblock \bibinfo{title}{Characterizing and overcoming the greedy nature of
  learning in multi-modal deep neural networks}, in:
  \bibinfo{booktitle}{Proceedings of the 39th International Conference on
  Machine Learning (ICML)}, pp. \bibinfo{pages}{24043-24055}.
\newblock \URLprefix \url{https://proceedings.mlr.press/v162/wu22d.html}.
\bibitem[{Xi et~al.(2026)Xi, Hu, Wang, Ding, Li, He, Zhang, {Del Balzo}, Zhong,
  Hu, Liu, Liu and Yang}]{cf2seg}
\bibinfo{author}{Xi, S.}, \bibinfo{author}{Hu, S.}, \bibinfo{author}{Wang, S.},
  \bibinfo{author}{Ding, H.}, \bibinfo{author}{Li, Y.}, \bibinfo{author}{He,
  W.}, \bibinfo{author}{Zhang, K.}, \bibinfo{author}{{Del Balzo}, L.},
  \bibinfo{author}{Zhong, C.}, \bibinfo{author}{Hu, M.}, \bibinfo{author}{Liu,
  C.C.}, \bibinfo{author}{Liu, Y.}, \bibinfo{author}{Yang, X.},
  \bibinfo{year}{2026}.
\newblock \bibinfo{title}{Grounding radiology report findings into medical
  image segmentation}.
\newblock \bibinfo{journal}{npj Digit. Med.}
  \DOIprefix\doi{10.1038/s41746-026-03051-0}.
\bibitem[{Ye et~al.(2024)Ye, Meng, Li, Feng and Kim}]{sgseg}
\bibinfo{author}{Ye, S.}, \bibinfo{author}{Meng, M.}, \bibinfo{author}{Li, M.},
  \bibinfo{author}{Feng, D.}, \bibinfo{author}{Kim, J.}, \bibinfo{year}{2024}.
\newblock \bibinfo{title}{Enabling text-free inference in language-guided
  segmentation of chest {X}-rays via self-guidance}, in:
  \bibinfo{booktitle}{Medical Image Computing and Computer Assisted
  Intervention - {MICCAI} 2024}, \bibinfo{publisher}{Springer Nature
  Switzerland}, \bibinfo{address}{Cham}. pp. \bibinfo{pages}{242-252}.
\newblock \DOIprefix\doi{10.1007/978-3-031-72111-3_23}.
\bibitem[{Ye et~al.(2025)Ye, Naseem, Meng and Kim}]{prolearn}
\bibinfo{author}{Ye, S.}, \bibinfo{author}{Naseem, U.}, \bibinfo{author}{Meng,
  M.}, \bibinfo{author}{Kim, J.}, \bibinfo{year}{2025}.
\newblock \bibinfo{title}{Alleviating textual reliance in medical
  language-guided segmentation via prototype-driven semantic approximation},
  in: \bibinfo{booktitle}{Proceedings of the IEEE/CVF International Conference
  on Computer Vision (ICCV)}, pp. \bibinfo{pages}{22316-22326}.
\newblock \DOIprefix\doi{10.1109/ICCV51701.2025.02072}.
\bibitem[{Yu et~al.(2025)Yu, Yang, Du, Huang, Li and Wang}]{fmiseg}
\bibinfo{author}{Yu, B.}, \bibinfo{author}{Yang, J.}, \bibinfo{author}{Du, Z.},
  \bibinfo{author}{Huang, Y.}, \bibinfo{author}{Li, C.}, \bibinfo{author}{Wang,
  L.}, \bibinfo{year}{2025}.
\newblock \bibinfo{title}{Frequency-domain multi-modal fusion for
  language-guided medical image segmentation}, in: \bibinfo{booktitle}{Medical
  Image Computing and Computer Assisted Intervention - {MICCAI} 2025},
  \bibinfo{publisher}{Springer Nature Switzerland}, \bibinfo{address}{Cham}.
  pp. \bibinfo{pages}{278-288}.
\newblock \DOIprefix\doi{10.1007/978-3-032-05114-1_27}.
\bibitem[{Zech et~al.(2018)Zech, Badgeley, Liu, Costa, Titano and
  Oermann}]{zech2018}
\bibinfo{author}{Zech, J.R.}, \bibinfo{author}{Badgeley, M.A.},
  \bibinfo{author}{Liu, M.}, \bibinfo{author}{Costa, A.B.},
  \bibinfo{author}{Titano, J.J.}, \bibinfo{author}{Oermann, E.K.},
  \bibinfo{year}{2018}.
\newblock \bibinfo{title}{Variable generalization performance of a deep
  learning model to detect pneumonia in chest radiographs: A cross-sectional
  study}.
\newblock \bibinfo{journal}{PLoS Med.} \bibinfo{volume}{15},
  \bibinfo{pages}{e1002683}.
\newblock \DOIprefix\doi{10.1371/journal.pmed.1002683}.
\bibitem[{Zhao et~al.(2025a)Zhao, Hou, Yan and Huo}]{zhao2025kd}
\bibinfo{author}{Zhao, P.}, \bibinfo{author}{Hou, Y.}, \bibinfo{author}{Yan,
  Z.}, \bibinfo{author}{Huo, S.}, \bibinfo{year}{2025}a.
\newblock \bibinfo{title}{Text-driven medical image segmentation with text-free
  inference via knowledge distillation}.
\newblock \bibinfo{journal}{IEEE Trans. Instrum. Meas.} \bibinfo{volume}{74},
  \bibinfo{pages}{1-15}.
\newblock \DOIprefix\doi{10.1109/TIM.2025.3545506}.
\bibitem[{Zhao et~al.(2025b)Zhao, Gu, Yang, Usuyama, Lee, Kiblawi, Naumann,
  Gao, Crabtree, Abel et~al.}]{biomedparse}
\bibinfo{author}{Zhao, T.}, \bibinfo{author}{Gu, Y.}, \bibinfo{author}{Yang,
  J.}, \bibinfo{author}{Usuyama, N.}, \bibinfo{author}{Lee, H.H.},
  \bibinfo{author}{Kiblawi, S.}, \bibinfo{author}{Naumann, T.},
  \bibinfo{author}{Gao, J.}, \bibinfo{author}{Crabtree, A.},
  \bibinfo{author}{Abel, J.}, et~al., \bibinfo{year}{2025}b.
\newblock \bibinfo{title}{A foundation model for joint segmentation, detection
  and recognition of biomedical objects across nine modalities}.
\newblock \bibinfo{journal}{Nat. Methods} \bibinfo{volume}{22},
  \bibinfo{pages}{166-176}.
\newblock \DOIprefix\doi{10.1038/s41592-024-02499-w}.
\bibitem[{Zhao et~al.(2024)Zhao, Li, Diao, Fan and Chen}]{cap2seg}
\bibinfo{author}{Zhao, W.}, \bibinfo{author}{Li, F.}, \bibinfo{author}{Diao,
  Y.}, \bibinfo{author}{Fan, P.}, \bibinfo{author}{Chen, Z.},
  \bibinfo{year}{2024}.
\newblock \bibinfo{title}{{Cap2Seg}: Leveraging caption generation for enhanced
  segmentation of {COVID-19} medical images}.
\newblock \bibinfo{journal}{Front. Phys.} \bibinfo{volume}{12},
  \bibinfo{pages}{1439122}.
\newblock \DOIprefix\doi{10.3389/fphy.2024.1439122}.
\bibitem[{Zhao et~al.(2025c)Zhao, Zhang, Wu, Zhang, Zhou, Zhang, Wang and
  Xie}]{sat}
\bibinfo{author}{Zhao, Z.}, \bibinfo{author}{Zhang, Y.}, \bibinfo{author}{Wu,
  C.}, \bibinfo{author}{Zhang, X.}, \bibinfo{author}{Zhou, X.},
  \bibinfo{author}{Zhang, Y.}, \bibinfo{author}{Wang, Y.},
  \bibinfo{author}{Xie, W.}, \bibinfo{year}{2025}c.
\newblock \bibinfo{title}{Large-vocabulary segmentation for medical images with
  text prompts}.
\newblock \bibinfo{journal}{npj Digit. Med.} \bibinfo{volume}{8},
  \bibinfo{pages}{566}.
\newblock \DOIprefix\doi{10.1038/s41746-025-01964-w}.
\bibitem[{Zhong et~al.(2023)Zhong, Xu, Liang, Chen and Wu}]{guidedecoder}
\bibinfo{author}{Zhong, Y.}, \bibinfo{author}{Xu, M.}, \bibinfo{author}{Liang,
  K.}, \bibinfo{author}{Chen, K.}, \bibinfo{author}{Wu, M.},
  \bibinfo{year}{2023}.
\newblock \bibinfo{title}{{Ariadne's} thread: Using text prompts to improve
  segmentation of infected areas from chest {X}-ray images}, in:
  \bibinfo{booktitle}{Medical Image Computing and Computer Assisted
  Intervention - {MICCAI} 2023}, \bibinfo{publisher}{Springer Nature
  Switzerland}, \bibinfo{address}{Cham}. pp. \bibinfo{pages}{724-733}.
\newblock \DOIprefix\doi{10.1007/978-3-031-43901-8_69}.

\end{thebibliography}

\end{document}